\documentclass[preprint,secnumarabic,amssymb, nobibnotes, 
aps, 
prl, 
]{revtex4-2}

\usepackage{graphicx,footmisc}
\usepackage[caption=false]{subfig}
\usepackage{color}
\usepackage{natbib,url}
\usepackage{amsmath}
\usepackage{float}
\usepackage[normalem]{ulem}

\newcommand*{\be}{\begin{equation}}
\newcommand*{\ee}{\end{equation}}

\def\begineq{\begin{equation}}
\def\endeq{\end{equation}}

\def\begineqn{\begin{equation*}}
\def\endeqn{\end{equation*}}

\def\beginar{\begin{eqnarray}}
\def\endar{\end{eqnarray}}
\def\beginarn{\begin{eqnarray*}}
\def\endarn{\end{eqnarray*}}

\def\lb{\left ( }
\def\rb{\right ) }
\def\lsq{\left [ }
\def\rsq{\right ] }

\def\ub{\mathbf{u}}

\def\Bb{\mathbf{B}}

\def\mfv{\overline{v}}

\def\bb{\mathbf{b}}

\def\dst{{\partial_t}}

\def\dsz{{\partial_z}}

\def\hz{{\bf\widehat z}}

\def\he{\widehat{{\boldsymbol{\eta}}}}

\def\hy{{\bf\widehat y}}
\def\hx{{\bf\widehat x}}

\def\lap{{\nabla_^2}}

\def\lap{{\nabla^2}}

\begin{document}

\title{Quasi-static magnetoconvection with a tilted magnetic field}%

\author{Justin A. Nicoski}
\affiliation{ 
Department of Physics, University of Colorado, Boulder, Colorado 80309, USA
}%
\author{Ming Yan}
\affiliation{ 
Department of Physics, University of Colorado, Boulder, Colorado 80309, USA
}%

\author{Michael A. Calkins}%
\affiliation{ 
Department of Physics, University of Colorado, Boulder, Colorado 80309, USA
}%

\begin{abstract}
A numerical study of convection with stress-free boundary conditions in the presence of an imposed magnetic field that is tilted with respect to the direction of gravity is carried out in the limit of small magnetic Reynolds number. The dynamics are investigated over a range of Rayleigh number $Ra$ and Chandrasekhar numbers up to $Q = 2\times10^6$, with the tilt angle between the gravity vector and imposed magnetic field vector fixed at $45^{\circ}$. For a fixed value of $Q$ and increasing $Ra$, the convection dynamics can be broadly characterized by three primary flow regimes:  (1) quasi-two-dimensional convection rolls near the onset of convection; (2) isolated convection columns aligned with the imposed magnetic field; and (3) unconstrained convection reminiscent of non-magnetic convection.
The influence of varying $Q$ and $Ra$ on the various fields is analyzed. Heat and momentum transport, as characterized by the Nusselt and Reynolds numbers, are quantified and compared with the vertical field case. Ohmic dissipation dominates over viscous dissipation in all cases investigated. Various mean fields are investigated and their scaling behavior is analyzed.
Provided $Ra$ is sufficiently large, all investigated values of $Q$ exhibit an inverse kinetic energy cascade that yields strong `zonal' flows. Relaxation oscillations, as characterized by a quasi-periodic shift in the predominance of either the zonal or non-zonal component of the mean flow, appear for sufficiently large $Ra$ and $Q$.
\end{abstract}

\maketitle

\section{Introduction}



Convection in the presence of externally imposed magnetic fields, or magnetoconvection (MC), is important in stars and planetary interiors \citep[][]{nW14}. Magnetic fields can lead to novel flow regimes relative to non-conducting fluids. Notably, imposed magnetic fields induce flow anisotropy. The resulting change in flow structure, however, is dependent on the direction and magnitude of the imposed magnetic field, implying that a rich variety of dynamics can be realized in MC. In the context of planets and stars, the magnetic field tends to be self-generated through dynamo action and is therefore spatially (and temporally) complex. It is therefore of interest to understand how field direction, in addition to field magnitude, influences the underlying convective motions. 

The periodic plane layer geometry provides a particularly simple system in which to study MC. The linear theory of MC for the plane layer is well established and provides a useful starting point for understanding the resulting nonlinear dynamics \citep{sC61}. The constant gravitational field is denoted by $\boldsymbol{g} = - g \hz$ (where $\hz$ is the unit vector pointing normal to the planar boundaries) and the uniform imposed magnetic field is $\Bb_0$. In the present work we focus solely on the limit in which the induced magnetic field is weak relative to the imposed field -- known as the quasi-static limit \citep[e.g.][]{bK08} -- we therefore limit our present discussion to this case. When $\boldsymbol{g}$ and $\Bb_0$ are aligned, which we refer to as vertical MC (VMC), convection is stabilized and the horizontal scale of the most unstable eigenmodes decreases with increasing field strength. When $\Bb_0$ is horizontal (HMC), the preferred mode consists of two-dimensional (2D) convection rolls with their axes aligned with $\Bb_0$. The general case of a tilted magnetic field (TMC) is essentially a mixture of these two previous cases in which the most unstable eigenmodes consist of two-dimensional rolls aligned with the horizontal component of the imposed magnetic field, but with a horizontal length scale that decreases with increasing magnetic field strength.

The nonlinear evolution of MC has been investigated both experimentally and numerically for a variety of field configurations. VMC dynamics has been studied for both the quasi-static limit, and for the case of arbitrarily large induced magnetic fields using numerical simulations \citep[e.g.][]{fC03}. In the discussion here we focus on those results pertaining to the quasi-static limit. The vertical field geometry tends to limit horizontal mixing when the field strength is sufficiently large given the preference for fluid to move along the magnetic field direction. In confined geometries, such as cylinders, distinct flows such as convective wall modes are also present \citep{wL18,tZ20,rA20}. Heat and momentum transport, while always weaker than non-magnetic Rayleigh-B\'enard convection (RBC), increases at a rate that is faster than RBC \citep{mY19,rA20}. For sufficiently strong buoyancy forcing, VMC data appears to approach the corresponding RBC data \citep{sC00,mY19,rA20}, which suggests an expected weakening dynamical role of the imposed magnetic field.

Studies of HMC show that 2D rolls persist over a significant range of parameter space and can yield heat transport that is more efficient than RBC \citep{sF81,uB02,aO03,tV21}. The flow eventually transitions to anisotropic 3D convection that can exhibit rich time dependent motions \citep{yT16,tV18,jY21}. In confined geometries the lateral walls can have a significant influence on the dynamics due to the formation of Hartmann boundary layers \citep{uB02}. These boundary layers stabilize the convection and lead to an increase in the temperature gradient required to initiate convection.

In comparison to VMC and HMC, less is presently known about the nonlinear behavior of TMC. Previous 2D studies of TMC have found mean flows and traveling wave solutions that are generated by the broken symmetry associated with the tilted magnetic field \citep{nH96}. This previous work suggests that the mean flows tend to travel in the direction of the tilt, though the dependence of the amplitude of this mean flow on the input parameters, namely the imposed field strength, has not been investigated in detail. In addition, to our knowledge the efficiency of heat and momentum transport has been unexplored for TMC. 

Rotating convection (RC) has been studied in great detail due to its relevance for planetary and stellar applications \citep{jmA15}. Like MC, RC with a vertical rotation vector stabilizes the convection and yields anisotropic structures \citep{sC61}. The inverse kinetic energy cascade is prevalent in RC provided that the influence of rotation is strong; the resulting flows are characterized by vortices that span the horizontal length of the system and are approximately invariant in the direction of the rotation axis \citep{kJ12,aR14,cG14,bF14}. When the two horizontal dimensions are unequal, the inverse cascade is instead manifested by the presence of large scale horizontally-directed jets \citep{kJ18}. Recent studies of RC with a tilted rotation axis find that both jets and vortices are present, depending on the tilt of the rotation axis (and likely also the relative importance of rotation and inertia) \citep{lN19,lC20}.

In the present work we carry out a systematic investigation of three-dimensional TMC using direct numerical simulations. Flow regimes are delineated, and heat and momentum transport are quantified over a range of imposed field strengths and buoyancy forcing. When possible, a comparison is made with recent VMC simulations; we find that the primary difference between TMC and VMC is the presence of magnetically constrained turbulent states in the former. We find that TMC, like rotating convection, yields an inverse cascade of kinetic energy, characterized by energetic (relative to the convection) mean flows that tend to be dominated by a meandering, alternating jet structure. In some cases we also find that these jets can become unstable and give rise to relaxation oscillations.

\section{Methods}

\begin{figure*}
\centering
\includegraphics[width=0.5\textwidth]{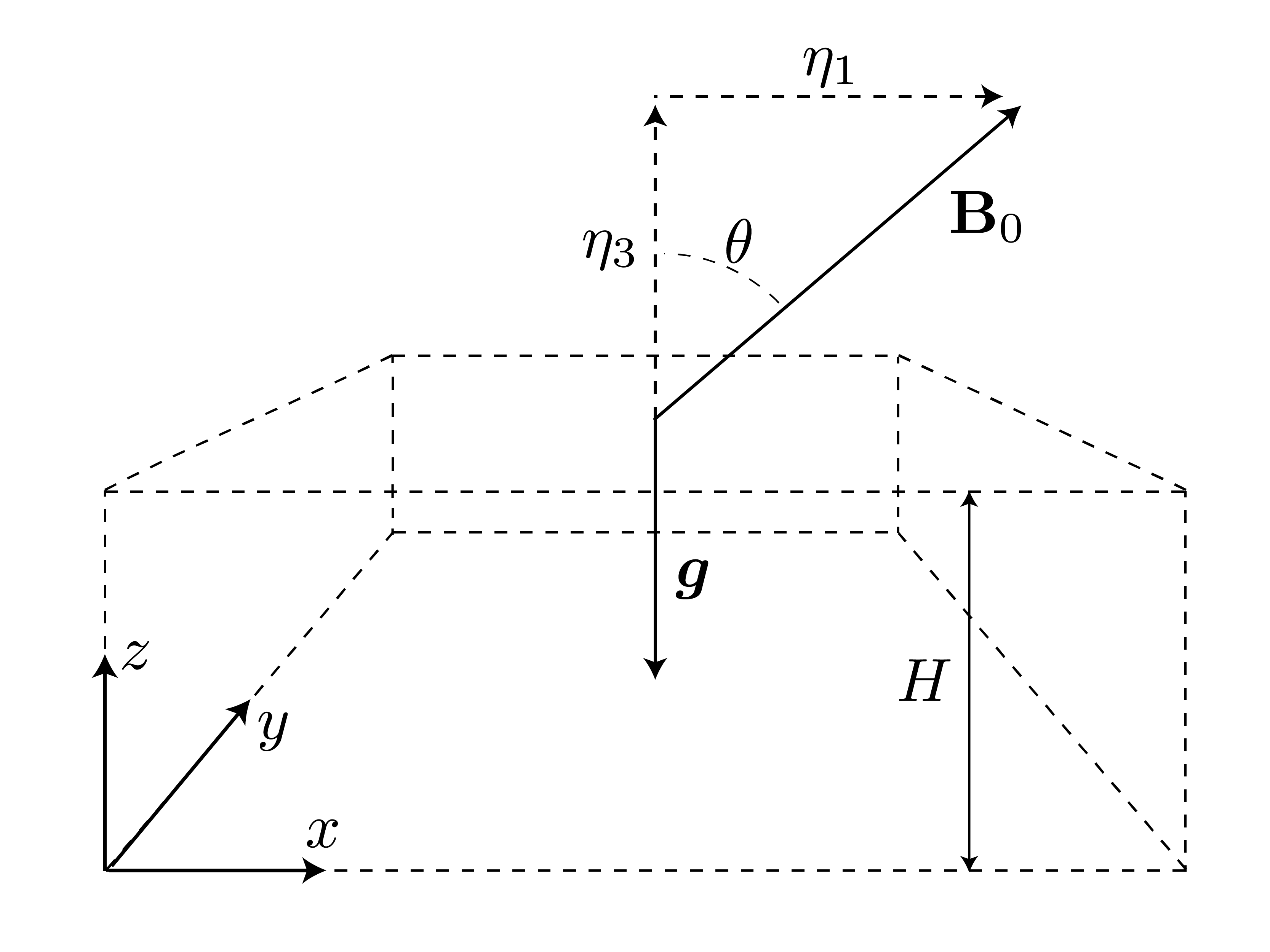}
\caption{Geometry used in the present study. The imposed magnetic field points in the direction $\he = \eta_1 \hx + \eta_3 \hz$, where $\eta_1 = \sin \theta$ and $\eta_3 = \cos \theta$. For the present study the angle is fixed to $\theta=45^{\circ}$.}
\label{F:geometry}
\end{figure*}

We employ a periodic plane layer geometry, as shown in Fig.~\ref{F:geometry}. The fluid layer has depth $H$ and the spatially uniform imposed magnetic field vector is given by
\be
\Bb_0 = B_0 \he,
\ee
where we define the vector
\be
\he = \eta_1 \, \hx + \eta_3 \, \hz,
\ee
with $\eta_1 = \sin \theta$ and $\eta_3 = \cos \theta$. The angle $\theta$ is measured relative to the vertical and fixed in the present study to $\theta = 45^{\circ}$.

The fluid is Oberbeck-Boussinesq with density $\rho$, kinematic viscosity $\nu$, thermal diffusivity $\kappa$, thermal expansion coefficient $\alpha$, magnetic diffusivity $\lambda$, and vacuum permeability $\mu$. The dimensional temperature difference between the bottom and top boundaries is denoted by $\Delta 
\mathcal{T} > 0$. The governing equations are non-dimensionalized with the viscous diffusion timescale $H^2/\nu$ and are given by
\be
D_t \ub  =   - \nabla P + \frac{Ra}{Pr} T \, \hz  + Q \, \he \cdot \nabla \bb +  \nabla^2 \ub,
\label{eq7}
\ee
\be
0=  \he \cdot \nabla \ub + \nabla^2\bb  ,
\label{eq8}
\ee
\be
D_t T = \frac{1}{Pr} \lap T,
\label{E:heat}
\ee
\be
\nabla \cdot \ub = 0,
\ee
\be
\nabla \cdot \bb = 0,
\label{E:db}
\ee
where the material derivative is denoted by $D_t ( \cdot ) = \dst ( \cdot ) + \ub \cdot \nabla ( \cdot )$,  $\mathbf{u} = \left(u,v,w\right)$ is the velocity field,  
$\mathbf{b}$ is the induced magnetic field, $T$ is the temperature, and $P$ is the reduced pressure.

The non-dimensional parameters appearing in the above equations are the Rayleigh number, the thermal Prandtl number, and the Chandrasekhar number defined by, respectively
\be
Ra = \frac{\alpha g \Delta \mathcal{T} H^3}{\kappa \nu}, \quad Pr = \frac{\nu}{\kappa}, \quad Q = \frac{B_0^2H^2}{\rho \nu \mu \lambda} .
\ee
In all cases presented here the thermal Prandtl number is fixed at $Pr=1$, whereas $Ra$ and $Q$ are both varied.

We use the quasi-static magnetohydrodynamic approximation in which the induced magnetic field is asymptotically smaller than the imposed magnetic field. This approximation is equivalent to assuming that $Rm=RePm \rightarrow 0$, where $Rm$ is the magnetic Reynolds number, $Re$ is the Reynolds number and $Pm = \nu/\lambda$ is the magnetic Prandtl number. The quasi-static approximation is therefore valid when $Re \ll Pm^{-1}$; this condition is typically satisfied in laboratory experiments that use liquid metals \citep[e.g.][]{jmA01,uB02,tV18}.

The mechanical boundary conditions are impenetrable and stress-free, constant temperature and electrically insulating, which can be written as
\be
w = \frac{\partial u}{\partial z} =  \frac{\partial v}{\partial z} = 0 \quad \mbox{at} \quad z = 0,1.
\label{E:bc}
\ee
The thermal boundary conditions are constant temperature
\be
T = 1 \quad \mbox{at} \quad z = 0, \qquad T = 0 \quad \mbox{at } \quad z = 1. 
\ee
Electrically insulating electromagnetic boundary conditions are used such that the magnetic field is matched to a potential field at $z=0$ and $z=1$ \citep[e.g.~see][]{cJ00b}.

The most unstable eigenmodes consist of longitudinal rolls oriented parallel to the imposed magnetic field with marginal Rayleigh number
\be
Ra_m = \frac{\lb \pi^2 + k^2 \rb}{k^2} \lsq \lb \pi^2 + k^2 \rb^2  + \pi^2 \eta_3^2 Q  \rsq .
\ee
Minimizing the above expression gives
\be
2{k_c}^6+3\pi^2{k_c}^4 - \pi^4Q \eta_3^2 - \pi^6 = 0, 
\ee
where $k_c$ is the critical horizontal wavenumber. For the field strengths used in the present study, $Q=(2\times10^3, 2\times10^5, 2\times10^6)$, we find $k_c \approx (5.6842, 12.8343, 18.9823)$ and $Ra_c \approx (1.5207\times10^4, 1.0784\times10^6, 1.0281\times10^7)$.  In the asymptotic limit $Q \rightarrow \infty$, the critical wavenumber and the critical Rayleigh number are given by, respectively,
\be
k_c^{(a)} \rightarrow \left(\frac{1}{2}\pi^4Q\eta_3^2\right)^{1/6}, \qquad Ra_c^{(a)} \rightarrow \pi^2 \eta_3^2 Q. \qquad \lb Q \rightarrow \infty \rb
\ee

Both the velocity and magnetic fields are represented in terms of poloidal and toroidal scalars such that the solenoidal conditions are satisfied exactly. The resulting equations are solved numerically with a pseudo-spectral algorithm using Fourier series in the horizontal directions and Chebyshev polynomials in the vertical direction. A third-order implicit-explicit Runge-Kutta time stepping method is used. Further details of the code are given in \citep{pM16}.

We scale the horizontal length of the computational domain, $L$, in integer multiples of the critical horizontal wavelength, $\lambda_c=2\pi/k_c$, such that the aspect ratio is given by
\be
\Gamma = n \lambda_c.
\ee
We found that $n=10$ is sufficient for convergence of bulk quantities such as the Nusselt number and the Reynolds number. Thus, for the imposed field strengths used in the present study, the corresponding aspect ratios are given by $\Gamma \approx  (11.1, 4.9, 3.3)$.

\section{Results}

\subsection{Parametric overview and flow regimes}

\begin{figure}
 \begin{center}
\includegraphics[width=0.5\textwidth]{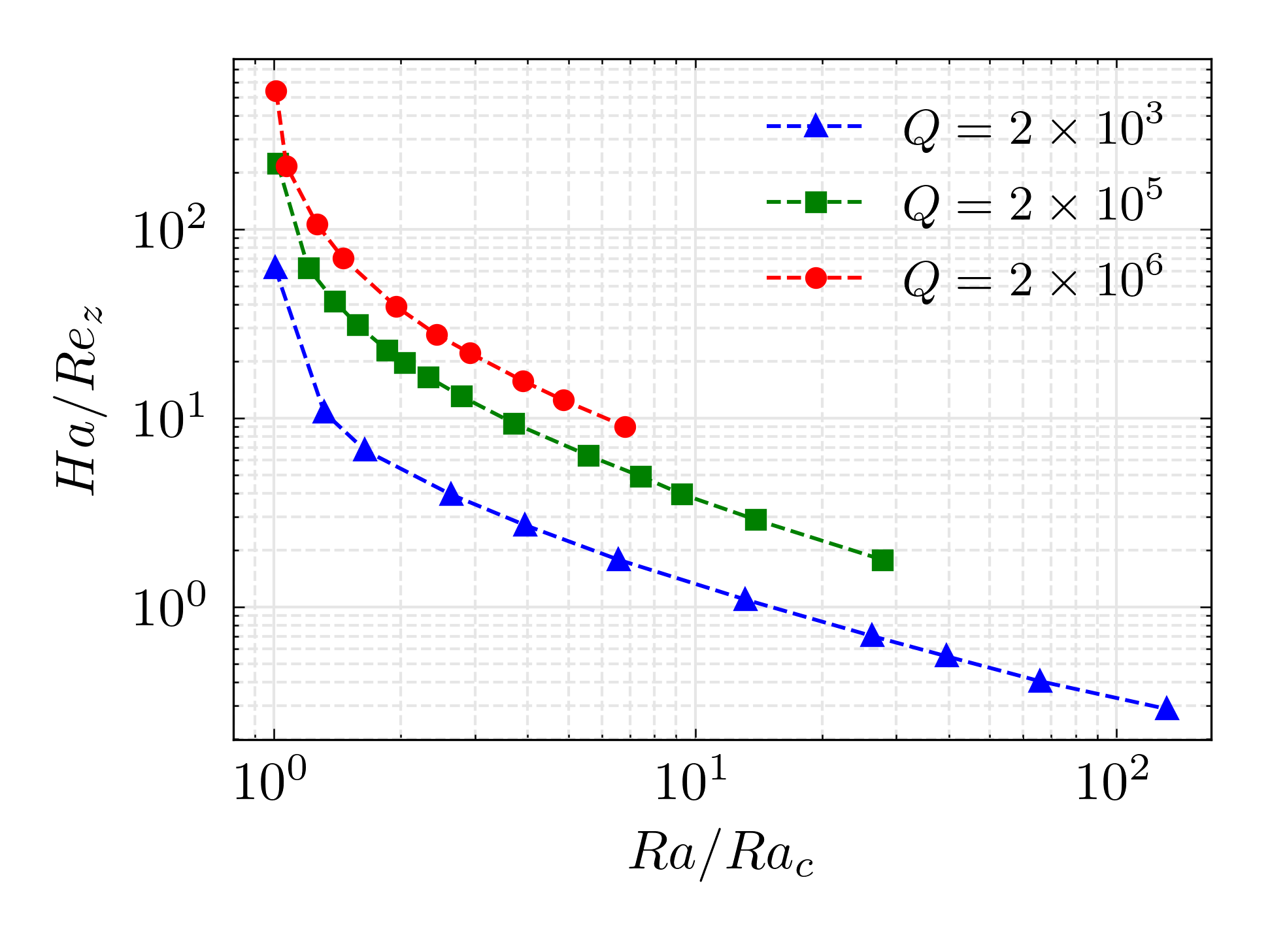}
	 \caption{Parametric overview of the simulations, as characterized by the ratio $Ha/Re_z$, where $Ha = \sqrt{Q}$ is the Hartmann number and $Re_z$ is the Reynolds number based on the rms of the vertical component of the velocity field. The critical Rayleigh number is denoted by $Ra_c$. The ratio $Ha/Re_z$ represents the relative size of the Lorentz force and inertia: cases with $Ha/Re_z \gtrsim O(1)$ are considered magnetically constrained. }
      \label{F:HartRey}
\end{center}
\end{figure}

We investigate imposed magnetic field strengths of $Q= (2\times 10^3, 2\times 10^5, 2\times 10^6)$, which were chosen so that the critical Rayleigh number, $Ra_c$, for each  field strength corresponds with those used in recent VMC simulations \citep{mY19}. We explore Rayleigh numbers up to $Ra \approx ( 132Ra_c, 28Ra_c, 7 Ra_c)$ for each of the three values of $Q$. Data for the simulations is summarized in Table \ref{tab:Data}. While spatial resolution requirements are significant for many of the simulated flows, we find that the primary limiting factor for the computations done at larger values of $Q$ and $Ra$ is the presence of slowly evolving relaxation oscillations that require prohibitively long computation times to obtain converged statistics. 

A useful measure for characterizing the relative importance of the imposed magnetic field is the ratio, $Ha/Re_z$, where the Hartmann number is defined as $Ha = \sqrt{Q}$ and $Re_z$ is the Reynolds number based on the rms of the vertical velocity component. The ratio $Ha/Re_z$ represents the relative influence of the Lorentz force to inertial forces in the quasi-static limit \citep[e.g.][]{jH71,bK08}. By restricting the Reynolds number to only include the vertical component of the velocity field, we are attempting to better characterize the relative influence of the magnetic field on the small scale convection. As discussed later, mean flows develop that have magnitudes significantly larger than the vertical component of the velocity field. Fig.~\ref{F:HartRey} shows this ratio for all of the simulations. Cases in which $Ha/Re_z \gg 1$ are considered magnetically constrained in the sense that the Lorentz force enters the leading order force balance \citep[cf.][]{mY19}. For values of $Ha/Re_z \lesssim 1$, inertia plays a leading order dynamical role and the resulting motions are only weakly influenced by the imposed magnetic field. In the present study we only find cases with $Ha/Re_z \lesssim 1$ for the smallest field strength of $Q = 2 \times 10^3$; for these cases we find that the convective structures are no longer aligned with the (tilted) magnetic field.

\begin{figure}
 \begin{center}
      \subfloat[][]{\includegraphics[width=0.32\textwidth]{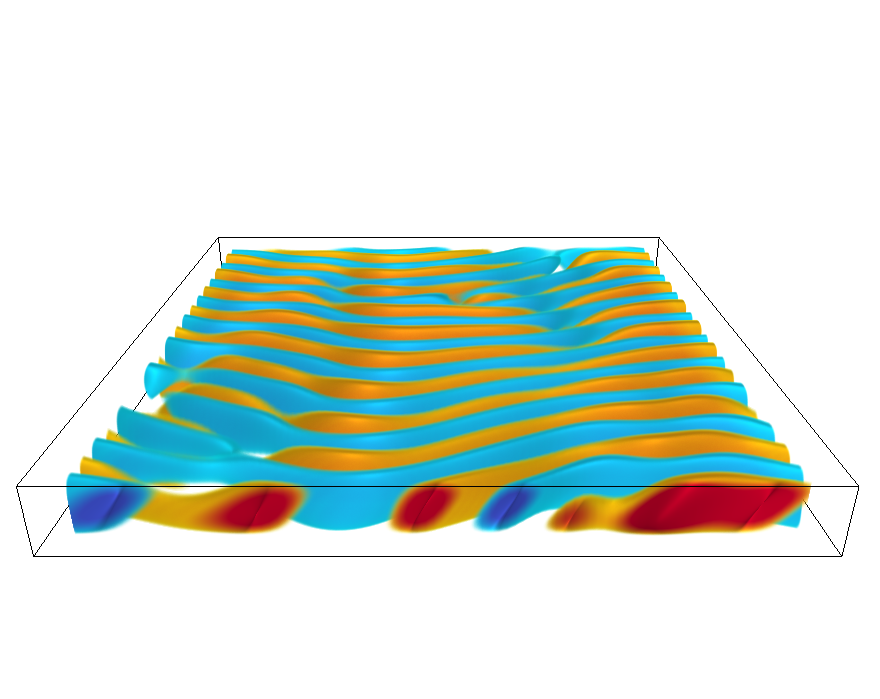}} 
      \subfloat[][]{\includegraphics[width=0.32\textwidth]{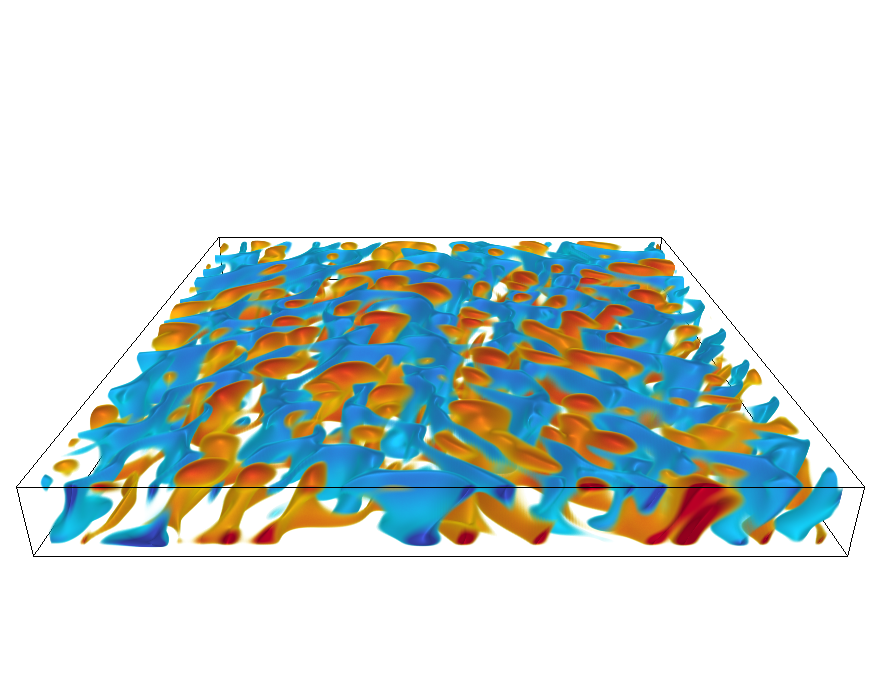}} 
      \subfloat[][]{\includegraphics[width=0.32\textwidth]{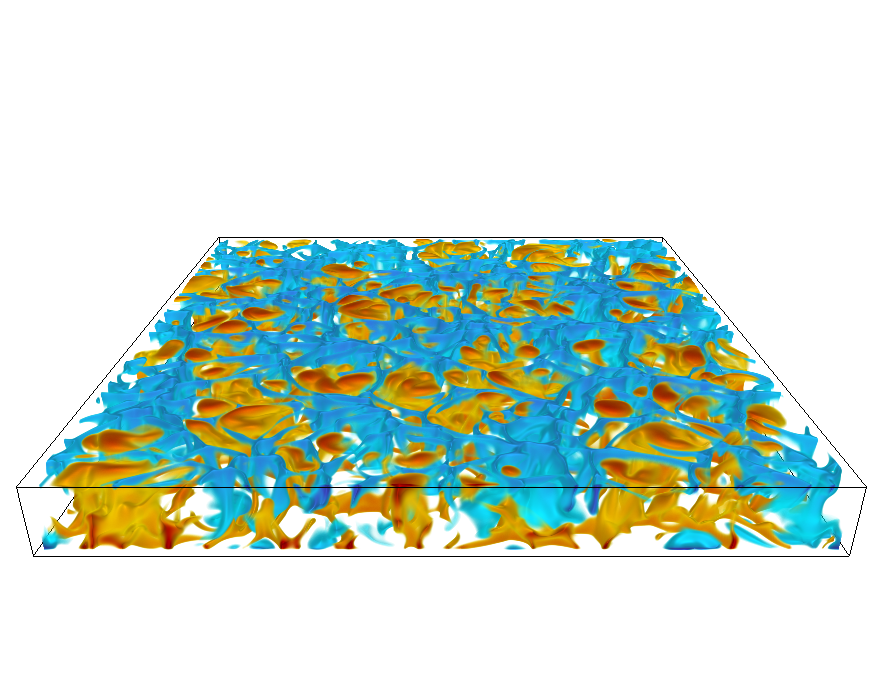}} \\ 
      
      \subfloat[][]{\includegraphics[width=0.32\textwidth]{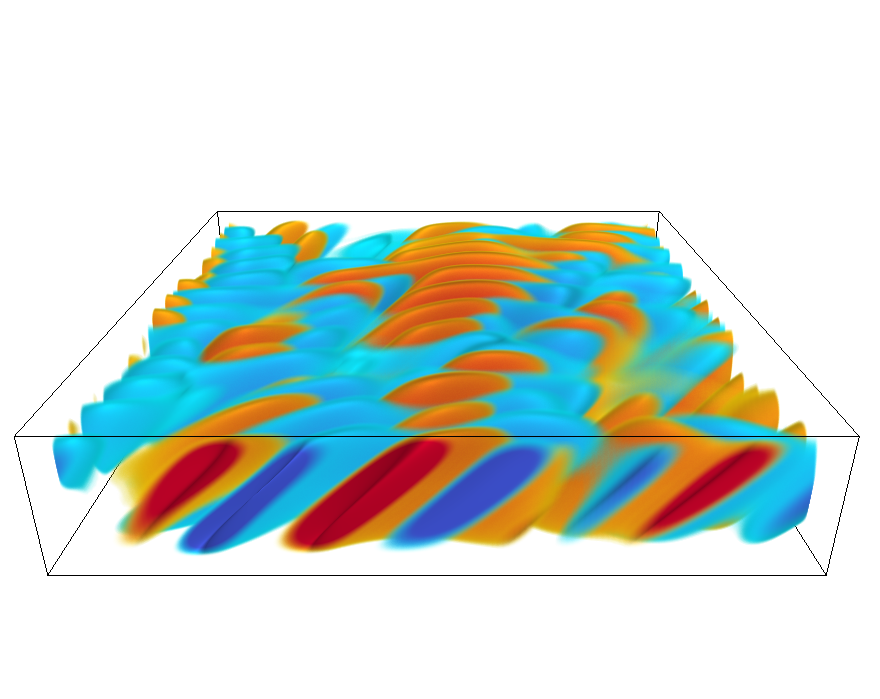}} 
      \subfloat[][]{\includegraphics[width=0.32\textwidth]{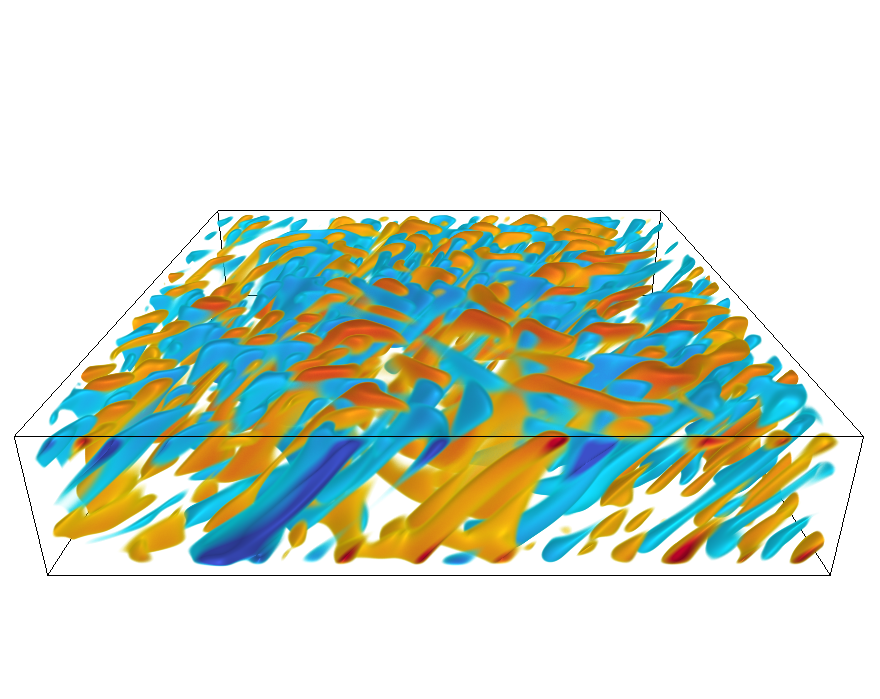}} 
      \subfloat[][]{\includegraphics[width=0.32\textwidth]{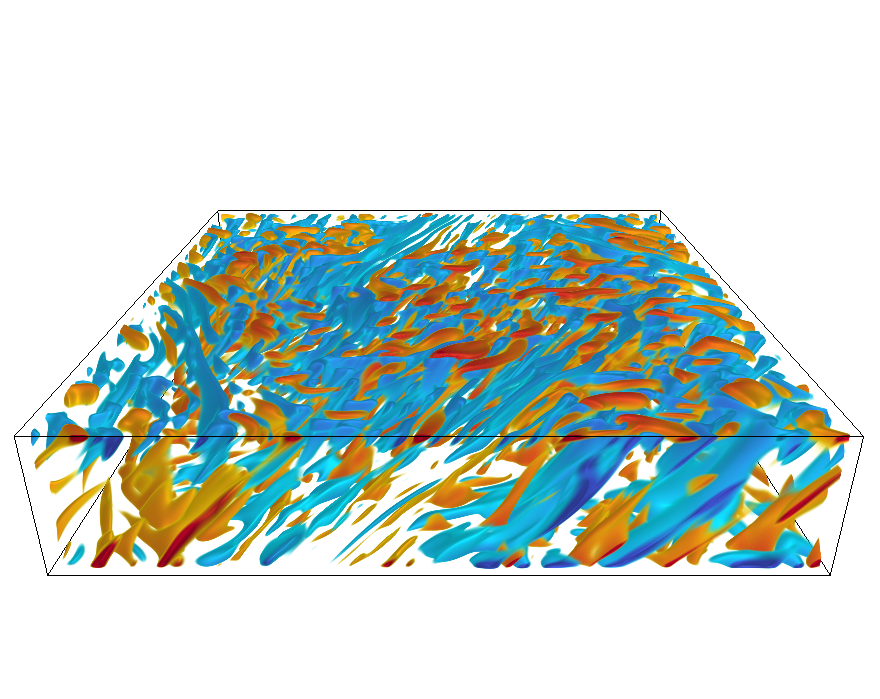}} \\  
        
      \subfloat[][]{\includegraphics[width=0.32\textwidth]{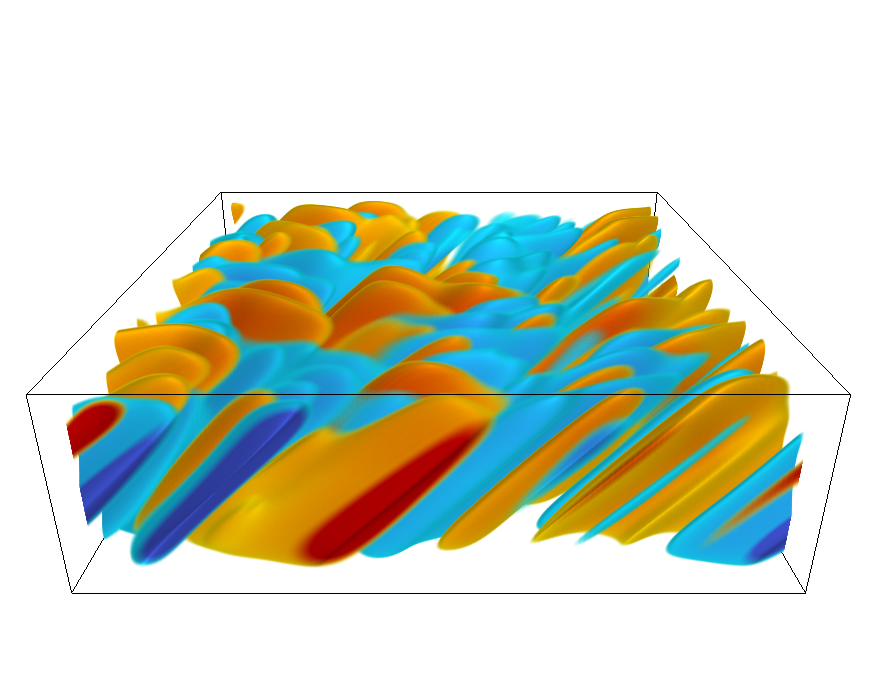}} 
      \subfloat[][]{\includegraphics[width=0.32\textwidth]{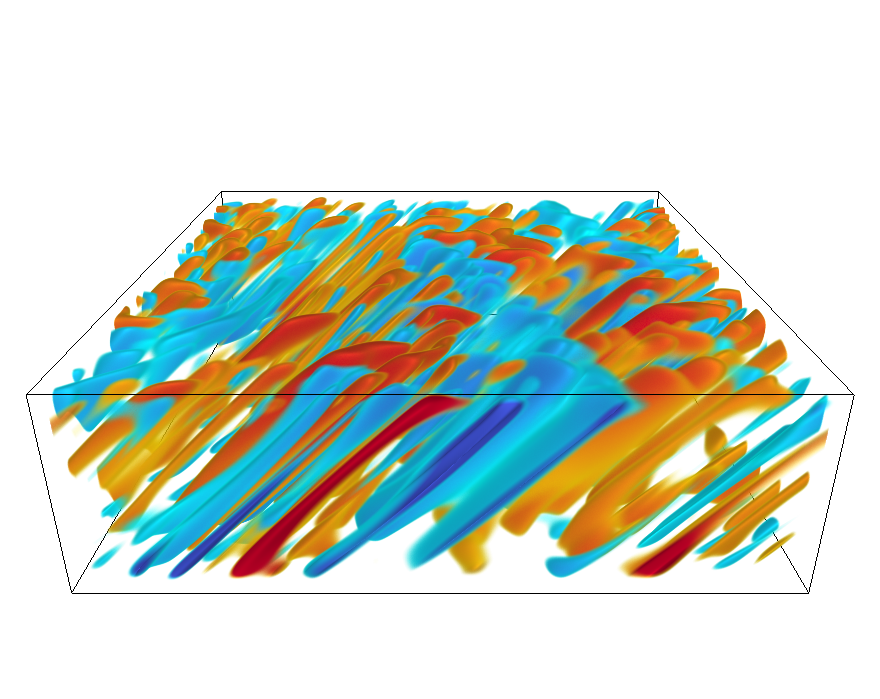}} 
      \subfloat[][]{\includegraphics[width=0.32\textwidth]{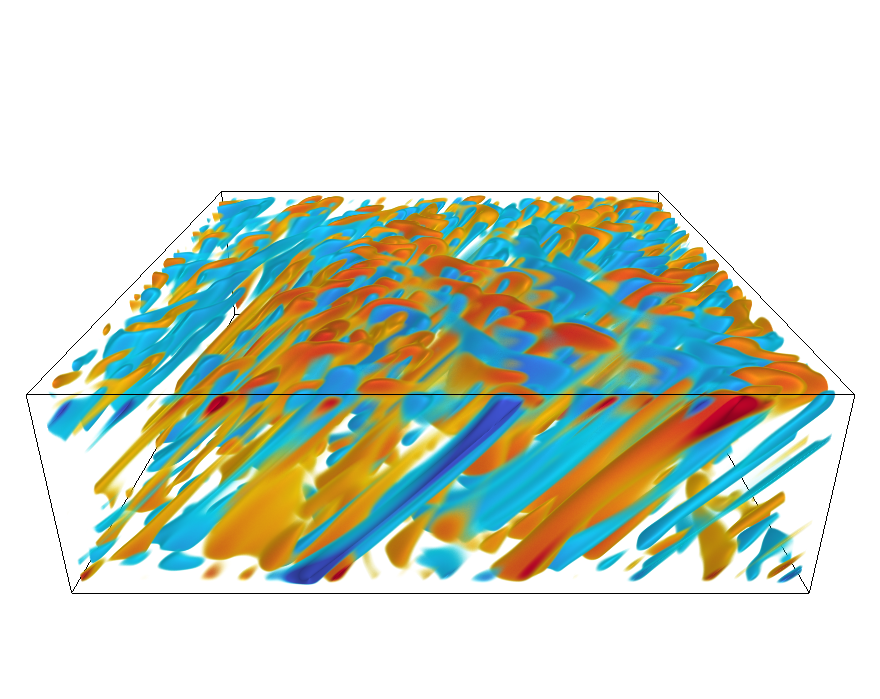}}  
	 \caption{Volumetric renderings of the fluctuating temperature for various cases. Top row ($Q=2\times 10^3$): (a) $Ra/Ra_c = 1.3$; (b) $Ra/Ra_c = 13.2$; (c) $Ra/Ra_c = 132$. Middle row ($Q=2\times 10^5$): (d) $Ra/Ra_c = 1.4$; (e) $Ra/Ra_c = 7.4$; (f) $Ra/Ra_c = 27.8$. Bottom row ($Q=2\times 10^6$): (g) $Ra/Ra_c = 1.5$; (h) $Ra/Ra_c = 3.9$; (i) $Ra/Ra_c = 6.8$. }
      \label{F:TempFluct}
\end{center}
\end{figure}

To illustrate the structure of the flow field as both $Q$ and $Ra$ are varied  (and therefore also $Ha/Re_z$) we show volumetric renderings of the fluctuating temperature in Fig.~\ref{F:TempFluct} for each of the three values of $Q$ and three particular values of $Ra$. As predicted by linear theory, for $Ra \approx Ra_c$, we observe anisotropic convective rolls that are predominantly aligned with the $x$-component of the imposed magnetic field -- these structures are evident in Fig.~\ref{F:TempFluct}(a) and to a lesser degree in panels (d) and (g). As the Rayleigh number is increased we find that the rolls develop a large scale, $k_x = 1$, modulation. This modulation interacts nonlinearly with the convective rolls and leads to the formation of a large scale mean flow that is discussed in more detail below. Further increases in $Ra$ lead to the development of convective columns that are aligned with the imposed magnetic field. These structures are particularly evident in the $Q=2 \times 10^5$ and $Q=2 \times 10^6$ cases, e.g.~Figures \ref{F:TempFluct}(e,f,h,i). These cases also show that the convective structures are elongated in the $x$-direction. For $Q=2\times10^3$ the tilt of the convective structures is less noticeable; in Fig.~\ref{F:TempFluct}(b) some tilt is observable, though for sufficiently large Rayleigh number the tilt is no longer obvious (Fig.~\ref{F:TempFluct}(c)). Despite the lack of constraint in simulations with $Ha/Re_z \lesssim 1$, we find that the magnetic field still plays an important dissipative role, as discussed in the next section.

\subsection{Heat and momentum transport}

\begin{figure}[]
 \begin{center}
      \subfloat[][]{\includegraphics[width=0.5\textwidth]{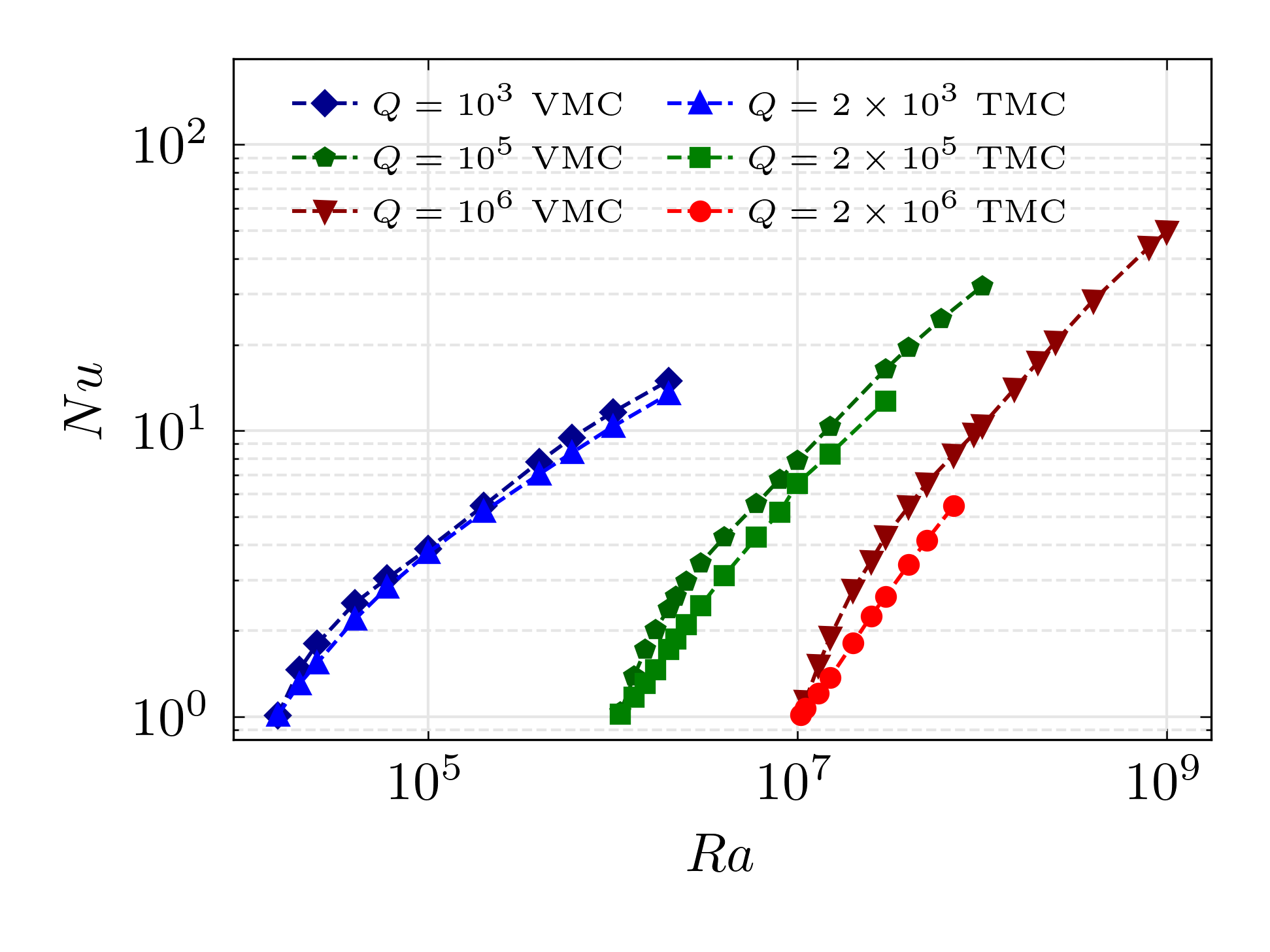}}
      \subfloat[][]{\includegraphics[width=0.5\textwidth]{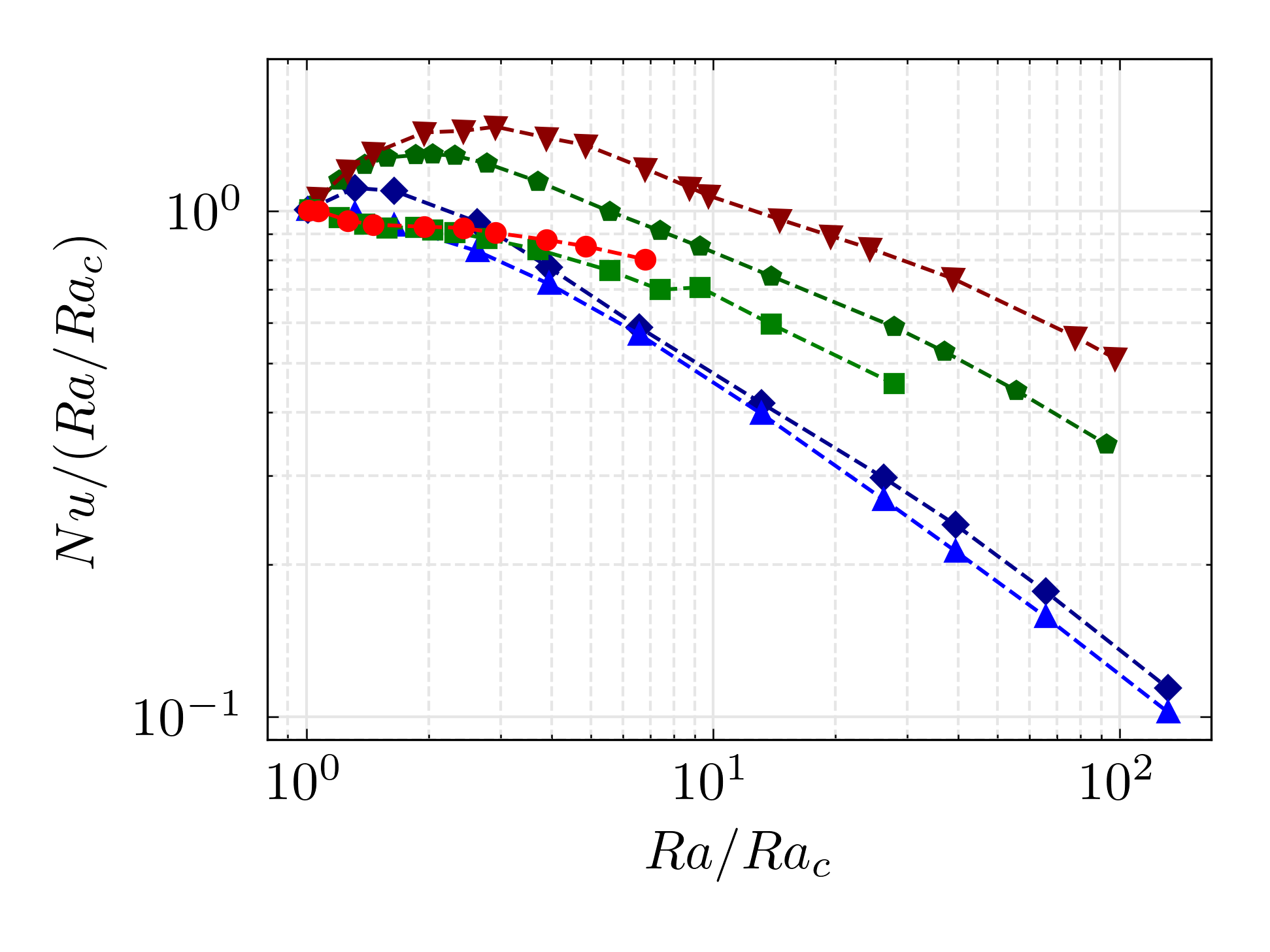}}
	 \caption{ Nusselt number data: (a) $Nu$ versus $Ra$; (b) compensated Nusselt number, $Nu (Ra_c/Ra)$, versus $Ra/Ra_c$. Tilted magnetoconvection cases are denoted by TMC, and the vertical field cases from Ref.~\citep{mY19} are denoted by VMC.}
      \label{F:Nu}
\end{center}
\end{figure}

Fig.~\ref{F:Nu} shows the Nusselt number, defined by
\be
Nu = 1 + Pr\langle wT'\rangle,
\ee
where the angled brackets, $\langle \cdot \rangle$, denote a volume and time average. Both the new TMC data and the VMC data from Ref.~\citep{mY19} are shown for comparison. The slope of the $Nu$-$Ra$ data increases with $Q$, similar to VMC. However, in VMC there is a steeper increase in $Nu$ near the onset of convection and a subsequent reduction in the growth rate of $Nu$ with increasing $Ra$. This difference between VMC and TMC is exhibited in the corresponding compensated value, $Nu/(Ra/Ra_c)$, as shown in Fig.~\ref{F:Nu}(b). The $Nu \sim Ra/Ra_c$ scaling behavior is independent of viscous dissipation, which might be expected in the limit $Q \rightarrow \infty$ \citep[e.g.][]{sB06}. However, we find that viscous dissipation becomes important in all cases just after the onset of convection, which suggests a possible explanation for why such a scaling is not observed. For sufficiently large values of $Ra/Ra_c$, both VMC and TMC show similar scaling behavior in $Nu$, at least for $Q=2\times10^3$ and $Q=2\times10^5$.


\begin{figure}[t]
 \begin{center}
      \subfloat[][]{\includegraphics[width=0.33\textwidth]{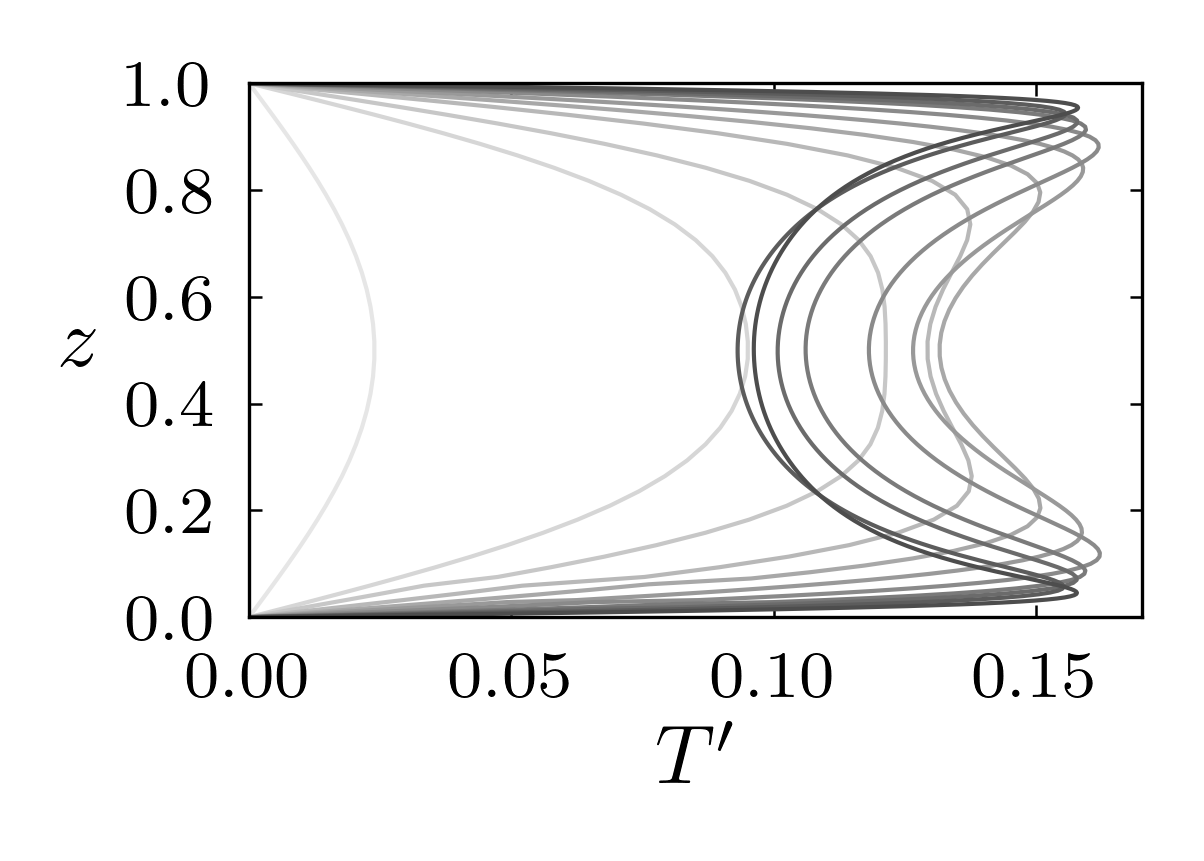}} 
      \subfloat[][]{\includegraphics[width=0.33\textwidth]{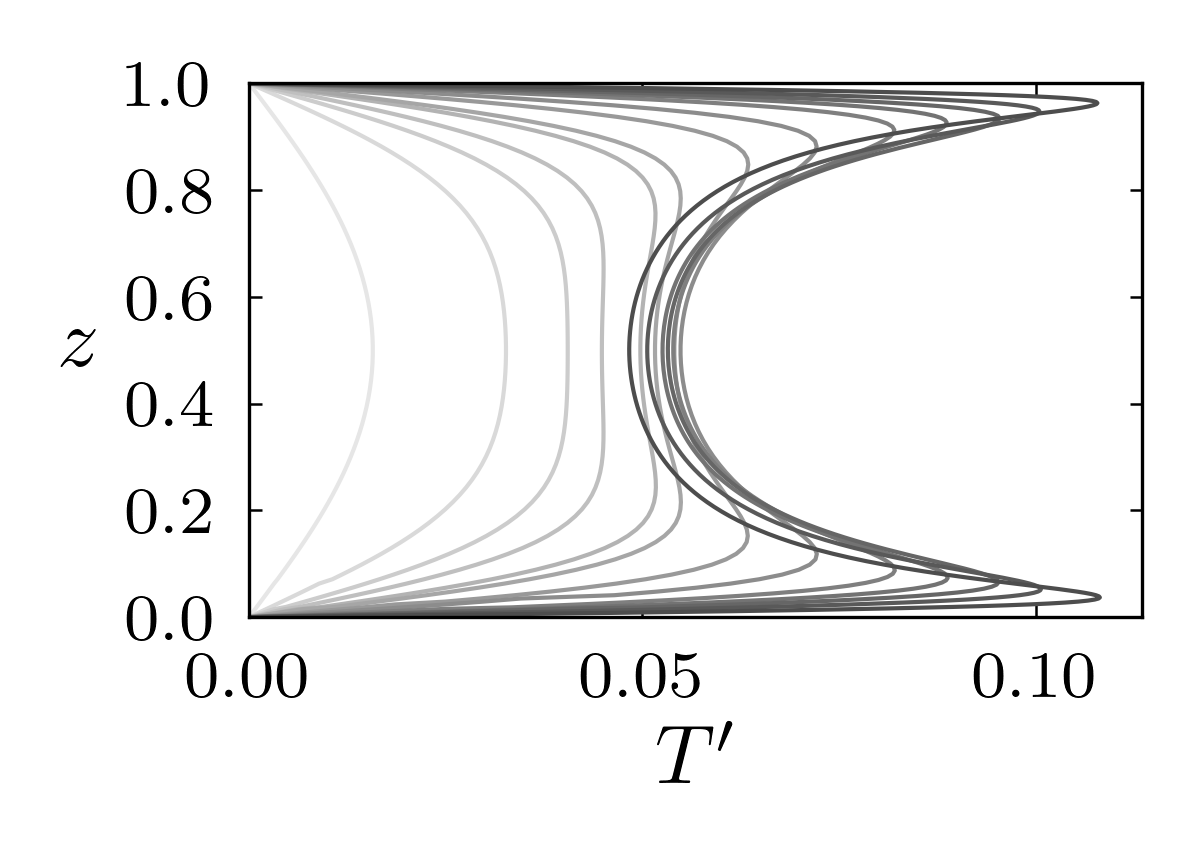}} 
      \subfloat[][]{\includegraphics[width=0.33\textwidth]{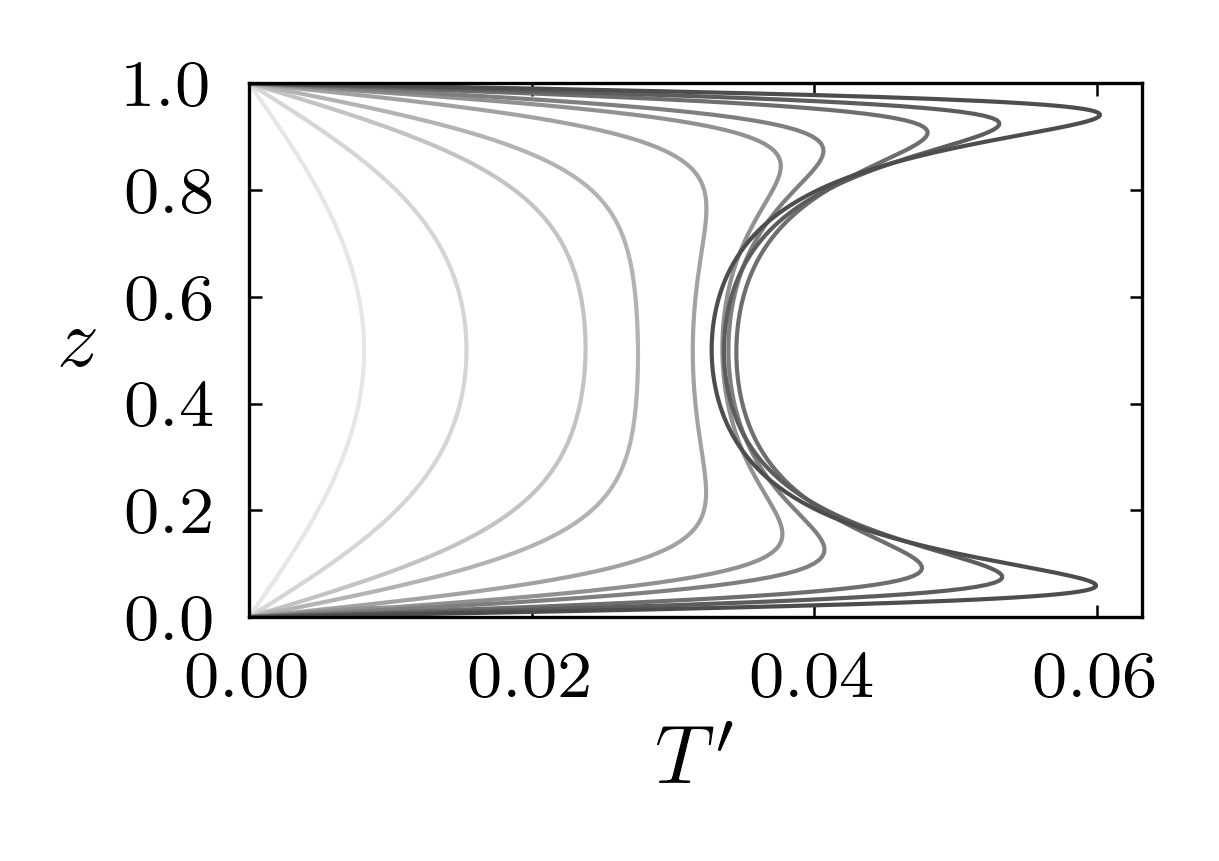}} \\
	 \caption{Vertical profiles of the rms temperature fluctuation: (a) $Q=2\times 10^3$; (b) $Q=2\times 10^5$; (c) $Q=2\times 10^6$. Darker greyscale lines correspond to larger Rayleigh numbers.}
      \label{F:T_profiles}
\end{center}
\end{figure}

Vertical profiles of the rms temperature fluctuation are shown for all values of $Q$ in Fig. \ref{F:T_profiles}. For a fixed value of $Q$ we find that the temperature fluctuation within the interior begins to decrease for sufficiently large $Ra/Ra_c$ once robust thermal boundary layers form; the growth of $T'$ with increasing $Ra/Ra_c$ is then achieved within the vicinity of the thermal boundary layer. Comparing the profiles for different values of $Q$ reveals that the rms temperature is generally a decreasing funtion of $Q$, as predicted by asymptotic theory \citep{pM99,kJ99c,kJ00}. Although not shown, we find that for all values of $Q$ the horizontally averaged temperature shows a trend toward a nearly isothermal bulk with well developed thermal boundary layers as $Ra$ is increased. These results indicate that, like RBC and VMC \citep{mY19}, the heat transfer is ultimately limited by the thermal boundary layers in TMC. This data suggests that the initial rapid growth in $Nu$ for small $Ra/Ra_c$ is due to the formation of thermal boundary layers.



\begin{figure}
 \begin{center}
      \subfloat[][]{\includegraphics[width=0.33\textwidth]{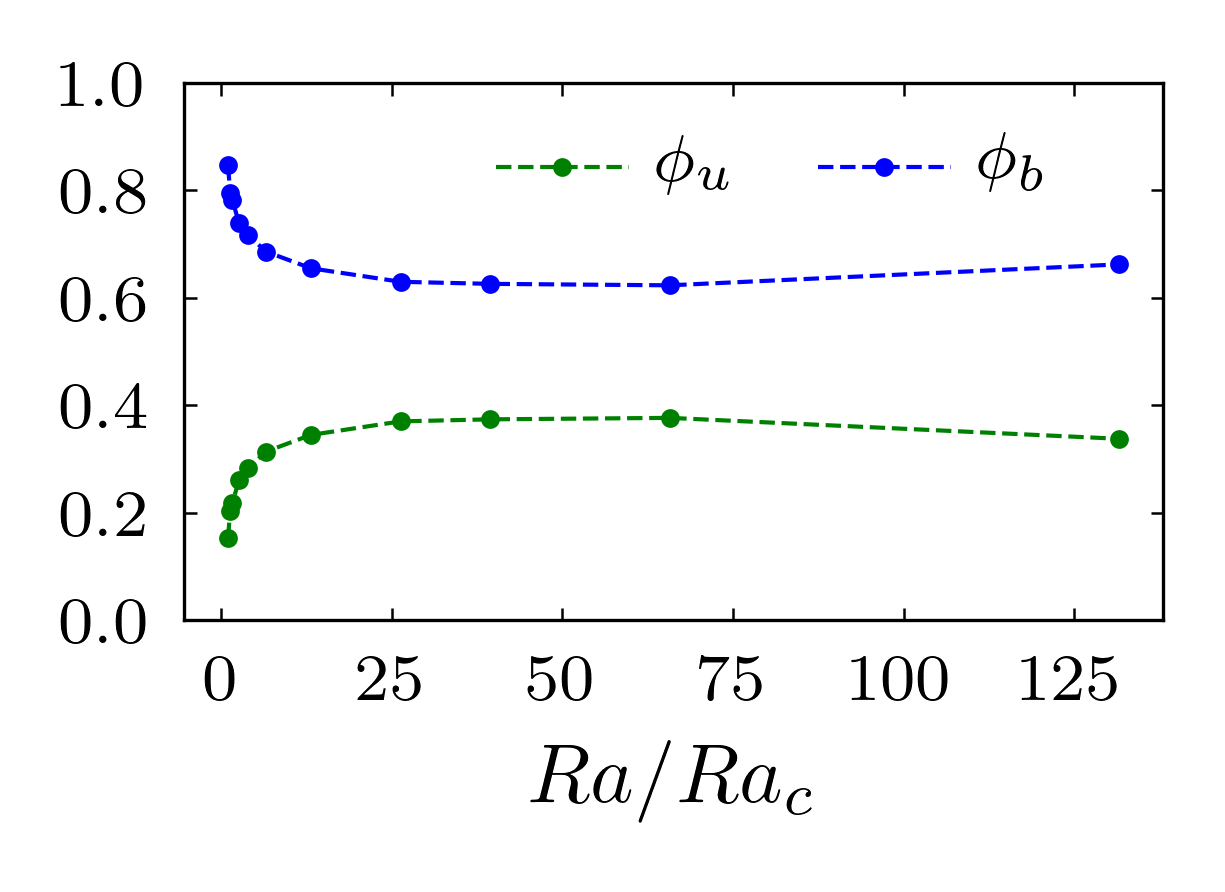}}
      \subfloat[][]{\includegraphics[width=0.33\textwidth]{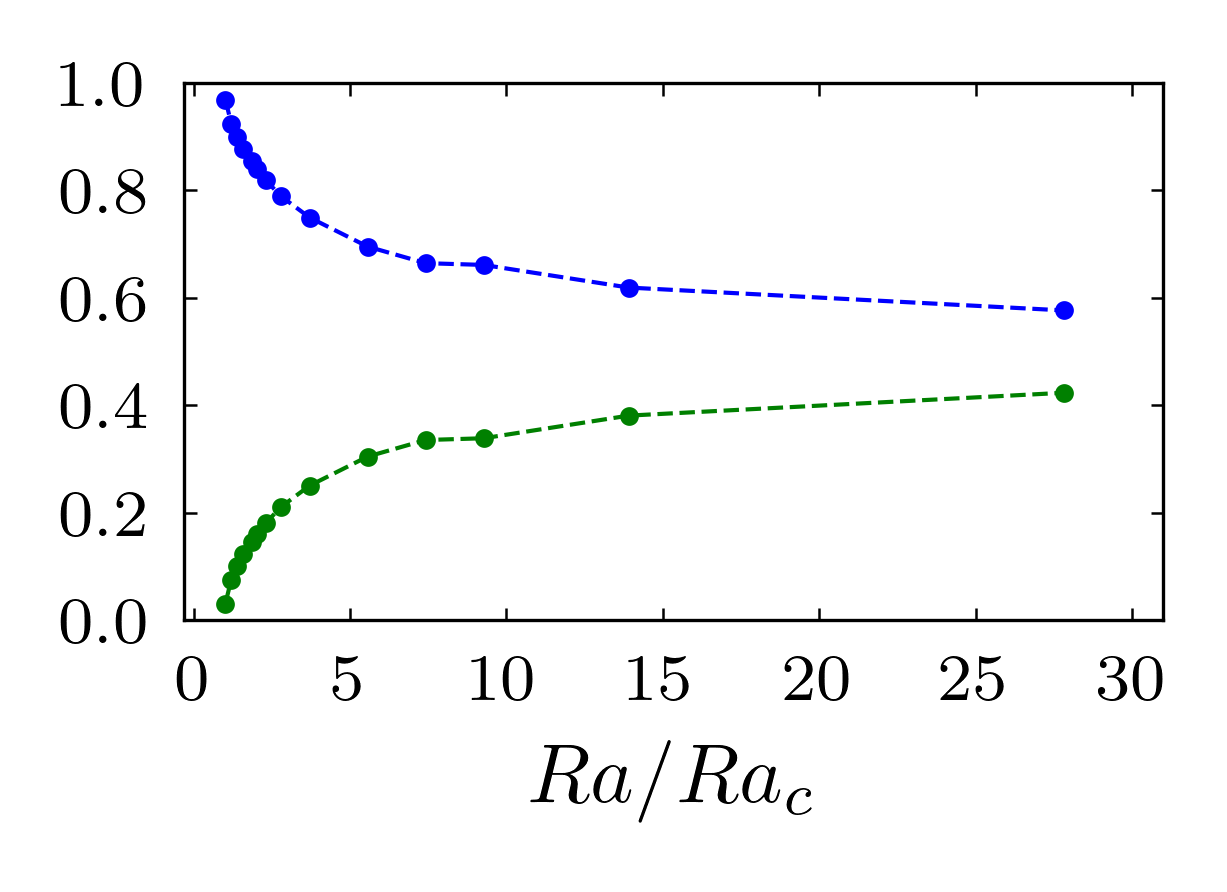}}
      \subfloat[][]{\includegraphics[width=0.33\textwidth]{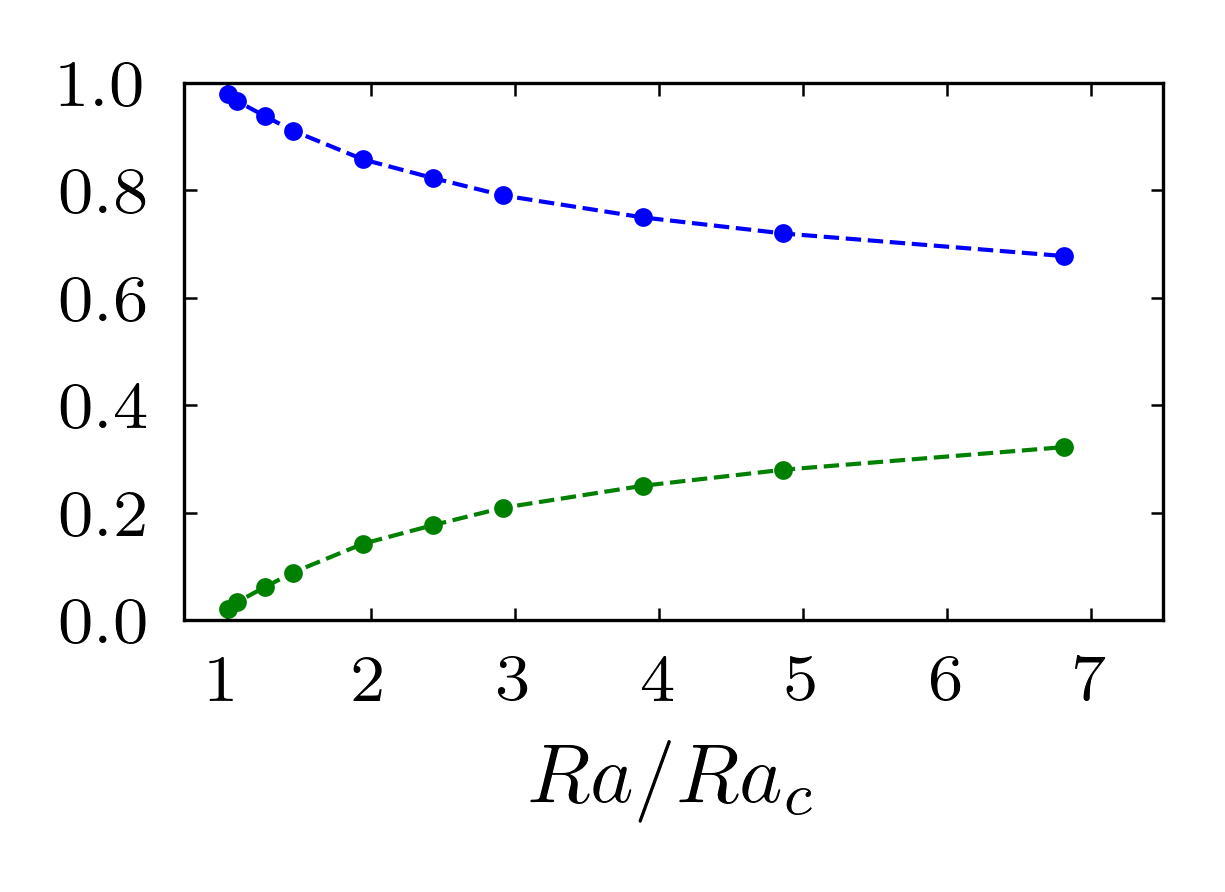}}      
\caption{Dissipation ratios: (a) $Q=2\times 10^3$; (b) $Q=2\times 10^5$; (c) $Q=2\times 10^6$. Note the difference in scale on the horizontal axis.}
\label{F:DissipationRatio}
\end{center}
\end{figure}

The Nusselt number and the energy dissipation are related via
\be
    \left(Nu - 1\right)\frac{Ra}{Pr^2} =  \varepsilon_u + \varepsilon_b ,
\ee
where the viscous and ohmic dissipation are given by $\varepsilon_u =  \langle \boldsymbol{\zeta}^2\rangle$ and
$\varepsilon_b = Q\langle\mathbf{j}^2\rangle$, respectively. The vorticity vector is denoted by $\boldsymbol{\zeta} =\nabla \times \mathbf{u}$ and the current density is $\mathbf{j}=\nabla \times \mathbf{b}$. 

It is helpful to define the ratios of viscous and ohmic dissipation according to
\be
   \phi_u = \frac{\varepsilon_u}{\varepsilon_b + \varepsilon_u}, \qquad \phi_b = \frac{\varepsilon_b}{\varepsilon_b + \varepsilon_u},
\ee
where we note that $\phi_u + \phi_b= 1$. The dissipation ratios are shown in Fig.~\ref{F:DissipationRatio} for all values of $Q$. For all cases, near the onset of convection we observe a rapid decrease (increase) in the ohmic (viscous) dissipation ratio as $Ra/Ra_c$ is increased. For $Q=2\times10^3$, there is an approximate saturation in both dissipation ratios in the range $25 \lesssim Ra/Ra_c \lesssim 75$. For the largest value of $Ra/Ra_c \approx 132$ we find a slight increase (decrease) in the ohmic (viscous) dissipation ratio. We note that within the saturated regime we find significant mean flows, though the mean flow is negligibly small at the largest value of $Ra/Ra_c \approx 132$. This increase (decrease) in the ohmic (viscous) dissipation ratio may be due to the strongly 3D flow that occurs, thus leading to significant induced magnetic field and associated current. The data suggests that viscous dissipation is important for all $Q$, though it is always smaller than ohmic dissipation.

\begin{figure}
 \begin{center}
      \subfloat[][]{\includegraphics[width=0.45\textwidth]{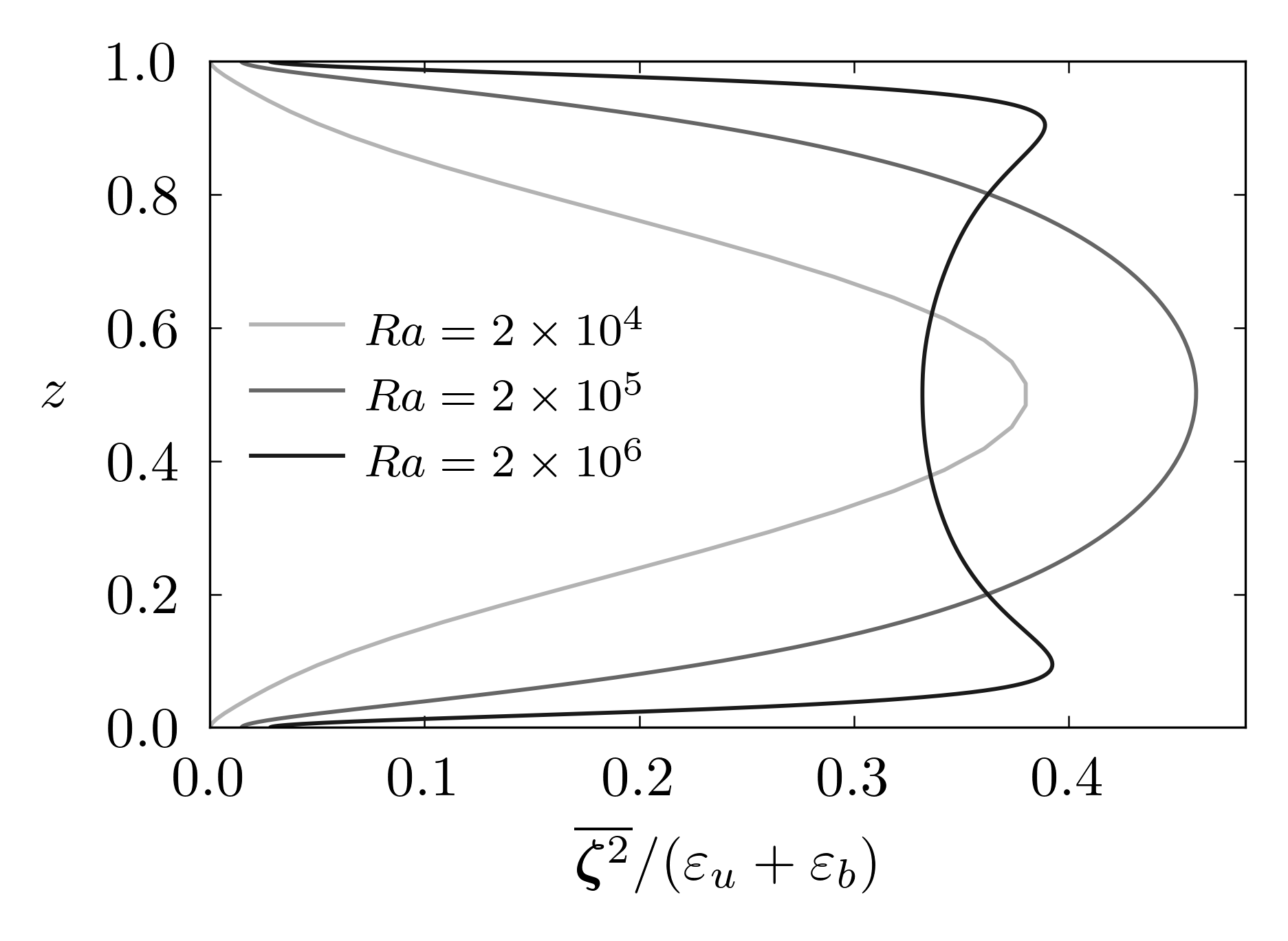}} \qquad
      \subfloat[][]{\includegraphics[width=0.45\textwidth]{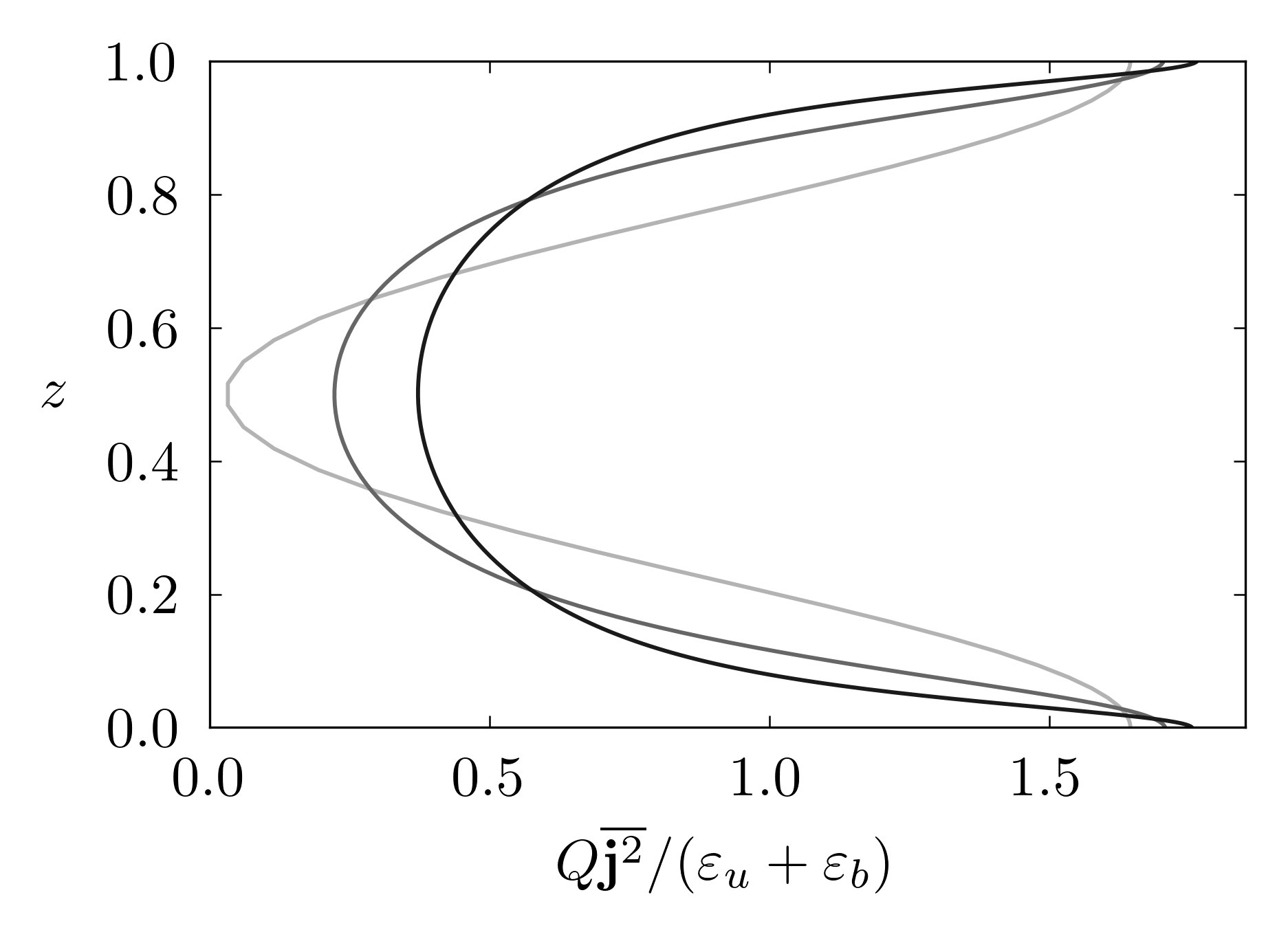}}      
\caption{Normalized dissipation profiles averaged in time for select cases from $Q=2\times 10^3$. (a) normalized viscous dissipation;  (b) normalized ohmic dissipation.}
\label{F:DissipationProfile}
\end{center}
\end{figure}

To examine the depth-dependence of the dissipation we compute horizontally averaged profiles of the squared vorticity and current density, i.e.~$\overline{\boldsymbol{\zeta}^2}$ and $Q \overline{\mathbf{j}^2}$. Fig.~\ref{F:DissipationProfile} shows these profiles, normalized by the total dissipation, for $Q=2\times10^3$ and three different values of $Ra$. Due to the stress-free mechanical boundary conditions we find that viscous dissipation is dominant within the interior of the domain and momentum boundary layers are evident. In contrast, we find that ohmic dissipation is dominant near the boundaries though we do not observe obvious boundary layer regions even for the largest values of $Ra/Ra_c$. As a result, it is not relevant to separate the ohmic dissipation into interior and boundary layer contributions.


\begin{figure}[]
 \begin{center}
      \subfloat[][]{\includegraphics[width=0.5\textwidth]{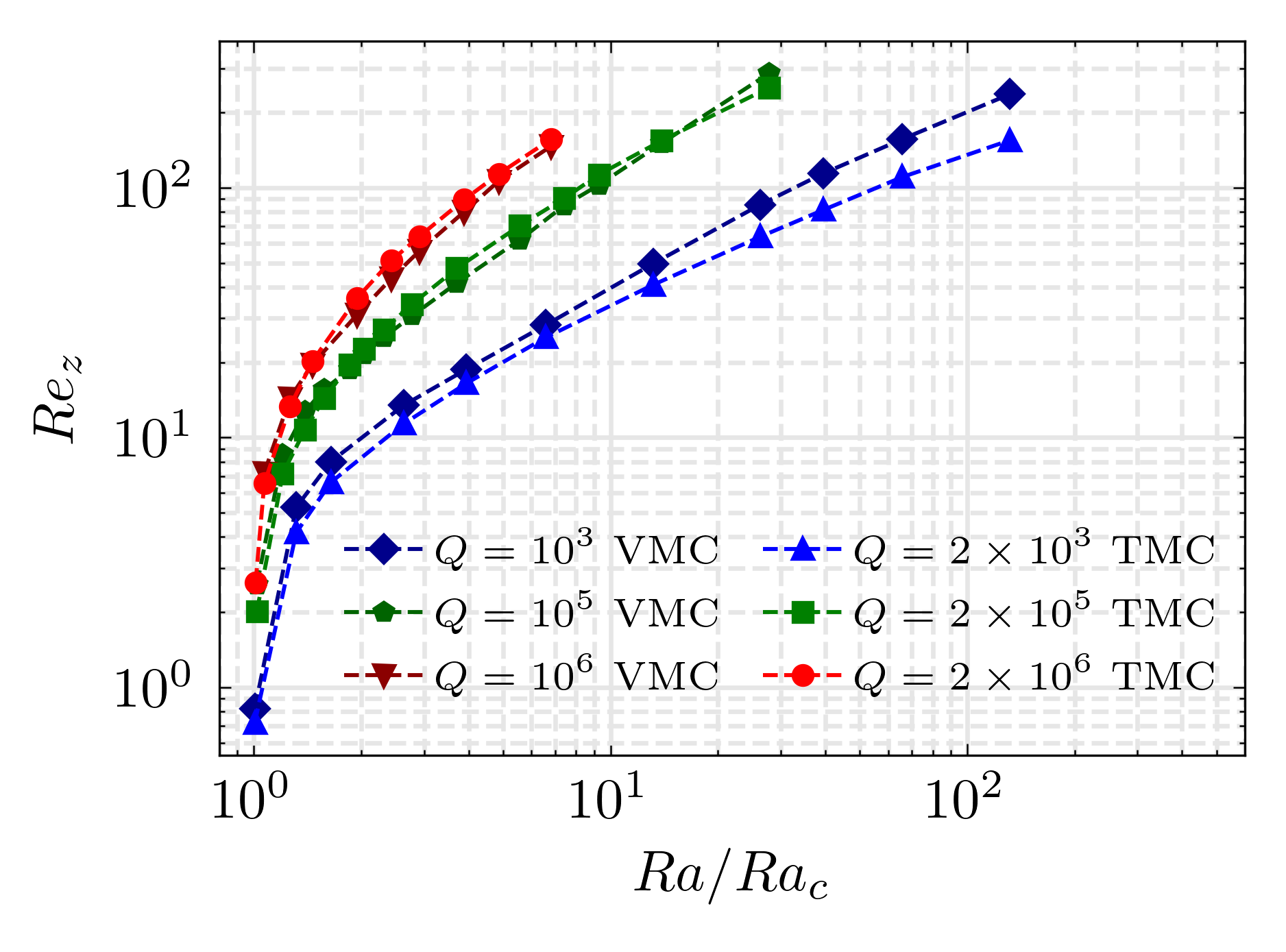}}
      \subfloat[][]{\includegraphics[width=0.5\textwidth]{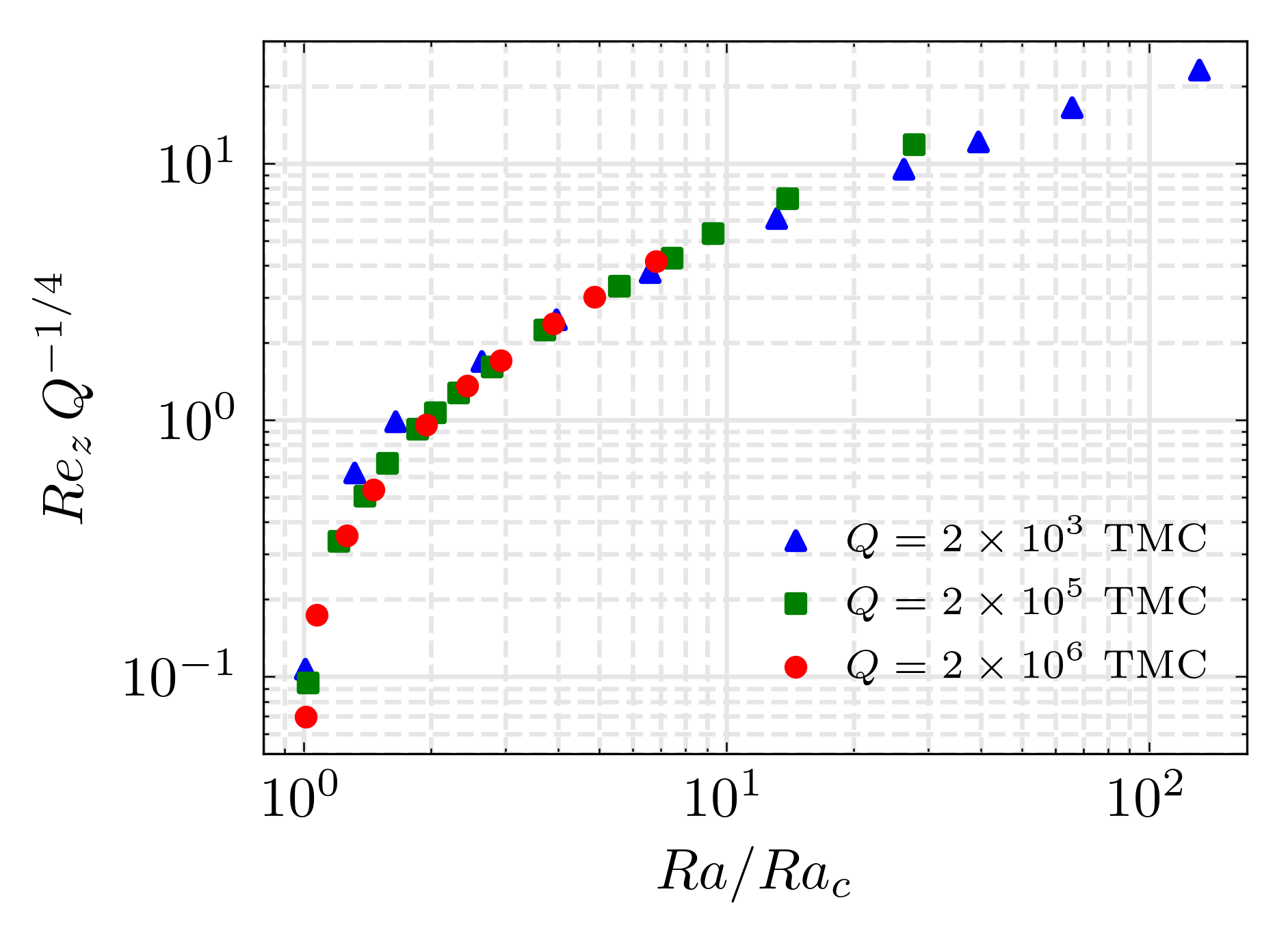}}
	 \caption{Reynolds number data for both TMC and VMC. $Re_z$ is shown for TMC and $Re$ is shown for VMC, where we note that the Reynolds number for VMC is dominated by the vertical component of velocity. (a) Reynolds number $Re_z$ versus $Ra/Ra_c$; (b) Rescaled Reynolds number, $Re_z \, Q^{-1/4}$, versus $Ra/Ra_c$.}
\label{F:Re}
\end{center}
\end{figure}

The convective flow speeds, as characterized by the Reynolds number based on the vertical component of the velocity, $Re_z$, are shown in Fig.~\ref{F:Re}(a). For the VMC data we show the total Reynolds number, as derived from Ref.~\citep{mY19}, though the vertical component of the velocity is larger than the corresponding horizontal components so long as the Lorentz force remains dominant \citep[e.g.][]{pM99}. We find that the scaling of the convective flow speeds are qualitatively similar in both TMC and VMC where there is a rapid rise in amplitude near the onset of convection, and a slower growth for larger $Ra/Ra_c$. There is a general trend of increasing $Re_z$ with increasing $Q$ for both data sets -- we find a good collapse of the data by rescaling the convective flow speeds according to $Re_z \, Q^{-1/4}$, as shown in Fig.~\ref{F:Re}(b). This $Q^{1/4}$ scaling was used in the asymptotic models of \citep{kJ99c,kJ00}, though only single mode (i.e.~single wavenumber) solutions were analyzed. Asymptotic behavior is only expected in the magnetically constrained  regime in which $Ha/Re_z \gg 1$; as previously mentioned, for $Q=2\times10^3$ the cases with $Ra/Ra_c \gtrsim 10$ are characterized by $Ha/Re_z =O(1)$.

\begin{figure}[]
 \begin{center}
      \subfloat[][]{\includegraphics[width=0.33\textwidth]{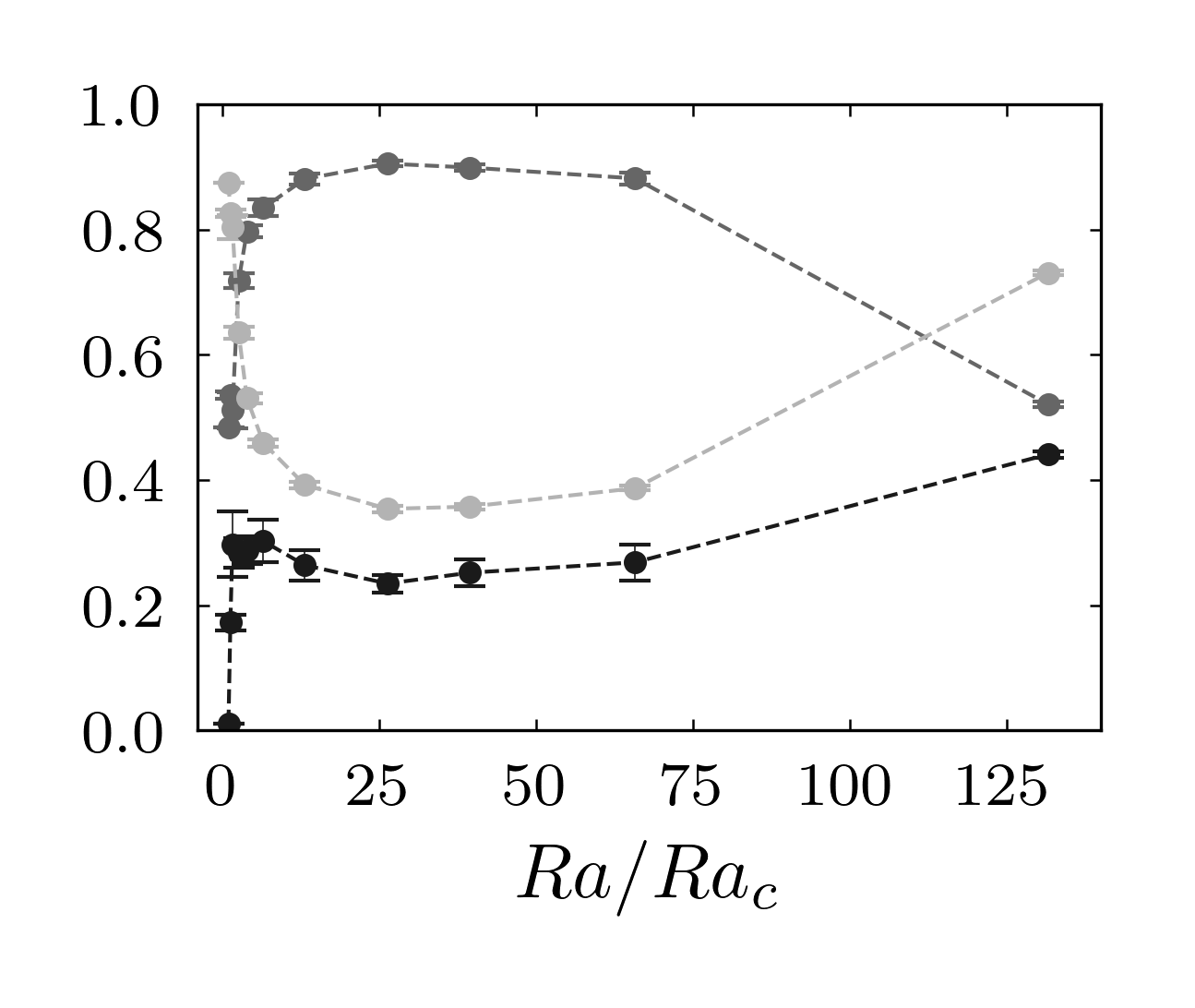}} 
      \subfloat[][]{\includegraphics[width=0.33\textwidth]{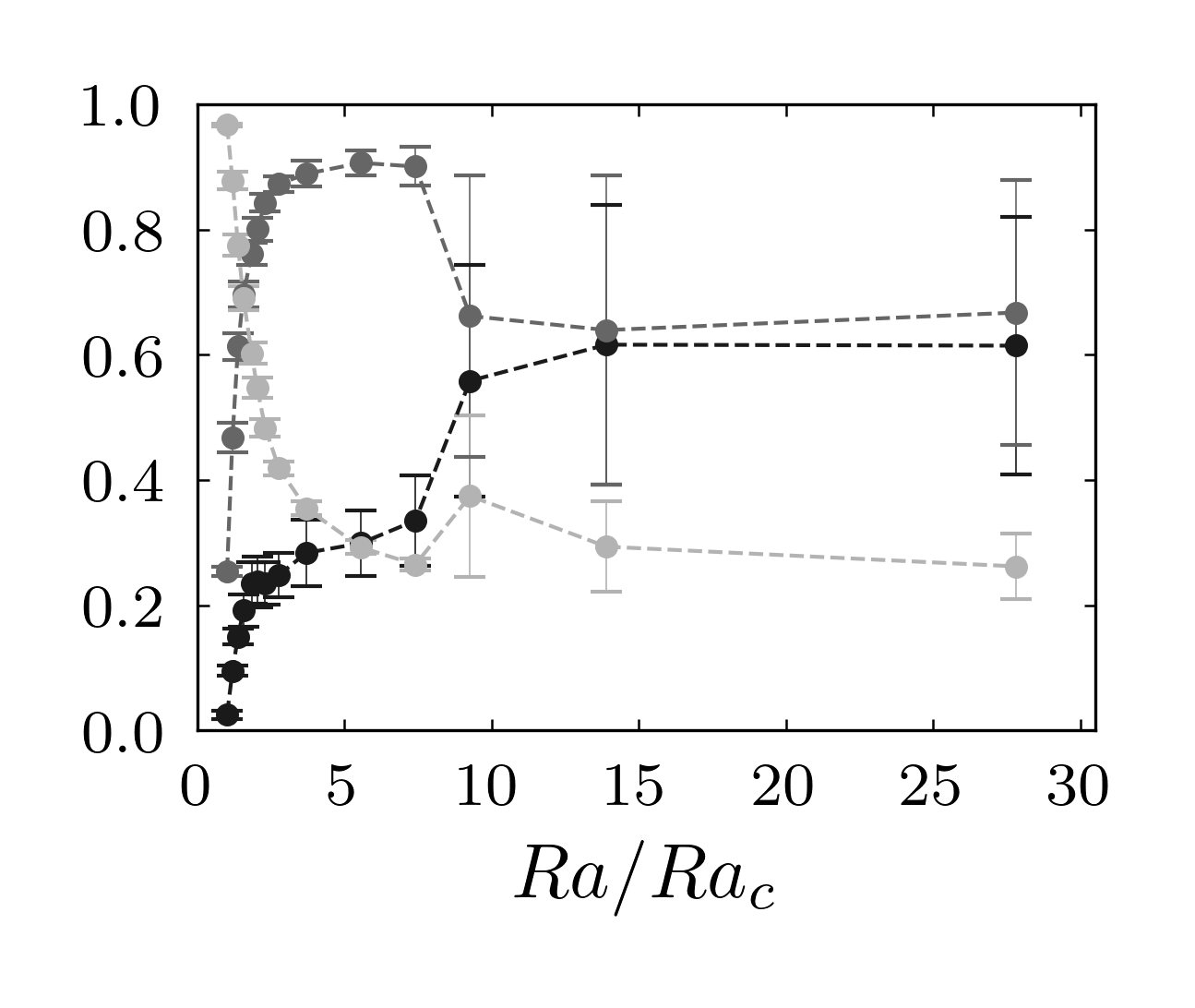}}
      \subfloat[][]{\includegraphics[width=0.33\textwidth]{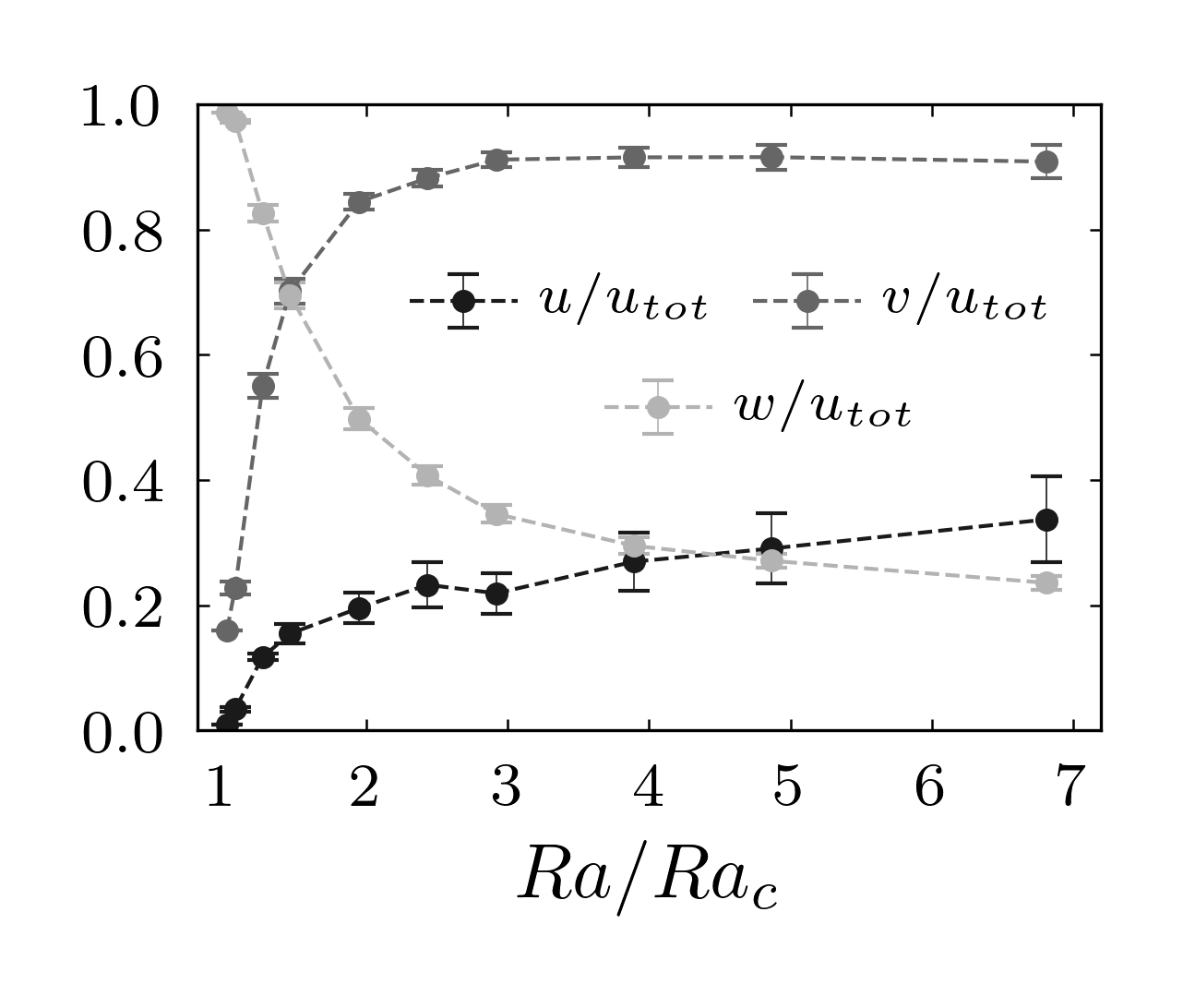}}      
	 \caption{Time-averaged velocity ratios: (a) $Q=2\times10^3$; (b) $Q=2\times10^5$; (c) $Q=2\times10^6$. All quantities are volumetric rms values and the error bars show the standard deviation.}
      \label{F:vel_ratio}
\end{center}
\end{figure}

The anisotropy in the velocity field can be characterized by computing the ratio of the volumetric rms of each velocity component to the total volumetric rms velocity \textcolor{red}{ which we denote $u_{tot}$}; the resulting data is shown in Fig.~\ref{F:vel_ratio}. Near the onset of convection, in which the flow consists of two-dimensional rolls, we find that the vertical component of the velocity dominates. However, we find a rapid decrease in the relative size of $w$ as the large scale flow forms and the $y$-component of the velocity field, $v$, dominates. All cases show a saturation in $v/u_{tot} \approx 0.9$. Moreover, this saturation is observed to occur at smaller values of $Ra/Ra_c$ as $Q$ increases. The relative size of $u/u_{tot}$ exhibits a slow but steady increase with $Ra/Ra_c$ and we find that the standard deviation also increases. For $Q=2\times10^5$ we find relaxation oscillations (discussed more below) for supercriticalities $Ra/Ra_c \gtrsim 7.5$ -- in this regime both $u$ and $v$ become of comparable magnitude and exhibit large amplitude variations with time.

\subsection{Horizontally averaged mean fields}

As previously noted, mean flows form in TMC for all values of $Q$ investigated here. These mean flows take various forms and are associated with corresponding mean magnetic fields. In the present section we analyze mean fields, both velocity and magnetic, that are defined by averages over the entire horizontal plane. In the next subsection we examine the dynamics of what we refer to as zonal flows, which are $y$-directed flows averaged only in the $y$-direction.  

\begin{figure}[]
 \begin{center}
      \subfloat[][]{\includegraphics[width=0.33\textwidth]{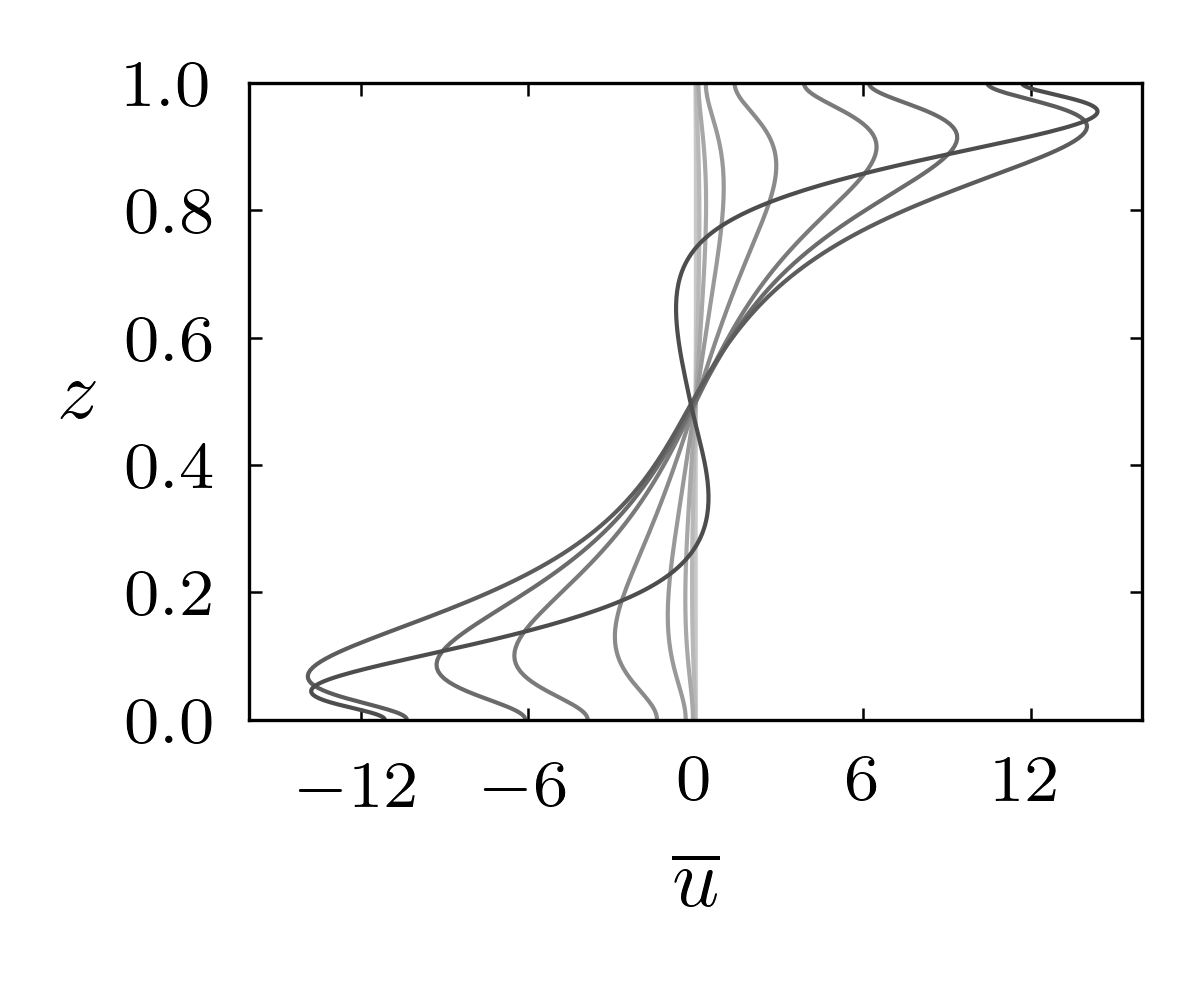}}
      \subfloat[][]{\includegraphics[width=0.33\textwidth]{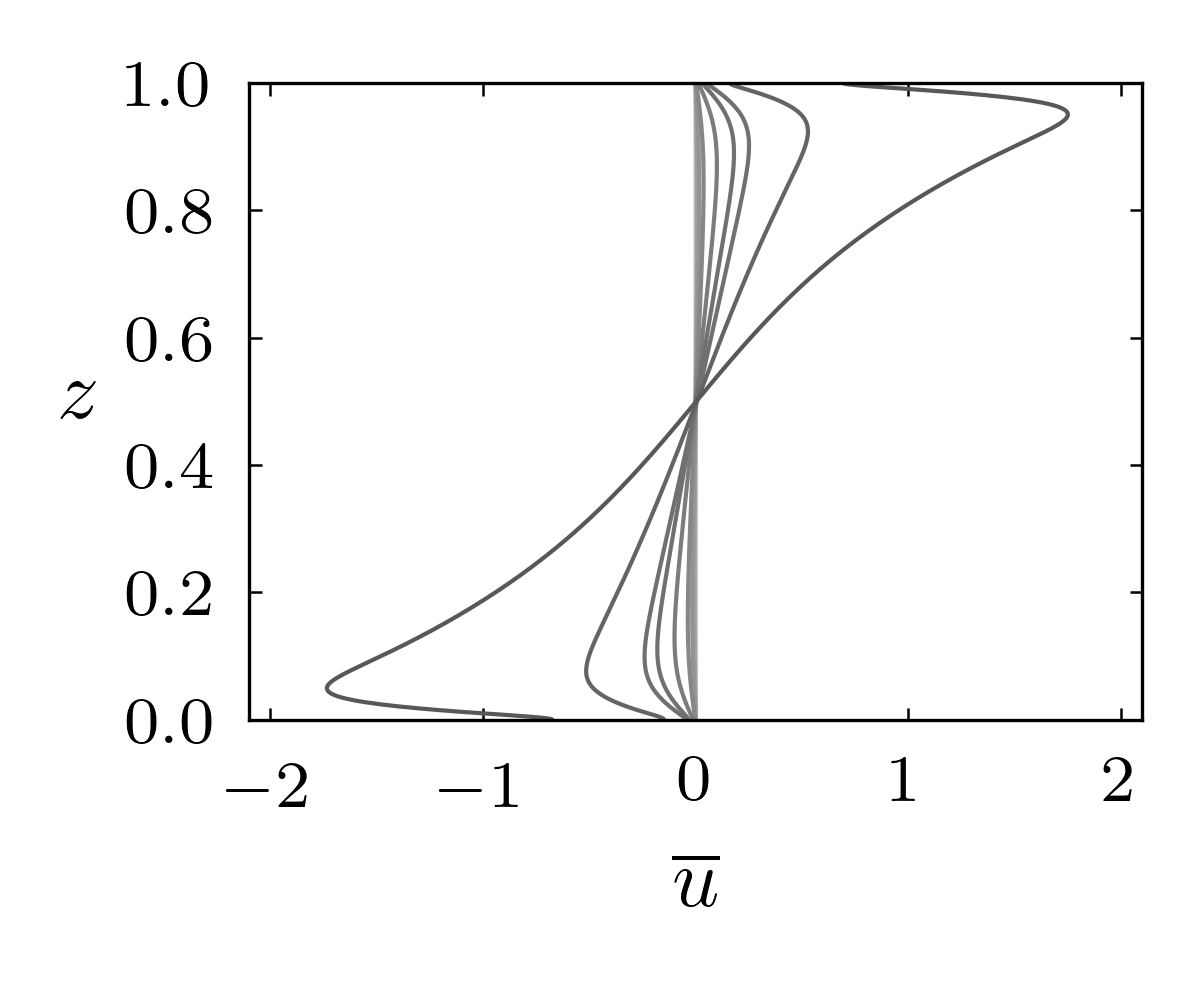}}
      \subfloat[][]{\includegraphics[width=0.33\textwidth]{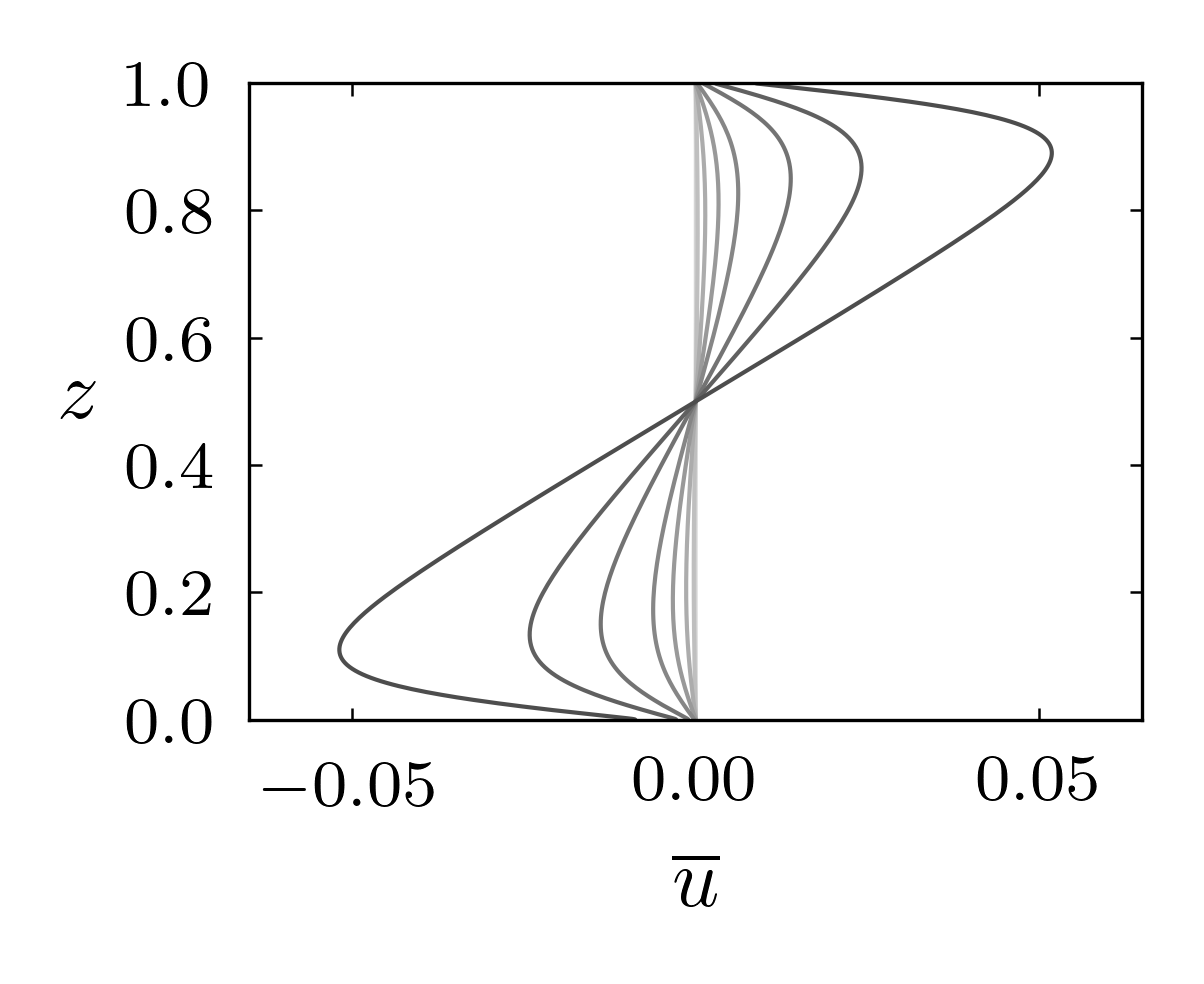}}      \\
      
      \subfloat[][]{\includegraphics[width=0.33\textwidth]{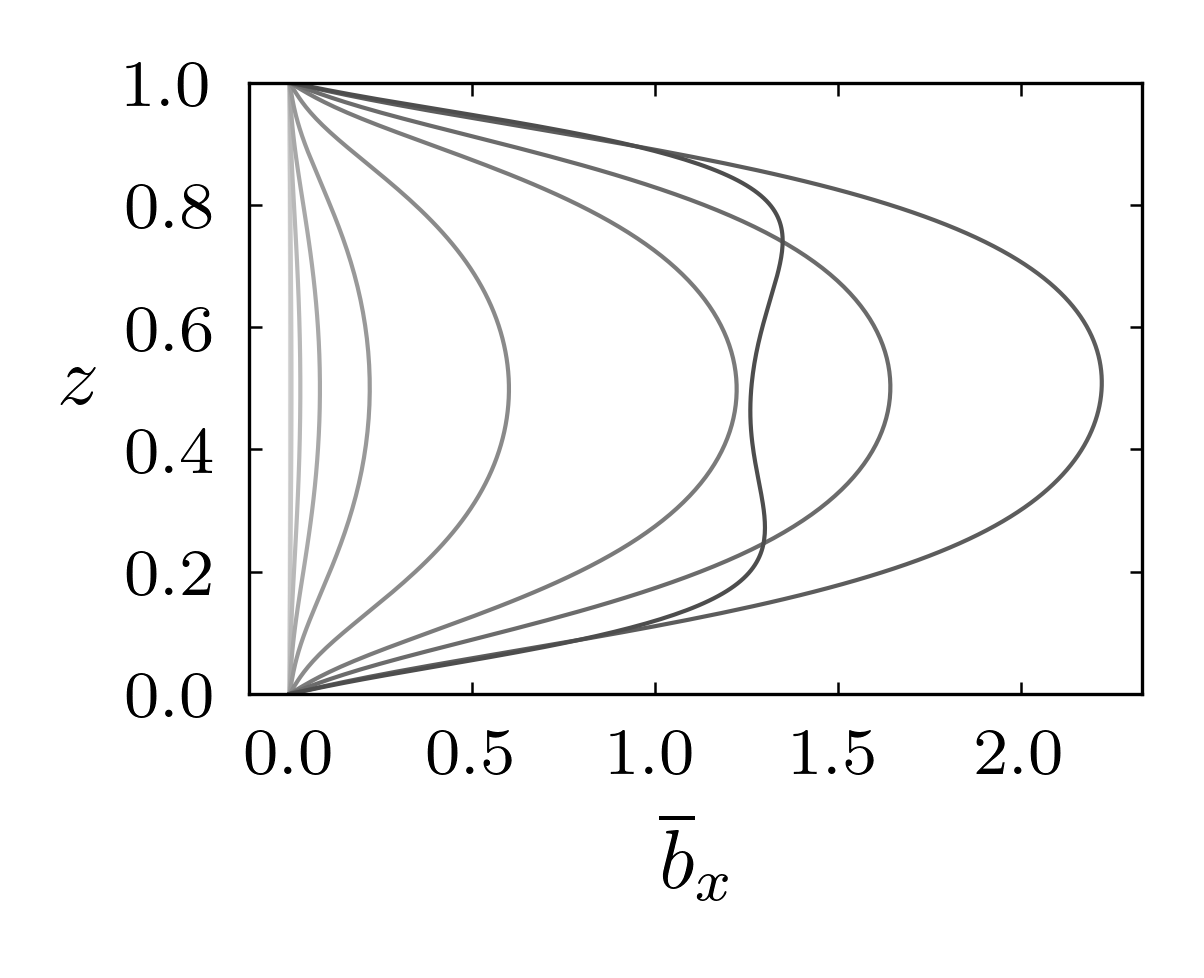}}
      \subfloat[][]{\includegraphics[width=0.33\textwidth]{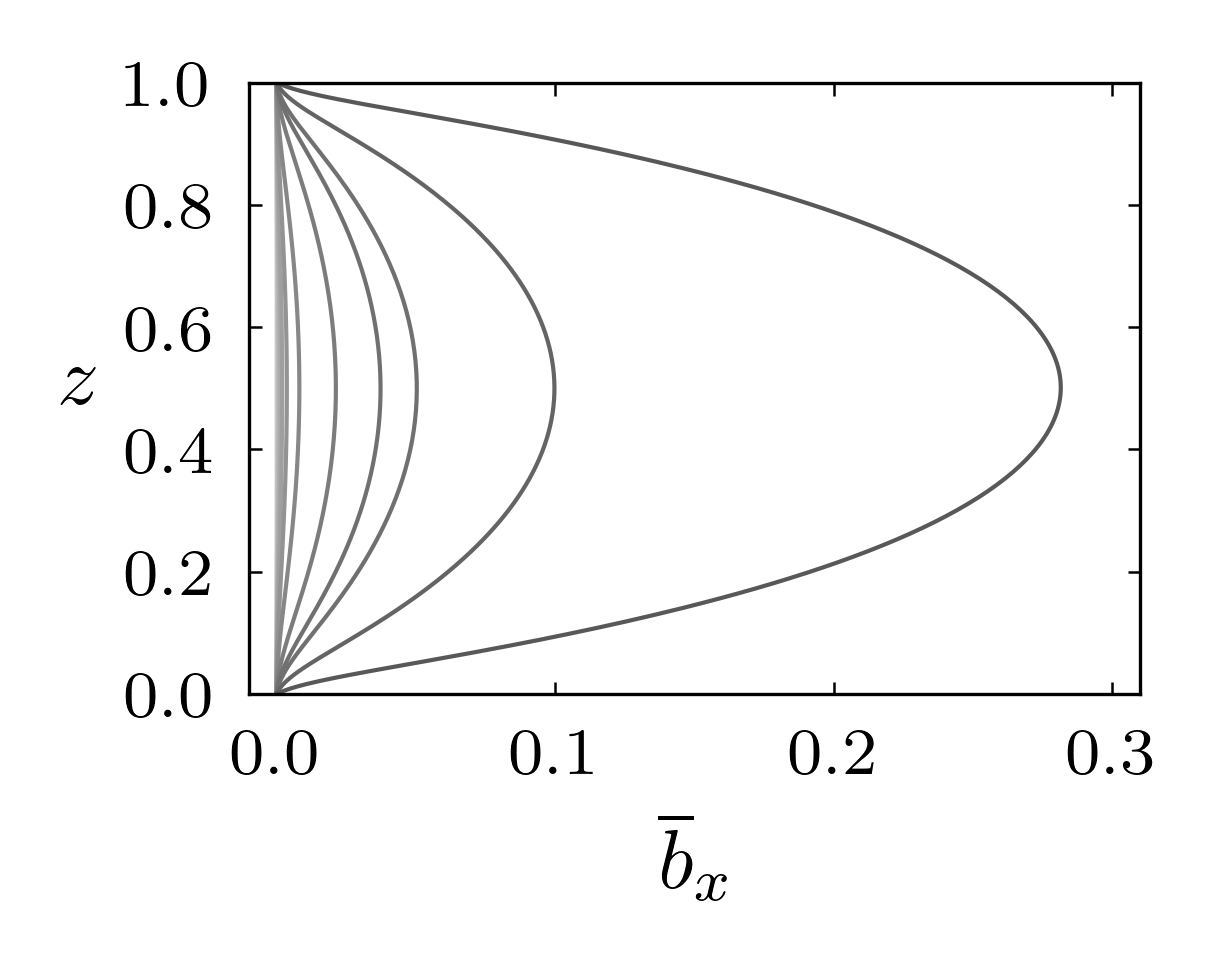}}
      \subfloat[][]{\includegraphics[width=0.33\textwidth]{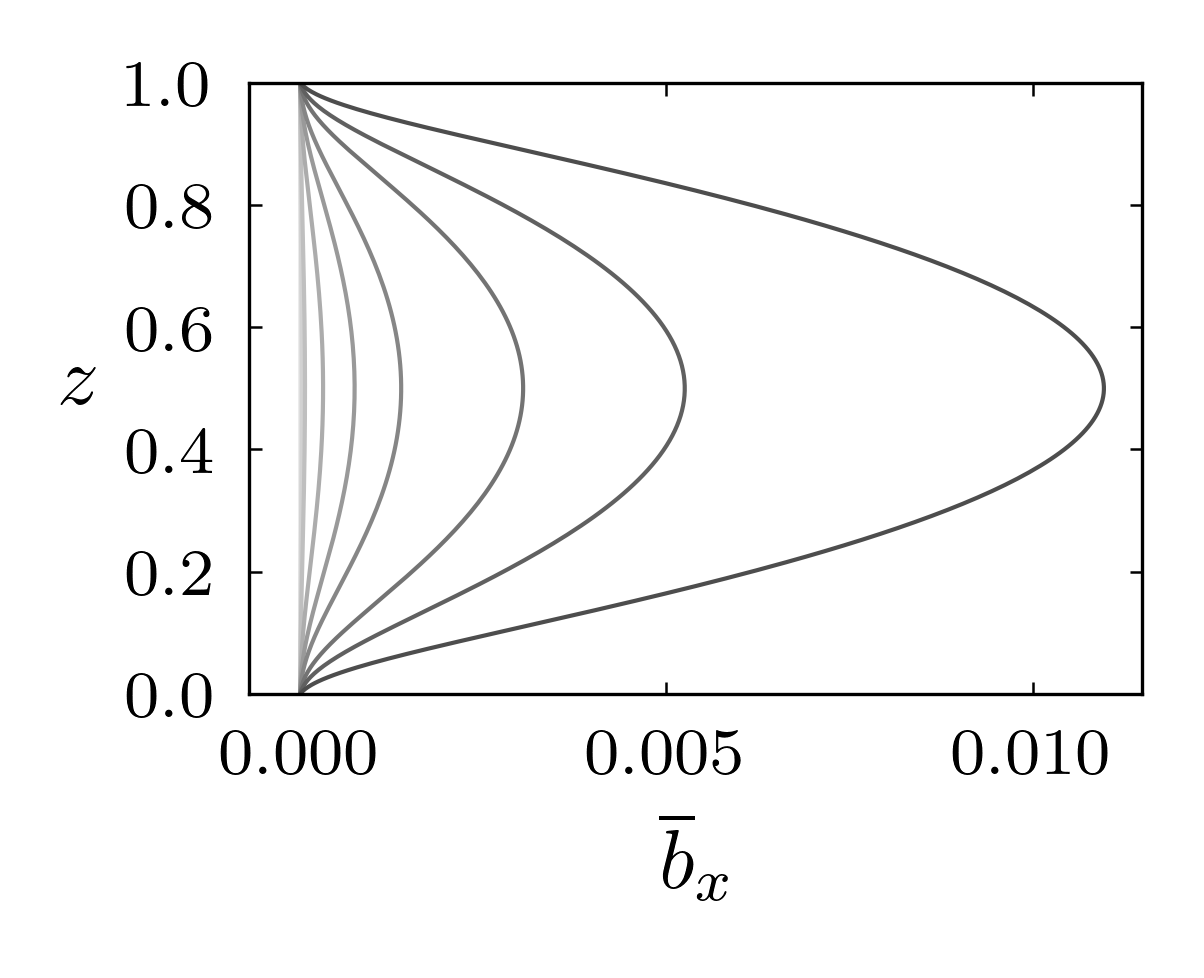}}           
	 \caption{Time and horizontally averaged x-component of the velocity field and magnetic field: (a,d) $Q=2\times 10^3$; (b,e) $Q=2\times 10^5$; (c,f) $Q=2\times 10^6$. Note that the volume averaged x-velocity was removed from the horizontally averaged x-velocity. Darker lines correspond to higher Rayleigh numbers.}
      \label{F:mean_flow}
\end{center}
\end{figure}

We restrict the present analysis to the $x$-component of the mean velocity field, $\overline{u}$, and the associated magnetic field $\overline{b}_x$, since the $y$-components are observed to be significantly smaller in magnitude and approach zero for sufficiently long time averages. The mean momentum equation in the $x$-direction is given by
\be
\dst \overline{u} + \dsz \overline{\lb u' w' \rb} = Q \eta_3 \dsz \overline{b}_x +  \partial^2_z \overline{u},
\ee
where $u' = u - \overline{u}$, etc. The corresponding mean magnetic field is governed by
\be
0 = \eta_3 \dsz \overline{u} + \partial^2_z \overline{b}_x ,
\label{E:mean_ind_x}
\ee
so that vertical shear in the mean flow is associated with a mean magnetic field. 

The computed $x$-components of the means fields, $\overline{u}$ and $\overline{b}_x$, are shown in Fig.~\ref{F:mean_flow} for a range of $Ra$ and all three values of $Q$. We find that the magnitude of the mean fields increases with $Ra/Ra_c$, with the singular exception of the case ($Q=2\times10^3$, $Ra/Ra_c = 132$), which shows a sudden decrease; profiles for this case are indicated by the darkest shade in panels (a) and (d). This latter case has reached sufficiently large Rayleigh number such that the magnetic field is no longer constraining the motion. We find that the $x$-components of the mean fields are robust in the sense that they maintain a similar structure with varying $Q$ and $Ra$, and evolve on a timescale much longer than the underlying convection. The mean velocity flows in the same direction as the imposed magnetic field; i.e.~it points in the negative (positive) $x$-direction in the lower (upper) half of the flow domain. We find that the magnitude of $\overline{u}$ is small relative to the convective fluctuations, $u'$ and $w'$, though it remains non-zero for all cases investigated. This non-zero vertical flux of $x$-momentum is associated with the tendency for the convection to align with the direction of the magnetic field.

Estimates for the sizes of the various terms in the above equations can be made, and help to explain some of the numerical findings shown in Fig.~\ref{F:mean_flow}. We scale vertical derivatives of mean quantities as $\dsz  \rightarrow \overline{\ell}_z^{-1}$, where $\overline{\ell}_z$ is some characteristic length scale in the vertical dimension. In what follows we assume $\eta_3 = O(1)$. Equation \eqref{E:mean_ind_x} then gives
\be
\overline{u} \sim \frac{\overline{b}_x}{\overline{\ell}_z} .
\label{E:mean_b_scale}
\ee
If, for the purpose of the present scale analysis, we interpret the overline as also including a time average, then we have three terms in the mean momentum equations to consider, which we scale as
\be
 \dsz \overline{\lb u' w' \rb} \rightarrow \frac{\overline{u' w'}}{\overline{\ell}_z}, \qquad 
Q \eta_3 \dsz \overline{b}_x \rightarrow \frac{Q \overline{b}_x}{\overline{\ell}_z}, \qquad
\partial^2_z \overline{u} \rightarrow \frac{\overline{u}}{\overline{\ell}_z^2} .
\ee
Using relationship \eqref{E:mean_b_scale} in the above Lorentz force term we then have
\be
\frac{Q \overline{b}_x}{\overline{\ell}_z} \rightarrow  Q \overline{u}  .
\ee
Balancing the mean Lorentz force with the mean viscous force would imply $\overline{\ell}_z \sim Q^{-1/2}$, which is the well-known Hartmann boundary layer scaling. We assume that such a strong dependence on $Q$ is only relevant within the Hartmann layer, and not in the bulk of the domain. We do observe a Hartmann boundary layer in the horizontally averaged fields, though its effects appear to be negligibly small. Outside of the Hartmann layer it is unclear that there should be any $Q$-dependence on the vertical length scale with regards to mean quantities. Therefore, as a first approximation we neglect any $Q$-dependence on this length scale and the mean Lorentz force can then only be balanced by the divergence of the Reynolds stress in the bulk. By directly integrating (in $z$) the resulting balance, the (time-averaged) mean momentum equation gives
\be
\overline{b}_x \approx \frac{1}{\eta_3 Q} \overline{u' w'} ,
\ee
where we have used the impenetrable and electrically insulating boundary conditions. In terms of the asymptotic dependence only, the above balance suggests
\be
\overline{b}_x = O\lb \frac{ \overline{ u' w' } }{Q} \rb,
\ee
This result shows that Reynolds stresses are directly responsible for the generation of a mean magnetic field. Using the mean induction equation then gives a relationship between the mean flow and the Reynolds stresses,
\be
\dsz \overline{u} \approx - \frac{1}{Q\eta_3^2} \partial_z^2 \overline{\lb u' w' \rb} .
\ee
Assuming $\overline{\ell}_z = O(1)$, the above balance leads to
\be
\overline{u} = O\lb \frac{ \overline{ u' w' } }{Q} \rb,
\ee
so that both the mean velocity and mean magnetic field scale similarly with $Q$. These relationships suggest that we require knowledge on the asymptotic size of the fluctuating velocity components in order to estimate the asymptotic size of both $\overline{u}$ and $\overline{b}_x$. Fig.~\ref{F:Re} suggests that $w' \sim Q^{1/4}$ and, though not shown, we also find that the $u' \sim Q^{1/4}$, which suggests that 
\be
\overline{u} = O\lb Q^{-1/2} \rb, \qquad \overline{b}_x = O\lb Q^{-1/2} \rb .
\ee

\begin{figure}[]
 \begin{center}
      \subfloat[][]{\includegraphics[width=0.45\textwidth]{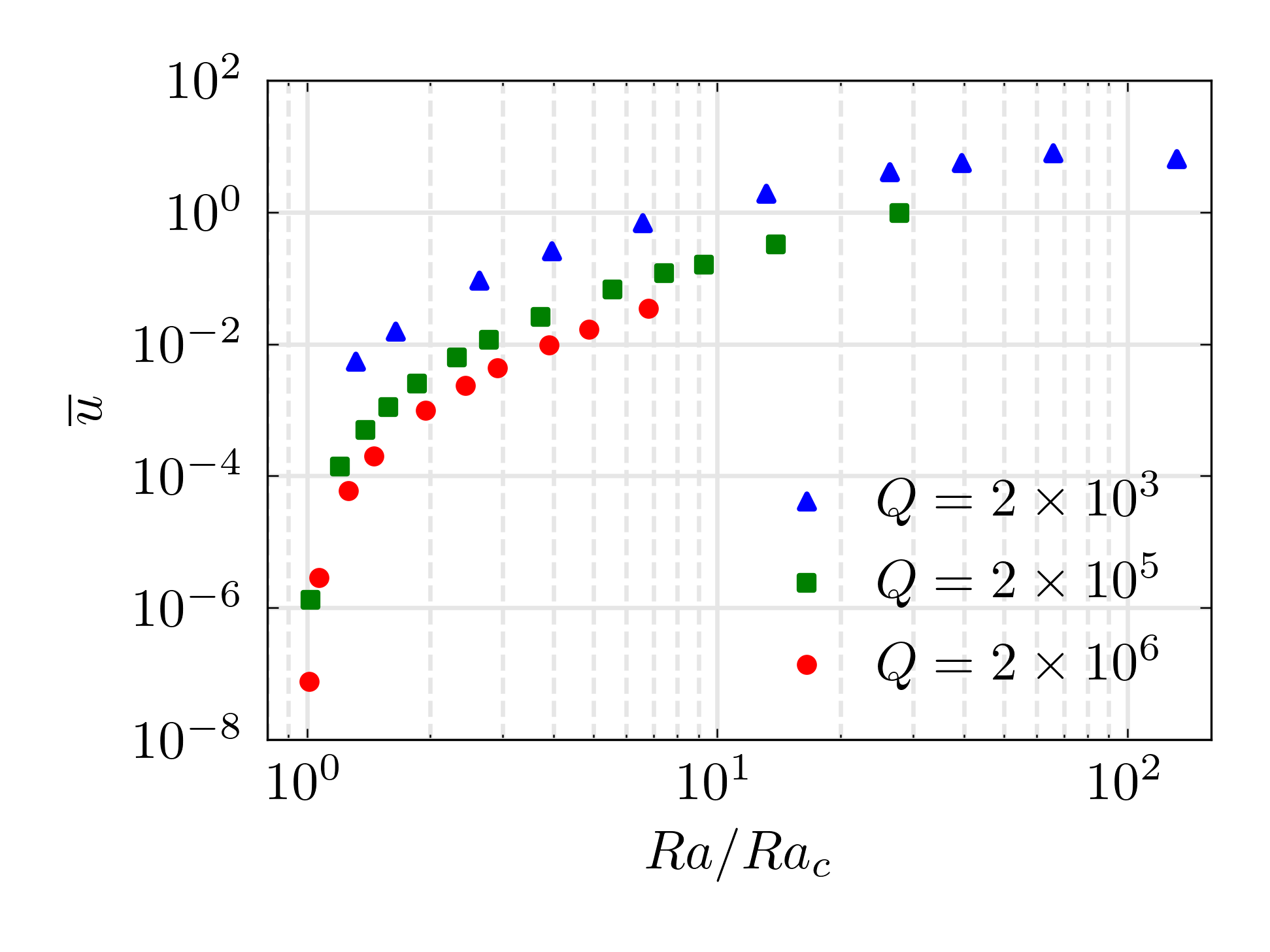}}  \qquad 
      \subfloat[][]{\includegraphics[width=0.45\textwidth]{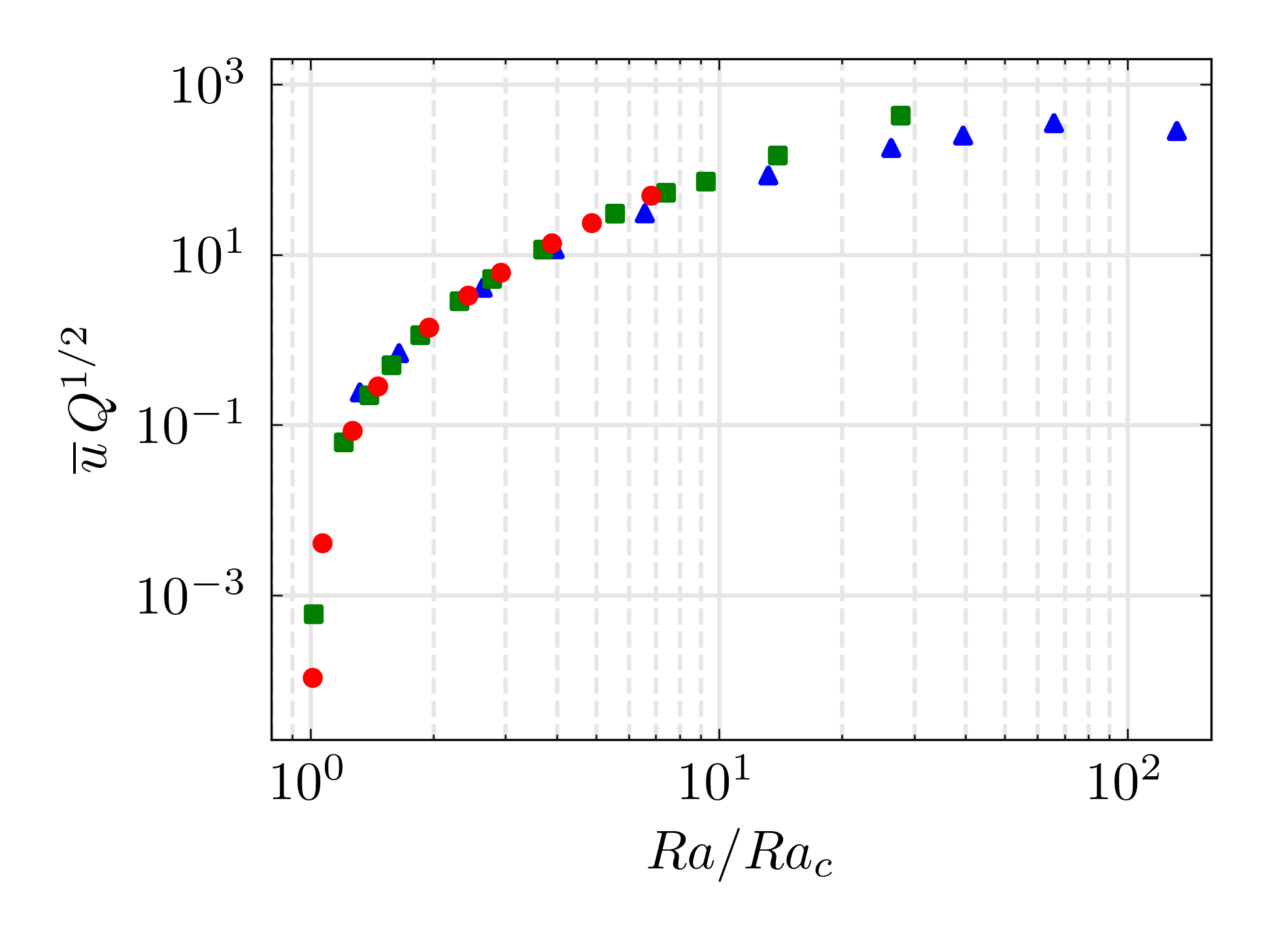}} \\
      \subfloat[][]{\includegraphics[width=0.45\textwidth]{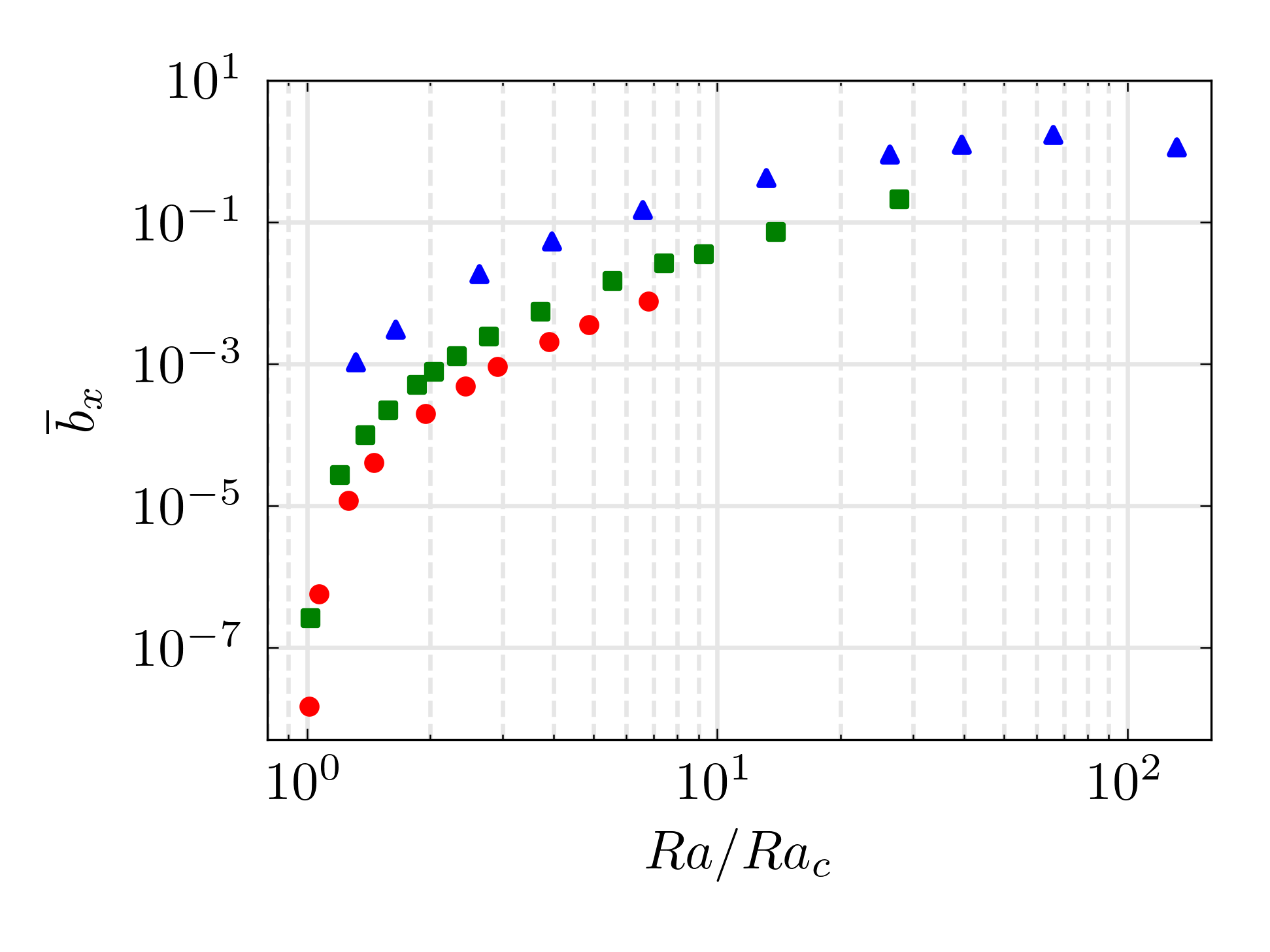}}  \qquad
      \subfloat[][]{\includegraphics[width=0.45\textwidth]{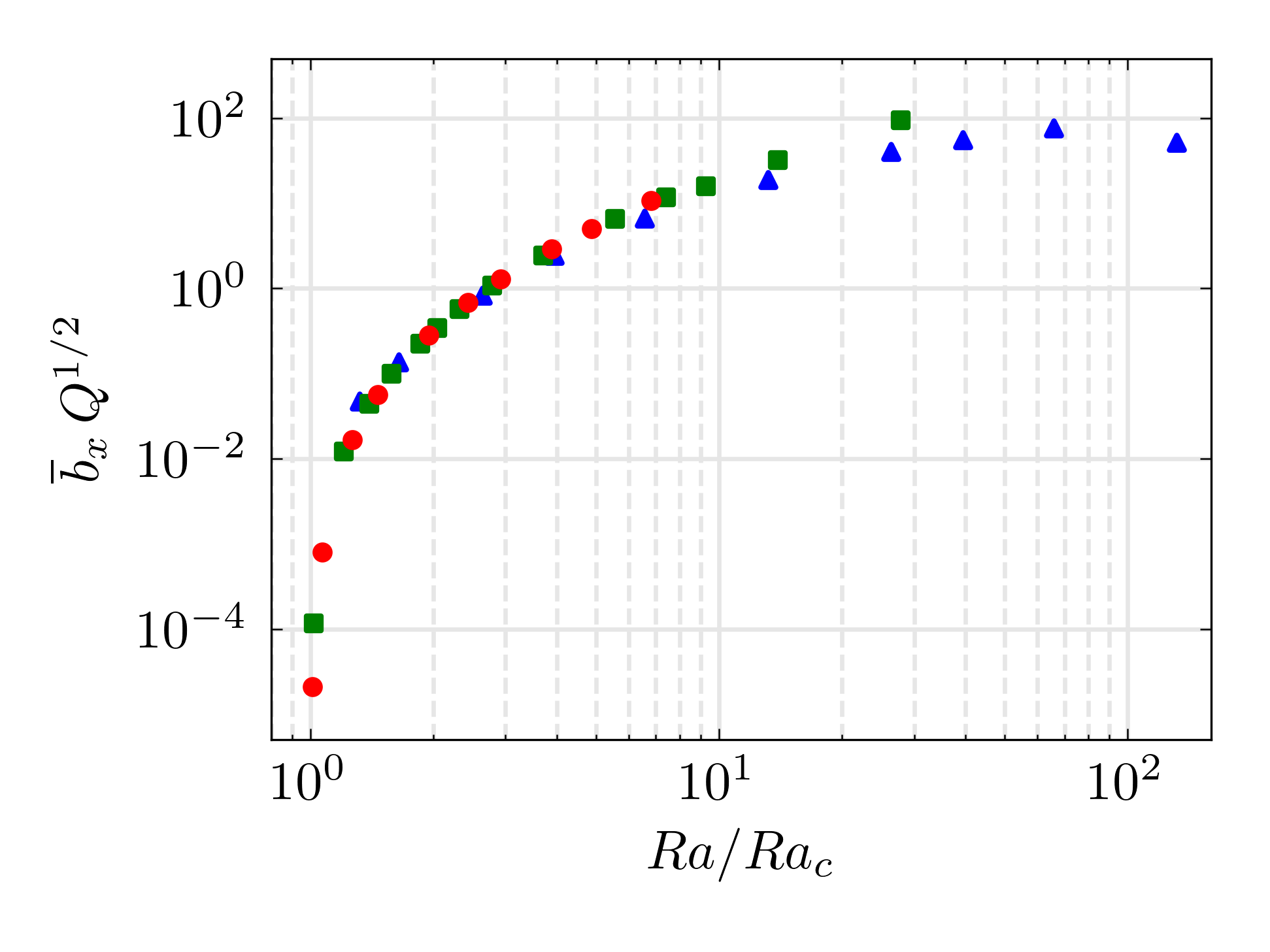}} \\
	 \caption{Scalings of the rms values for the horizontally averaged quantities: (a) $\overline{u}$; (b) $\overline{u} \, Q^{1/2}$; (c) $\overline{b}_x$; (d) $\overline{b}_x \, Q^{1/2}$. The rms is taken in the vertical direction, then a time-average is computed. Note that for $\overline{u}$, the volume-averaged x-velocity was subtracted off before the rms is calculated.}
      \label{F:avg_scalings}
\end{center}
\end{figure}

Figs.~\ref{F:avg_scalings}(a) and (c) show rms values of the mean velocity and magnetic field, respectively. The corresponding rescaled components are given in Figs.~\ref{F:avg_scalings}(b) and (d). The collapse of the rescaled quantities suggests that the scaling arguments given above lead to good estimates for the asymptotic dependence of these mean fields. These results suggest that the mean flow (and associated mean magnetic field) becomes less significant dynamically in the limit $Q \rightarrow \infty$.



\subsection{Zonal flow dynamics}

The horizontally averaged fields analyzed in the previous section represent one particular component of the mean flows that are observed in TMC. However, the strongest mean flows observed in the simulations are dominated by horizontal wavenumber $k_x =1$, and such motions are eliminated by a horizontal average over the entire horizontal plane. These mean flows are a combination of `zonal' ($y$-directed) flows and large scale vortices that result in a meandering jet structure. For a fixed value of $Q$, we find that the $y$-component of the flow dominates, though for sufficiently strong forcing for $Q=2\times10^5$ the non-zonal ($x$-directed) flow becomes comparable to the zonal component and relaxation oscillations occur. Given the directional dependence of the flow it is helpful to average in only a single horizontal direction, either $x$ or $y$. We denote such averages with an overline and a superscript that specifies the direction or directions of the average. Here we focus only on the zonal, or $y$-directed, mean flows since they are observed for all values of $Q$ and tend to dominate over much of the parameter space investigated here. A brief description of the relaxation oscillations is given in the next section.


\begin{figure}
 \begin{center}
 
\subfloat[][]{\includegraphics[width=0.33\textwidth]{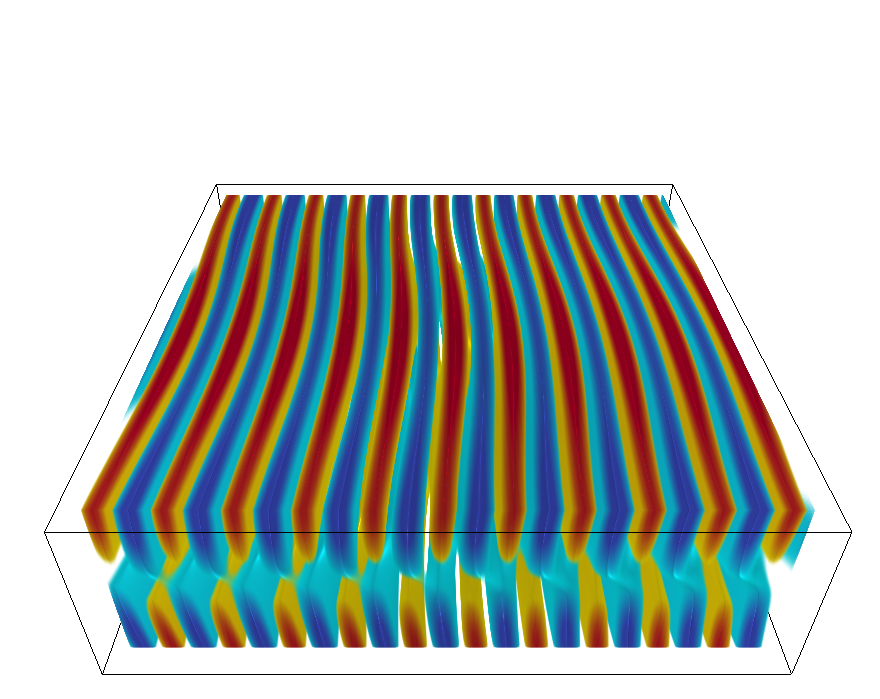}}      
\subfloat[][]{\includegraphics[width=0.33\textwidth]{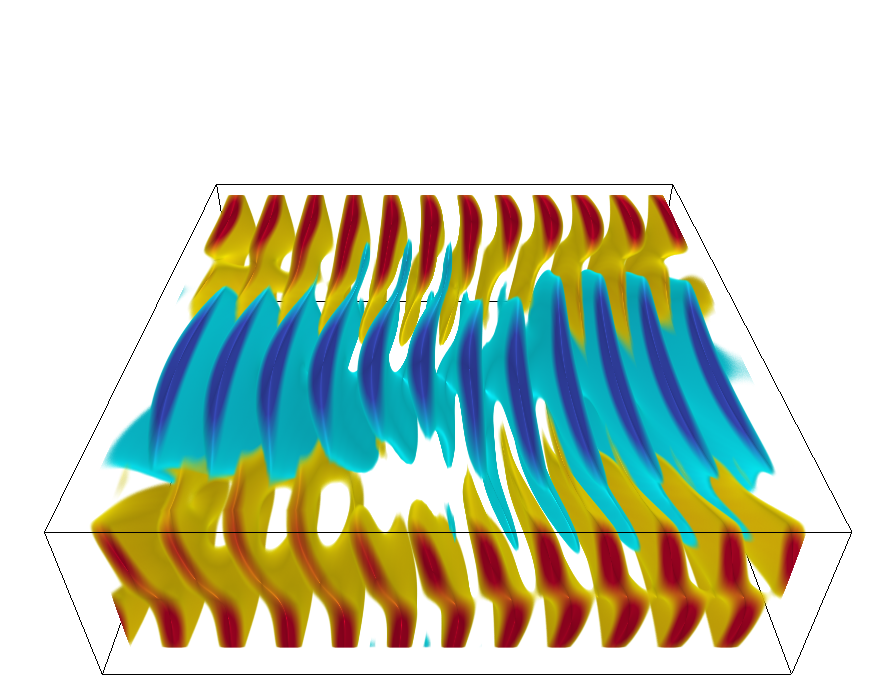}}      
\subfloat[][]{\includegraphics[width=0.33\textwidth]{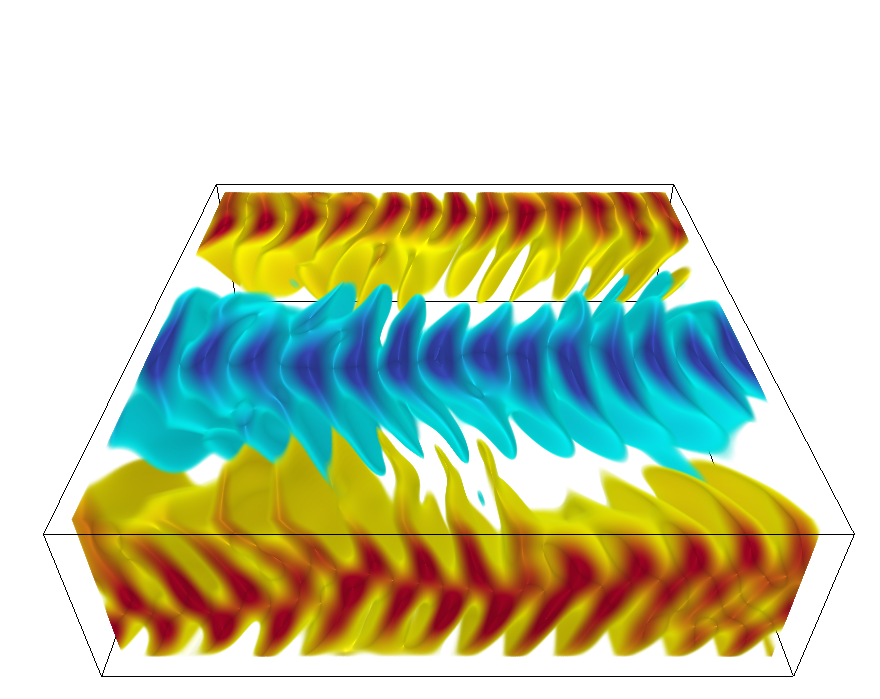}}    \\

   
	 \caption{Volumetric renderings of the $y$-component of the velocity, $v$, illustrating the development of the zonal flow for $Q=2\times10^6$ and increasing $Ra$: (a) $Ra = 1.04\times10^7$; (b) $Ra = 1.1\times10^7$; (c) $Ra = 1.3\times10^7$. The view is approximately along the imposed magnetic field.}
      \label{F:uy}
\end{center}
\end{figure}

As $Q$ is increased we find that the zonal flow quickly becomes the dominant component of the velocity. To illustrate this behavior, Fig.~\ref{F:uy} shows instantaneous volumetric renderings of the $y$-component of the velocity, $v$, for $Q=2\times10^6$ and three different values of $Ra$, increasing from left to right. Just beyond the onset of convection we find a mode with $k_x = 1$ appears, as shown in panel (a), which suggests the formation of the zonal flow. As $Ra$ increases in panels (b) and (c), this zonal flow becomes stronger and eventually becomes energetically dominant relative to the small scale convection by which it is driven.

\begin{figure}
 \begin{center}
 
\subfloat[][]{\includegraphics[width=0.33\textwidth]{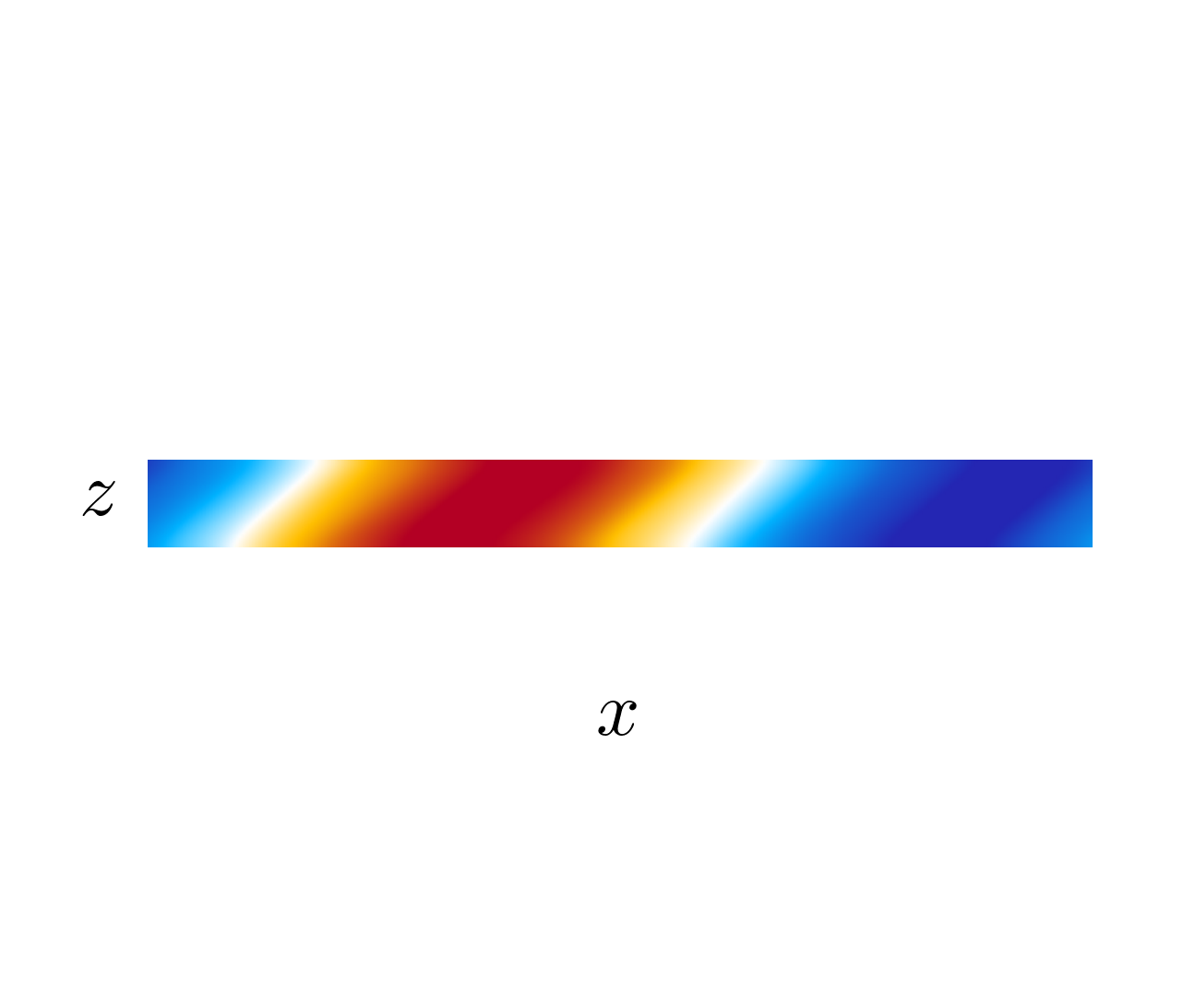}}      
\subfloat[][]{\includegraphics[width=0.33\textwidth]{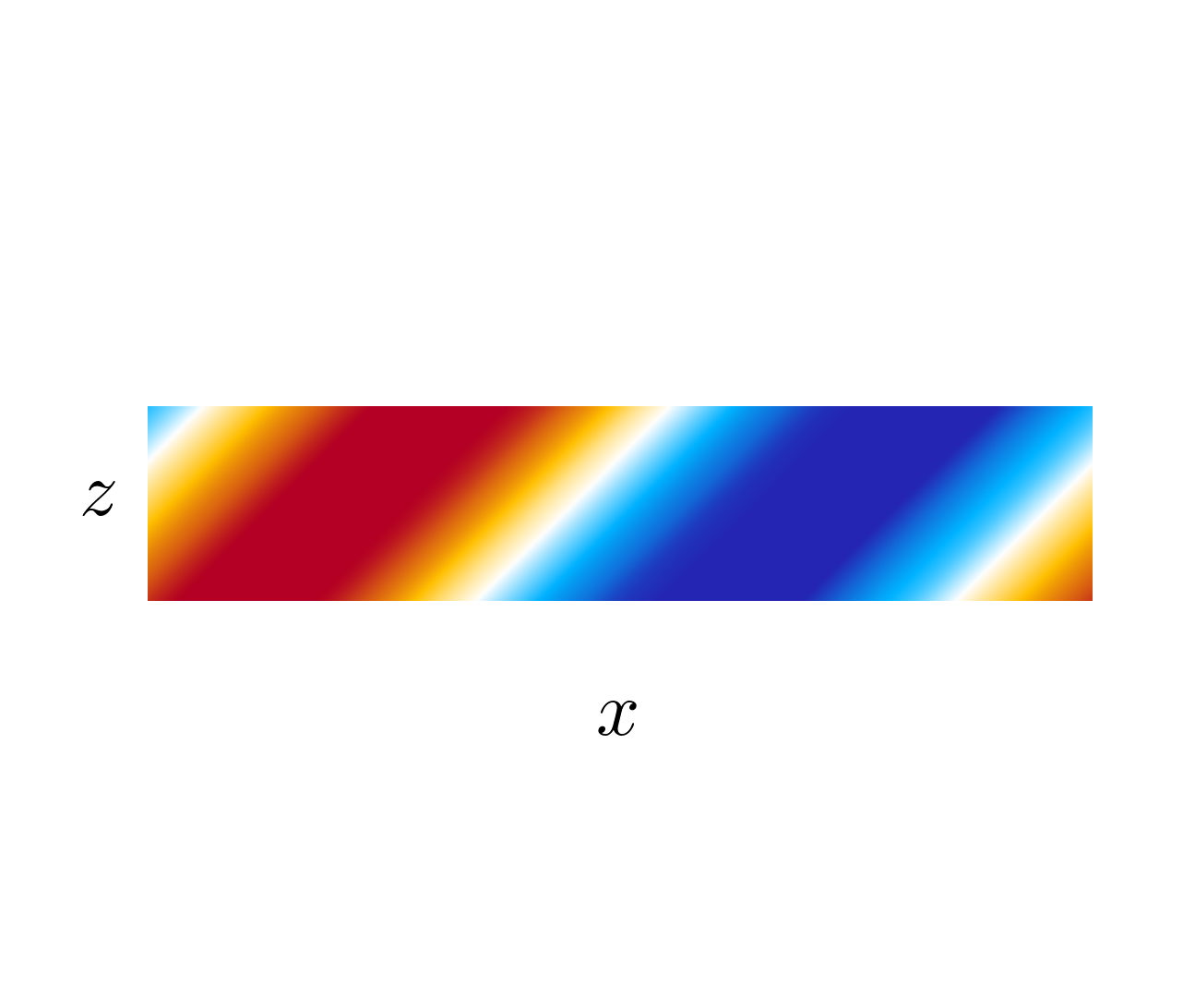}}      
\subfloat[][]{\includegraphics[width=0.33\textwidth]{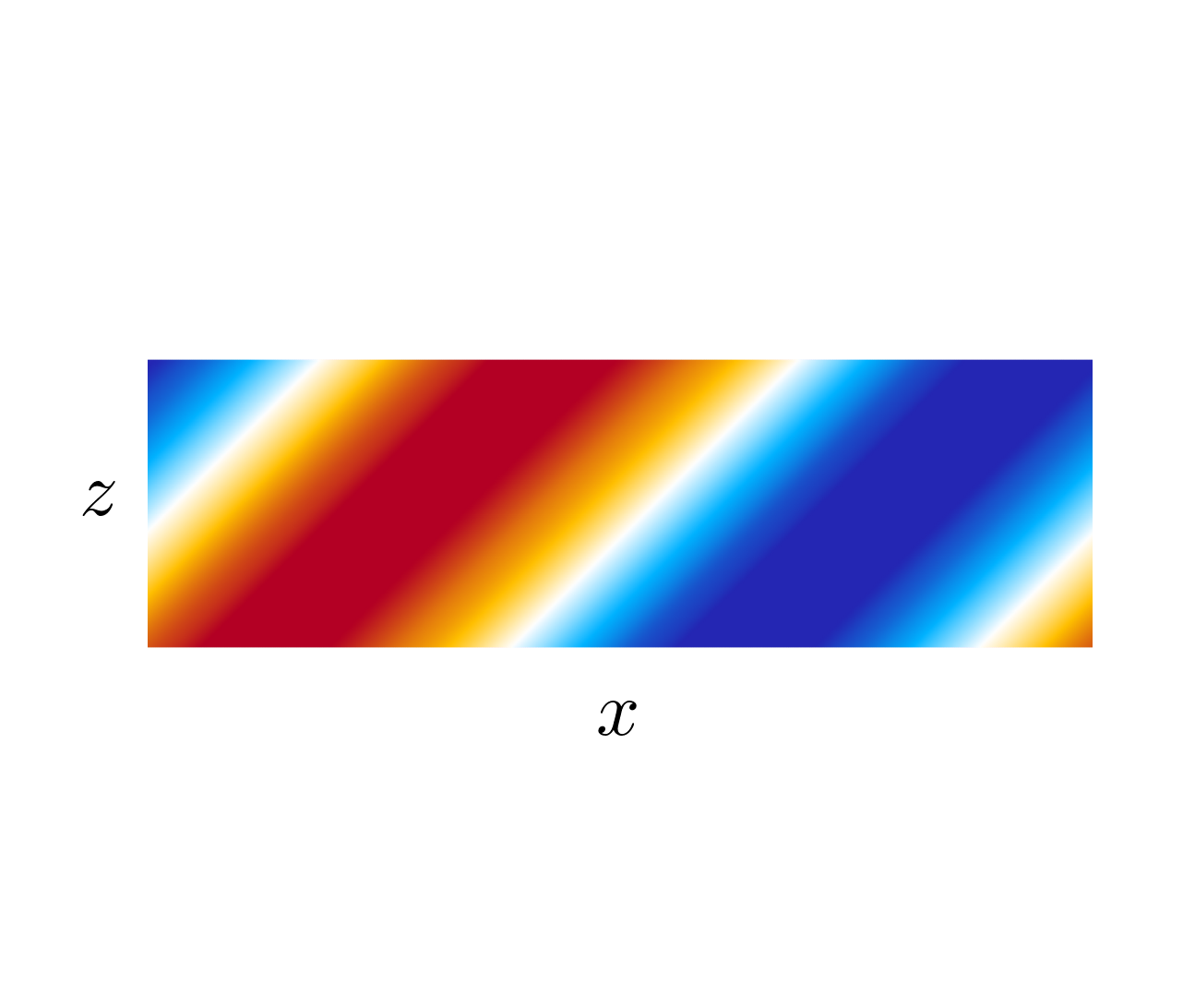}}    \\

\subfloat[][]{\includegraphics[width=0.33\textwidth]{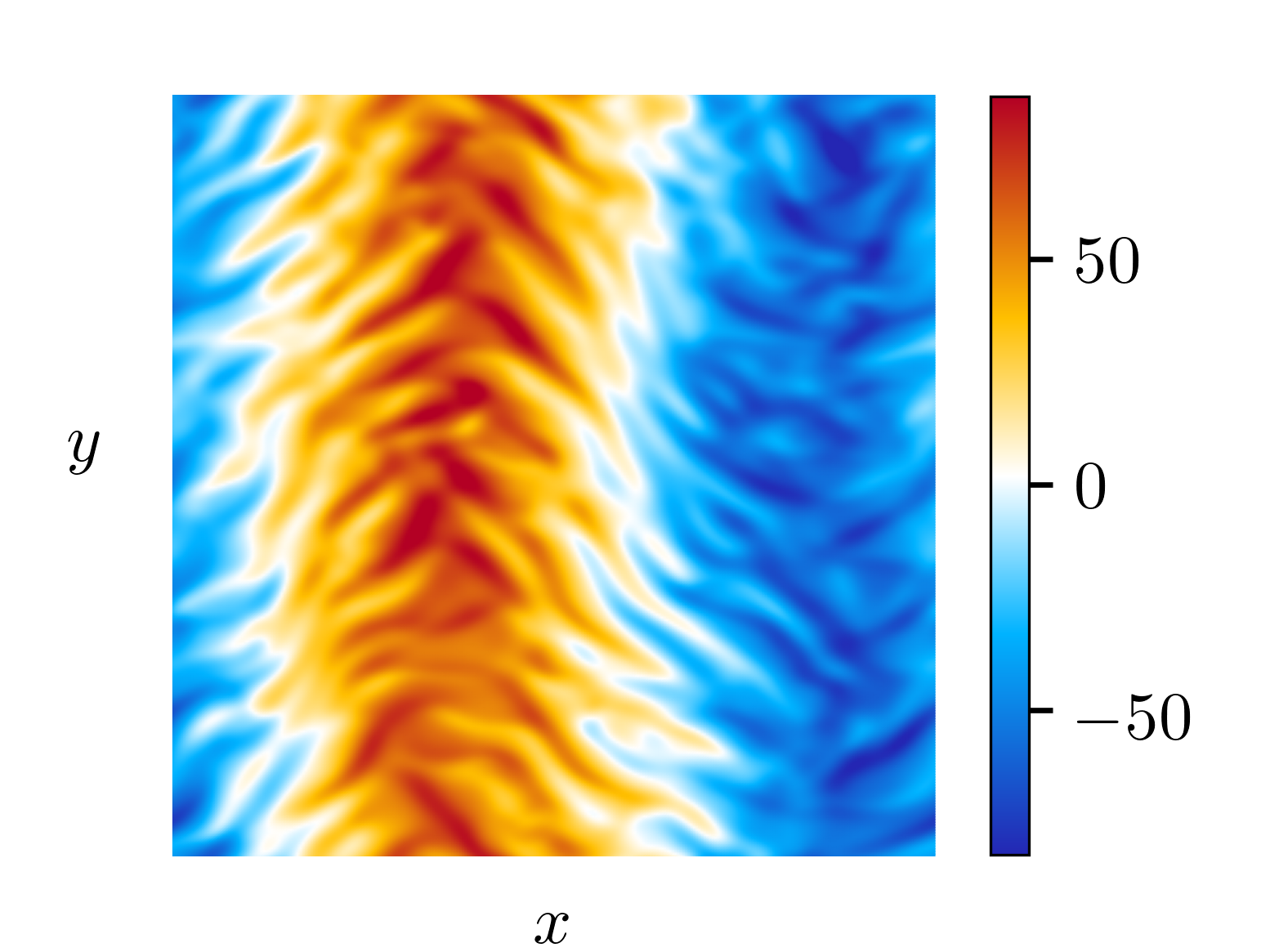}}      
\subfloat[][]{\includegraphics[width=0.33\textwidth]{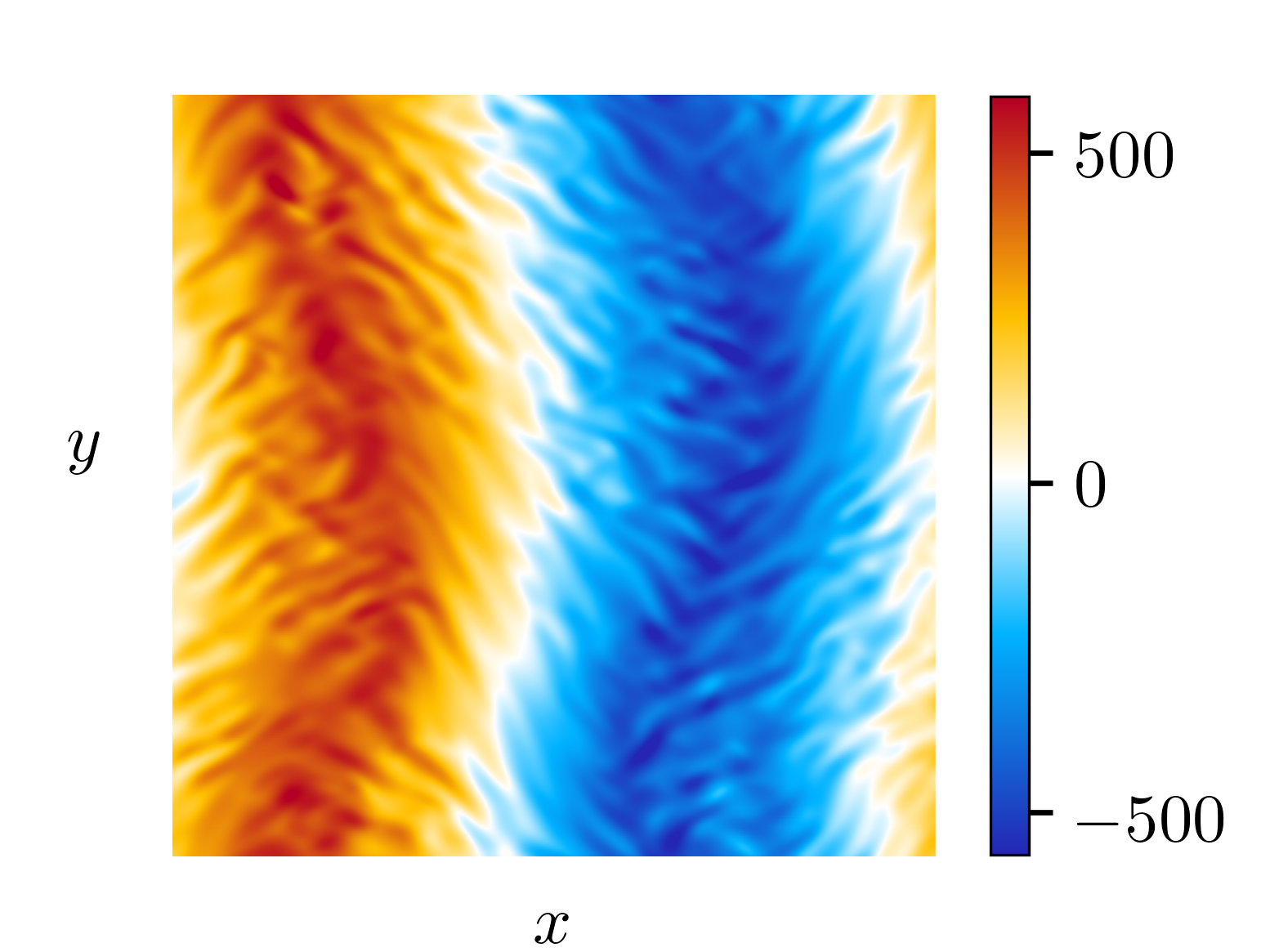}}      
\subfloat[][]{\includegraphics[width=0.33\textwidth]{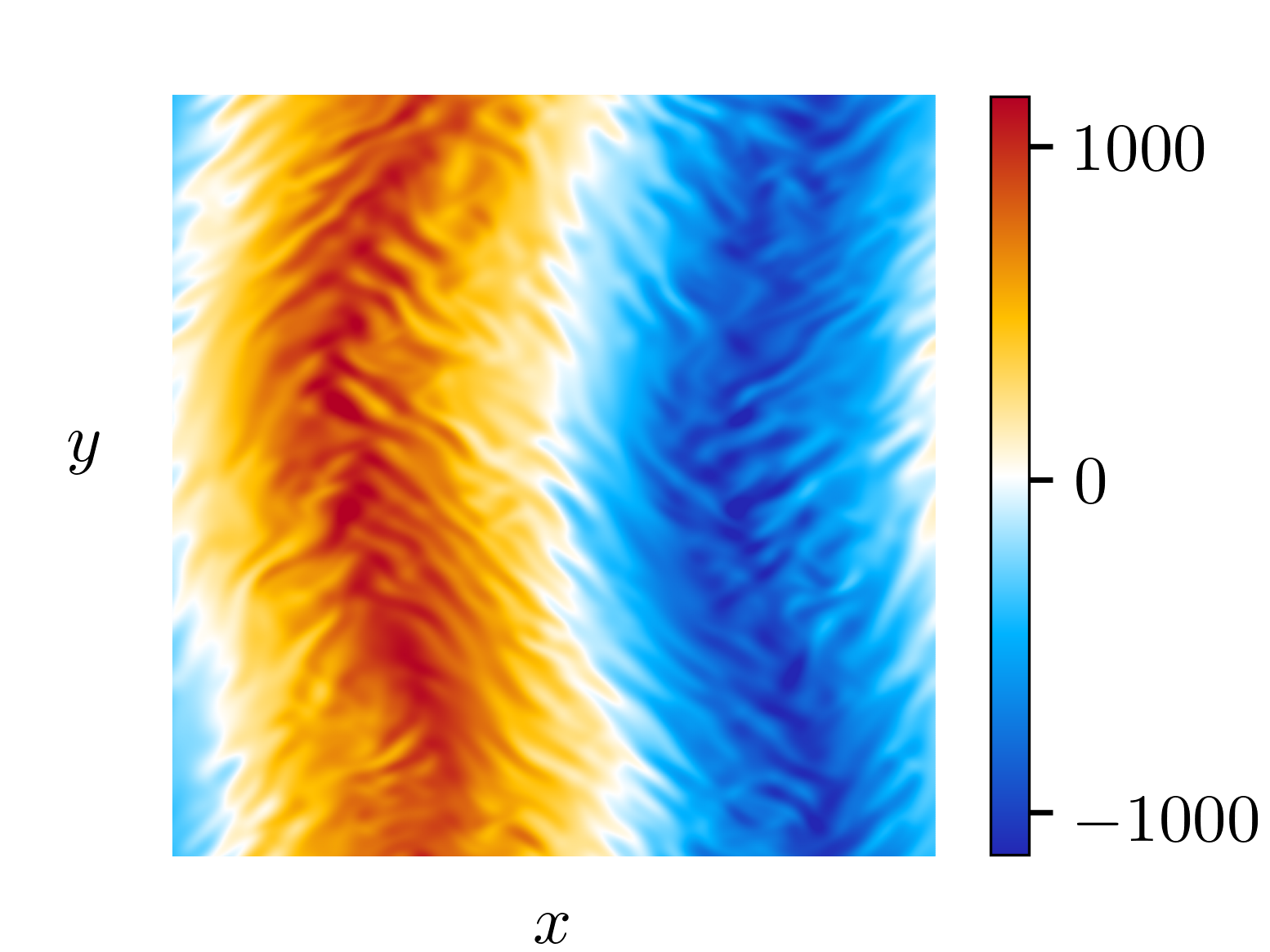}} 
   
	 \caption{Structure of the zonal flow from representative cases. (a,d) $Q=2\times 10^3$, $Ra=1\times 10^5$; (b,e) $Q=2\times 10^5$, $Ra=8\times 10^6$; (c,f) $Q=2\times10^6$, $Ra=7\times 10^7$. The top row shows instantaneous views of the zonal velocity, $\overline{v}^{y}$, and the bottom row shows instantaneous views of the $y$-component of the velocity, $v$, in the horizontal plane at depth $z \approx 0.5$.}
      \label{F:zonal}
\end{center}
\end{figure}

Figs. \ref{F:zonal}(a)-(c) show instantaneous views of $\mfv^y(x,z,t)$ in the $x$-$z$ plane for all three values of $Q$ with values of $Ra$ chosen such that the zonal flow is energetically dominant (relative to the convection). For all values of $Q$ we find that the zonal flow is aligned with the imposed magnetic field, and dominated by a $k_x=1$ structure at all depths. Corresponding top-down views of $v$ at depth $z \approx 0.5$ are shown in panels (d)-(f), where the meandering structure of this mean flow can be seen. Comparing the $y$-averaged flows shown in (a)-(c) with the full field in (d)-(f) shows how the action of $y$-averaging removes much of the small scale features present in (d)-(f).

\begin{figure}[]
 \begin{center}
      \subfloat[][]{\includegraphics[width=0.45\textwidth]{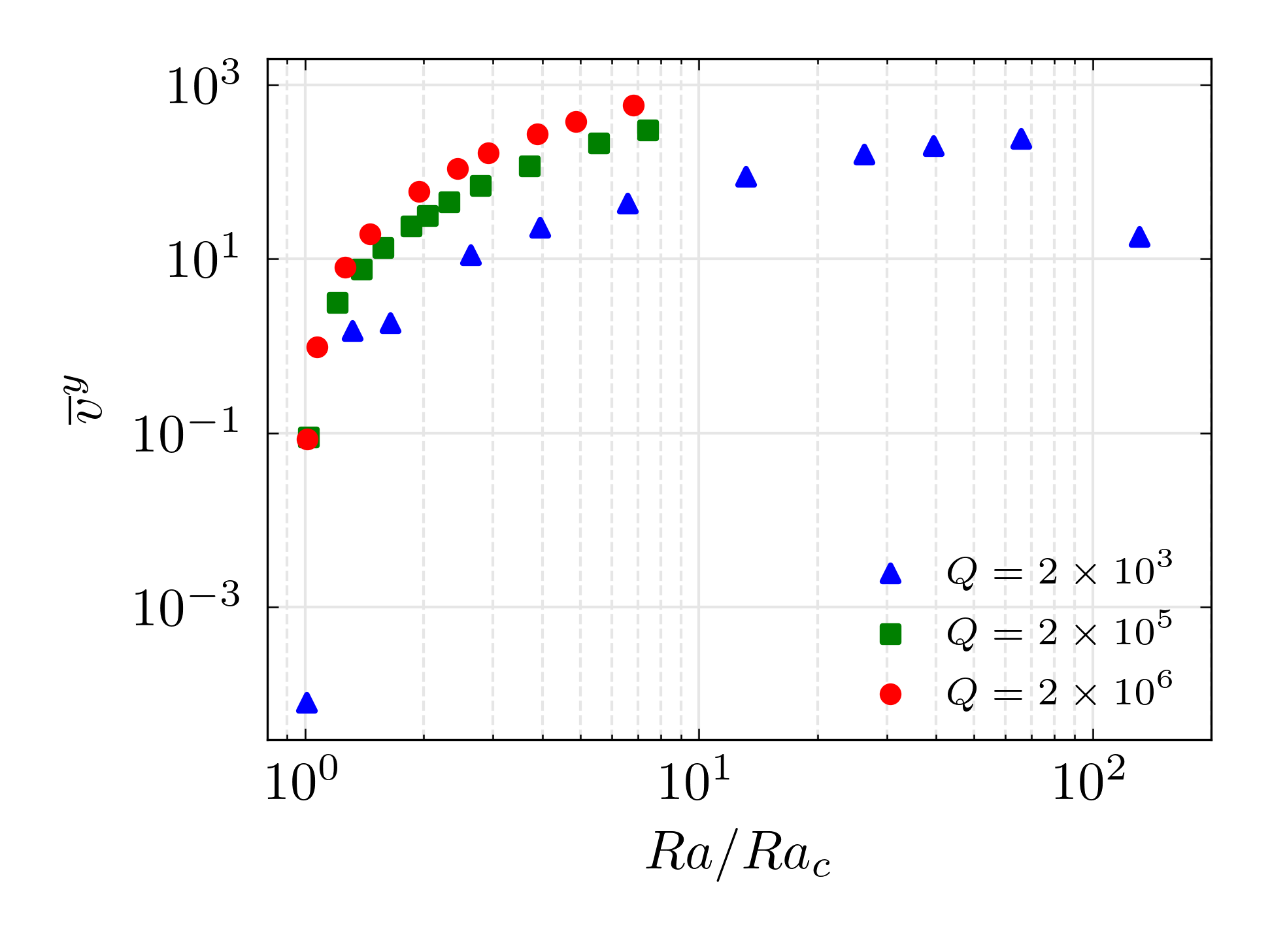}}  \qquad
      \subfloat[][]{\includegraphics[width=0.45\textwidth]{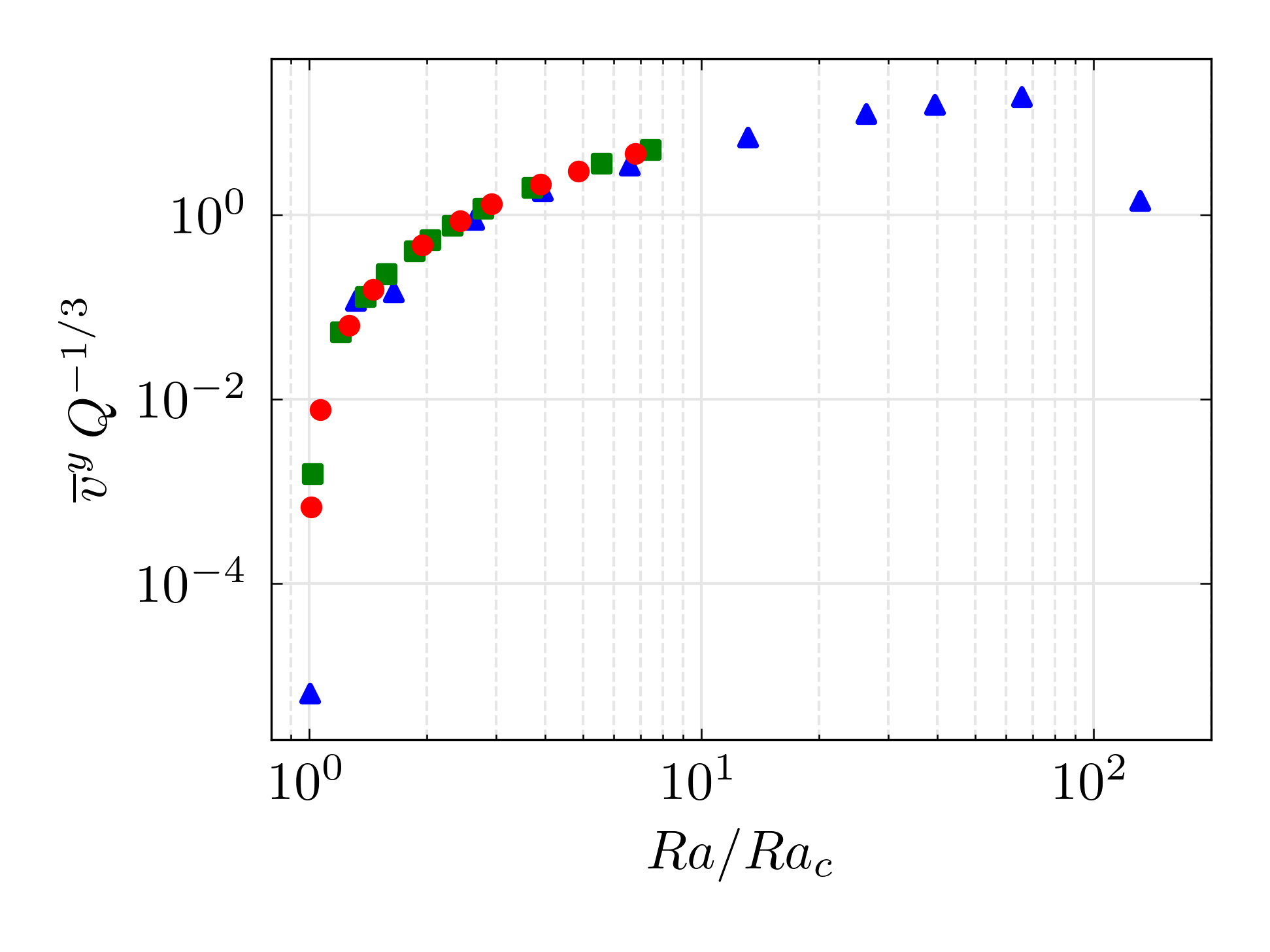}} 
	 \caption{Scaling behavior of the rms value for the zonal flow: (a) raw data; (b) rescaled data.}
      \label{F:zonal_scalings}
\end{center}
\end{figure}

The scaling behavior of the zonal velocity is shown in Fig.~\ref{F:zonal_scalings}(a) where rms values are plotted for all values of $Q$. We observe a trend of increasing magnitude with both increasing $Ra/Ra_c$ and increasing $Q$. As shown in Fig.~\ref{F:zonal_scalings}(b), the data can be collapsed by scaling the zonal velocity as $\overline{v}^y \sim Q^{1/3}$. As done previously for the horizontally averaged mean flows, the cause of this zonal flow scaling can be found by analyzing dominant balances in the zonal momentum equation. 

The alignment of the zonal flow with the imposed magnetic field suggests that the use of a non-orthogonal (i.e.~skewed) coordinate system is helpful. Such a coordinate system was used, for example, in Ref.~\cite{kJ06} for studying the asymptotic behavior of rapidly rotating convection with a tilted rotation axis. We follow Ref.~\cite{kJ06} and define the non-orthogonal coordinate system with unit vectors $\hx$, $\hy$, and $\he$ and coordinates $(\widetilde{x}, \widetilde{y}, \eta)$. The relevant transformations between the orthogonal and non-orthogonal variables are given by
\be
\widetilde{x} = x- \frac{\eta_1}{\eta_3}z, \qquad \widetilde{y}=y, \qquad \eta = \frac{1}{\eta_3} z
\ee
\be
\widetilde{u} = u-\frac{\eta_1}{\eta_3}w, \qquad \widetilde{v} = v, \qquad \widetilde{w} = \frac{1}{\eta_3}w .
\ee
Similarly we have $\widetilde{b}_y = b_y$.
The $y$-component of the momentum equation can now be written as
\be
    \frac{\partial v}{\partial t}  + \partial_{\widetilde{x}}\left(\widetilde{u}v\right) + \partial_y(v^2) + \partial_\eta\left(v\widetilde{w}\right) = -\partial_y P +\left(\frac{1}{\eta_3^2}\partial_{\widetilde{x}}^2+\partial_y^2 +\frac{1}{\eta_3^2}\partial^2_\eta-2\frac{\eta_1}{\eta_3}\partial^2_{ \widetilde{x} \eta}\right)v + Q\partial_\eta b_y.
\ee
Averaging in the $y$-direction, $\eta$-direction, and in time yields
\be
\overline{\overline{\widetilde{u}}^{y\eta} \partial_{\widetilde{x}} \overline{v}^{y\eta}}^t  +  \partial_{\widetilde{x}} \overline{\left(\widetilde{u}'v'\right)}^{y\eta t} = \left(\frac{1}{\eta_3} - 2\right)\partial_\eta \overline{v}^{yt}\bigg |^{1/\eta_3}_{0}+ \frac{1}{\eta_3^2}\partial_{\widetilde{x}}^2 \overline{v}^{y\eta t}.
\ee
Here the fluctuating terms are given by $v' = v - \overline{v}^{y\eta}$, etc. We note the absence of the Lorentz force in the above equation, indicating that it plays an indirect role in the zonal dynamics. The first term on the left side represents advection by the mean whereas the first term on the right side contains boundary terms associated with averaging the viscous force along $\eta$; the numerical simulations show that both of these terms are small in comparison to the two other terms present in the above equation. Advection by the mean is small because $\overline{\widetilde{u}}^{y}$ is not generally aligned with $\he$. In addition, because $\overline{v}^{y}$ is aligned with $\he$, the $\eta$-averaged value of $\partial_\eta \overline{v}^{y}$ is expected to be small.
Therefore, the largest terms are the Reynolds stress term and the viscous force,
\be
\partial_{\widetilde{x}} \overline{\lb \widetilde{u}' v' \rb}^{y\eta t} \sim  \frac{1}{\eta_3^2}\partial_{\widetilde{x}}^2 \overline{v}^{y\eta t}.
\ee
Furthermore, if we let $\overline{\ell}^{y \eta}$ denote a length scale associated with $y$ and $\eta$-averaged quantities, we find
\be
\mfv^{y \eta t} \sim \overline{\ell}^{y\eta} \, \overline{\lb \widetilde{u}' v' \rb}^{y\eta t} ,
\label{E:zonal scale}
\ee
where we have dropped the factor of $\eta_3^2$ since it is of order unity. In all simulations in which a zonal flow is observed we find that the zonal flow grows to fill the domain and is dominated by a $k_x = 1$ structure. As mentioned previously, as $Q$ is varied we fix the total number of horizontal critical wavelengths in the domain to be $n=10$. Since the critical wavelength changes with $Q$, i.e.~$\lambda_c \sim Q^{-1/6}$ as $Q$ becomes large, this implies that the horizontal (and vertical) scale of the zonal flow is also changing as we vary $Q$ in our simulations. Therefore, we can scale the characteristic zonal length scale as
\be
\overline{\ell}^{y\eta} \sim n \lambda_c \qquad \Rightarrow \qquad  \overline{\ell}^{y\eta}  \sim n Q^{-1/6}.
\label{E:zonal_length}
\ee
Combining this scaling with relationship \eqref{E:zonal scale} then yields
\be
\mfv^{y \eta t} = O \lb Q^{1/3} \rb,
\ee
where we have again used the numerical data that indicates $(\widetilde{u}', v') = O \lb Q^{1/4} \rb$.

The zonal length scaling relation \eqref{E:zonal_length} implies that for a fixed value of $Q$ the zonal flow magnitude increases linearly with the horizontal dimension (as quantified by $n$) of the simulation domain. Although not shown here, this scaling was confirmed in a set of numerical simulations in which $Q$ was fixed and $n$ was varied. This finding is in agreement with a previous investigation of the inverse kinetic energy cascade in rotating convection where the same scaling was observed \citep{sM21}.

\subsubsection{Relaxation Oscillations}

\begin{figure}
 \begin{center}
      \subfloat[][]{\includegraphics[width=0.45\textwidth]{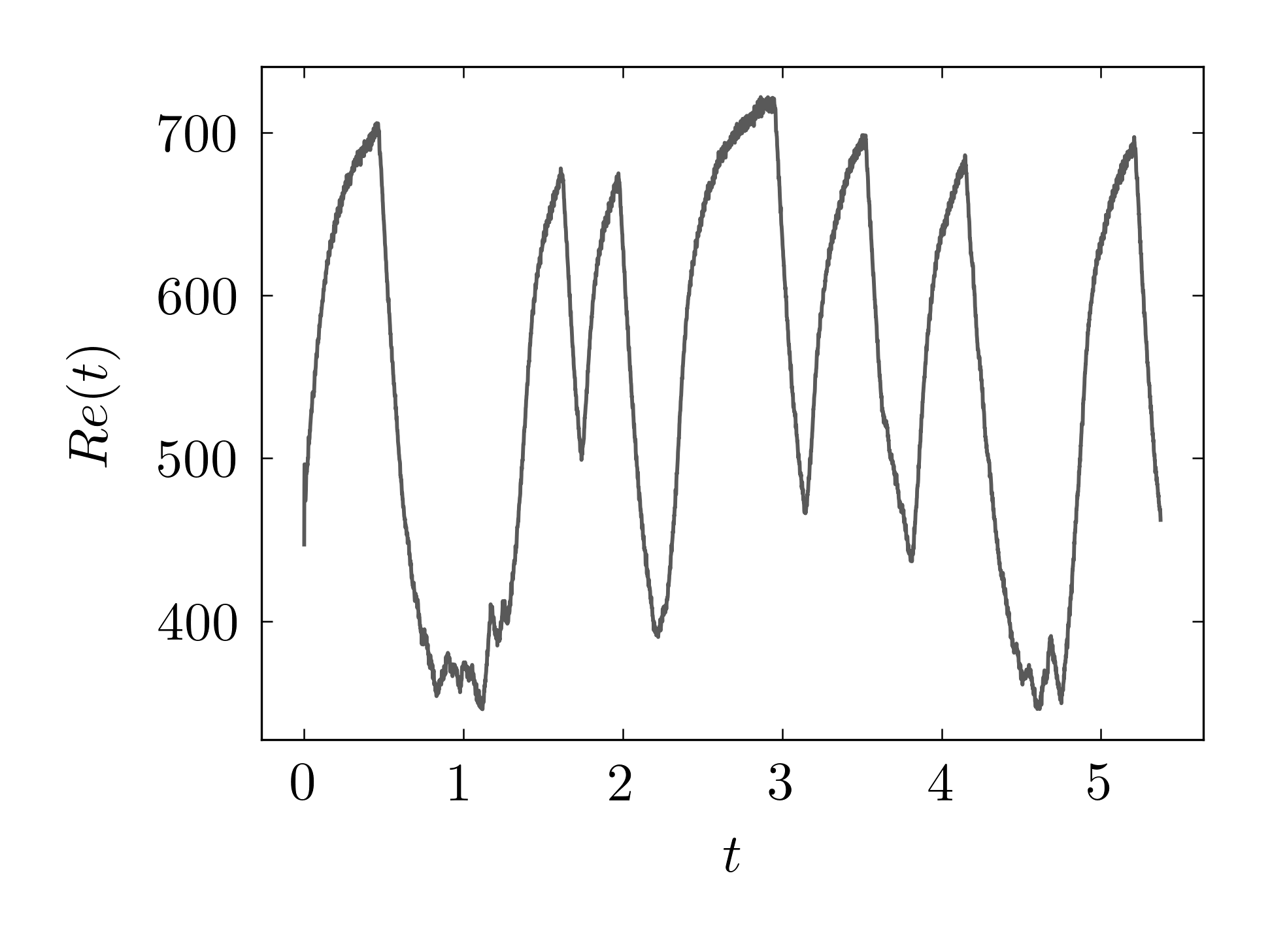}} \quad
      \subfloat[][]{\includegraphics[width=0.45\textwidth]{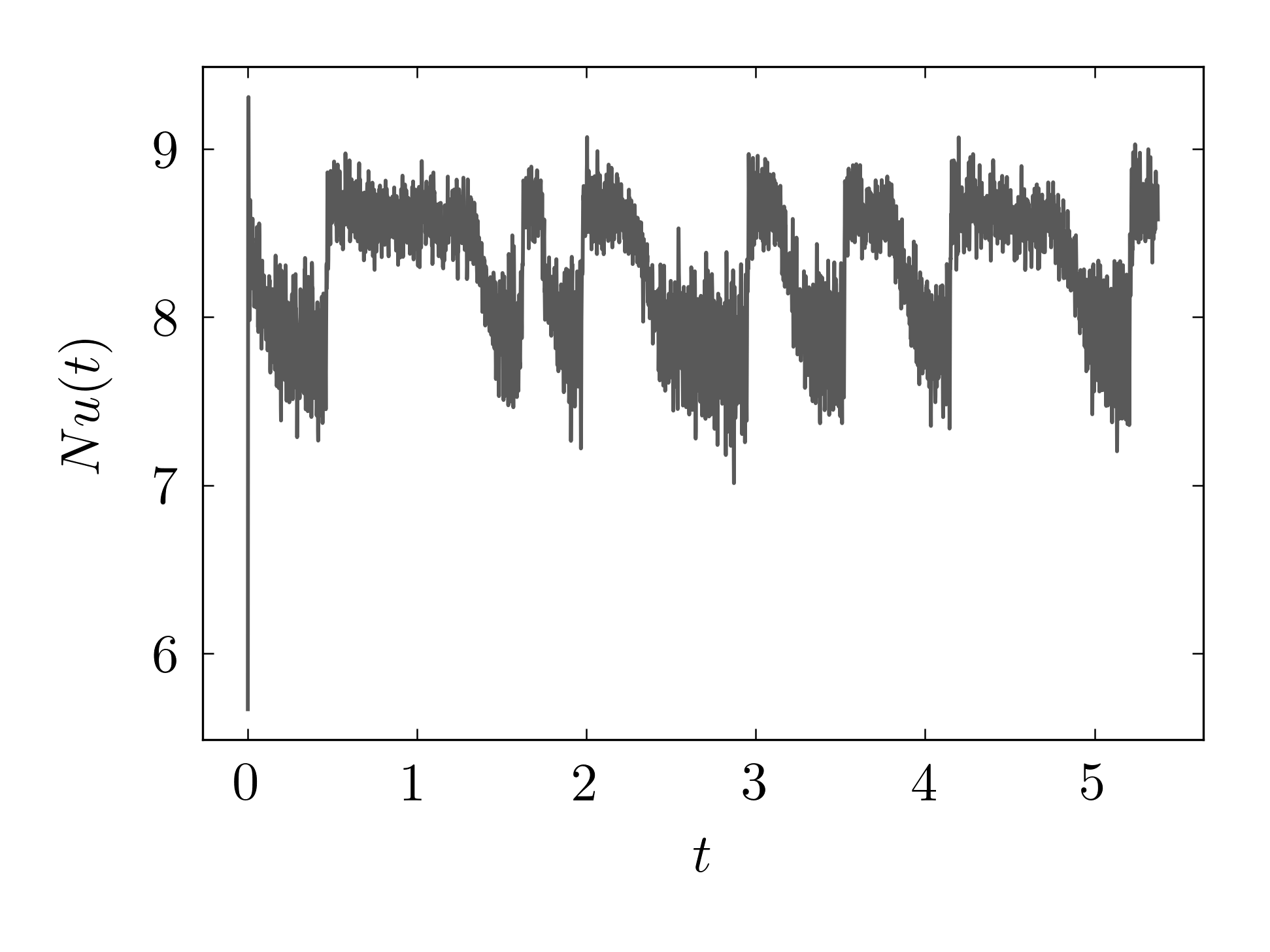}}
	 \caption{Times series data illustrating the temporal behavior of relaxation oscillations for $Q=2\times 10^5$ and $Ra=1.5\times 10^7$: (a) Reynolds number, $Re(t)$; (b) Nusselt number, $Nu(t)$.}
     \label{F:RelaxationOscillation}
\end{center}
\end{figure}

\begin{figure}
 \begin{center}
      \subfloat[][]{\includegraphics[width=0.4\textwidth]{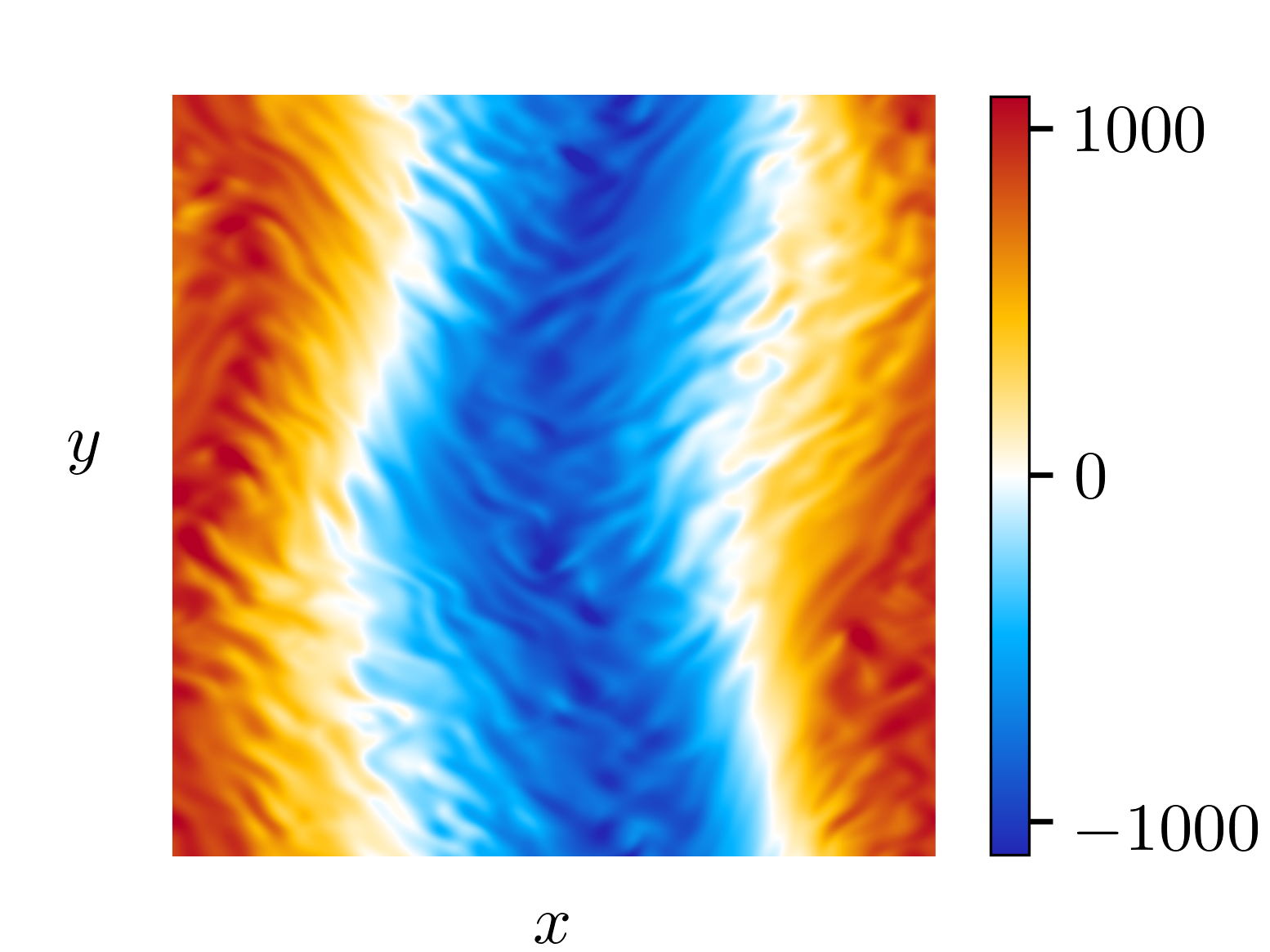}}
      \subfloat[][]{\includegraphics[width=0.4\textwidth]{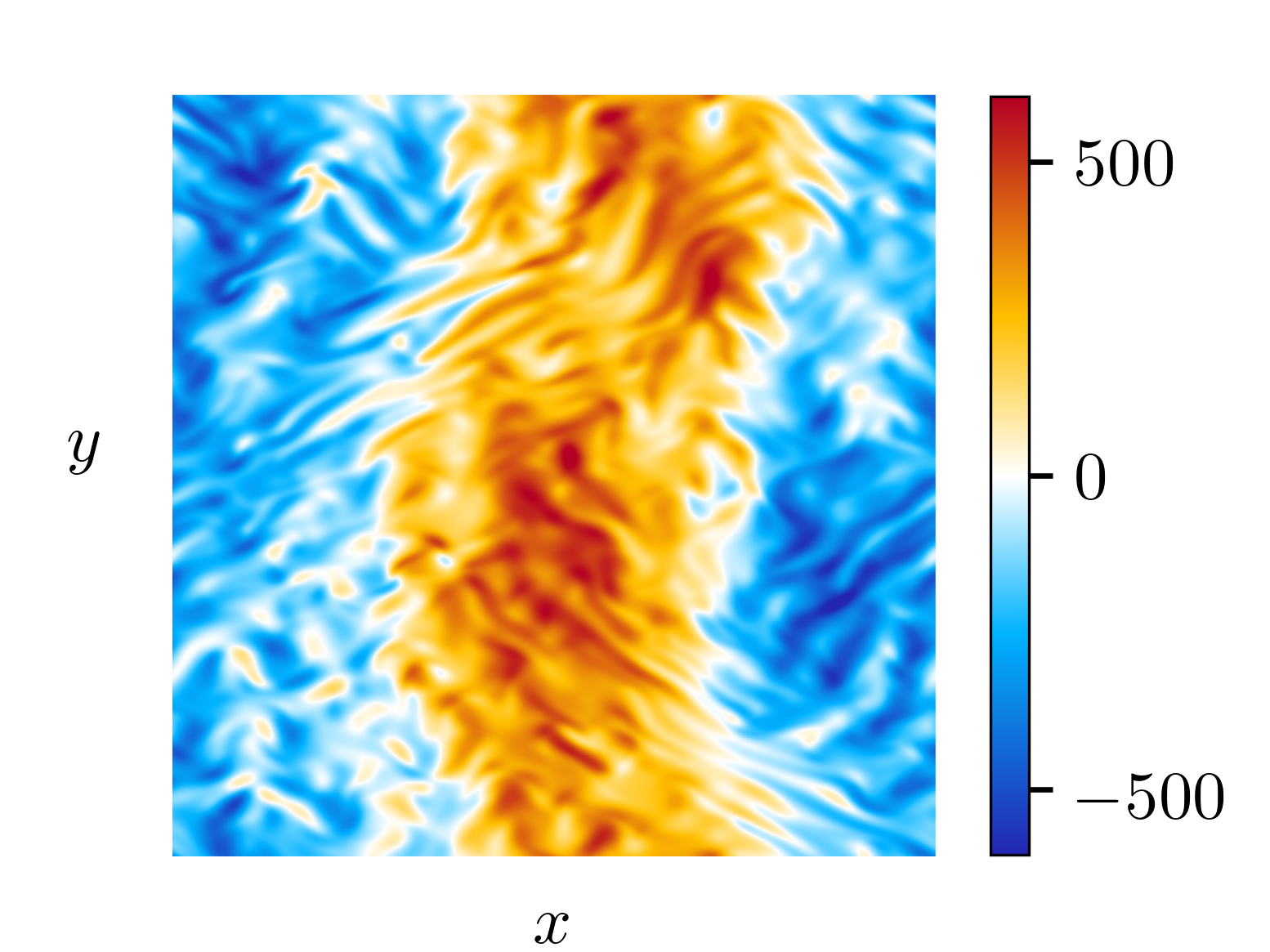}}
      \hspace{0mm}
      \subfloat[][]{\includegraphics[width=0.4\textwidth]{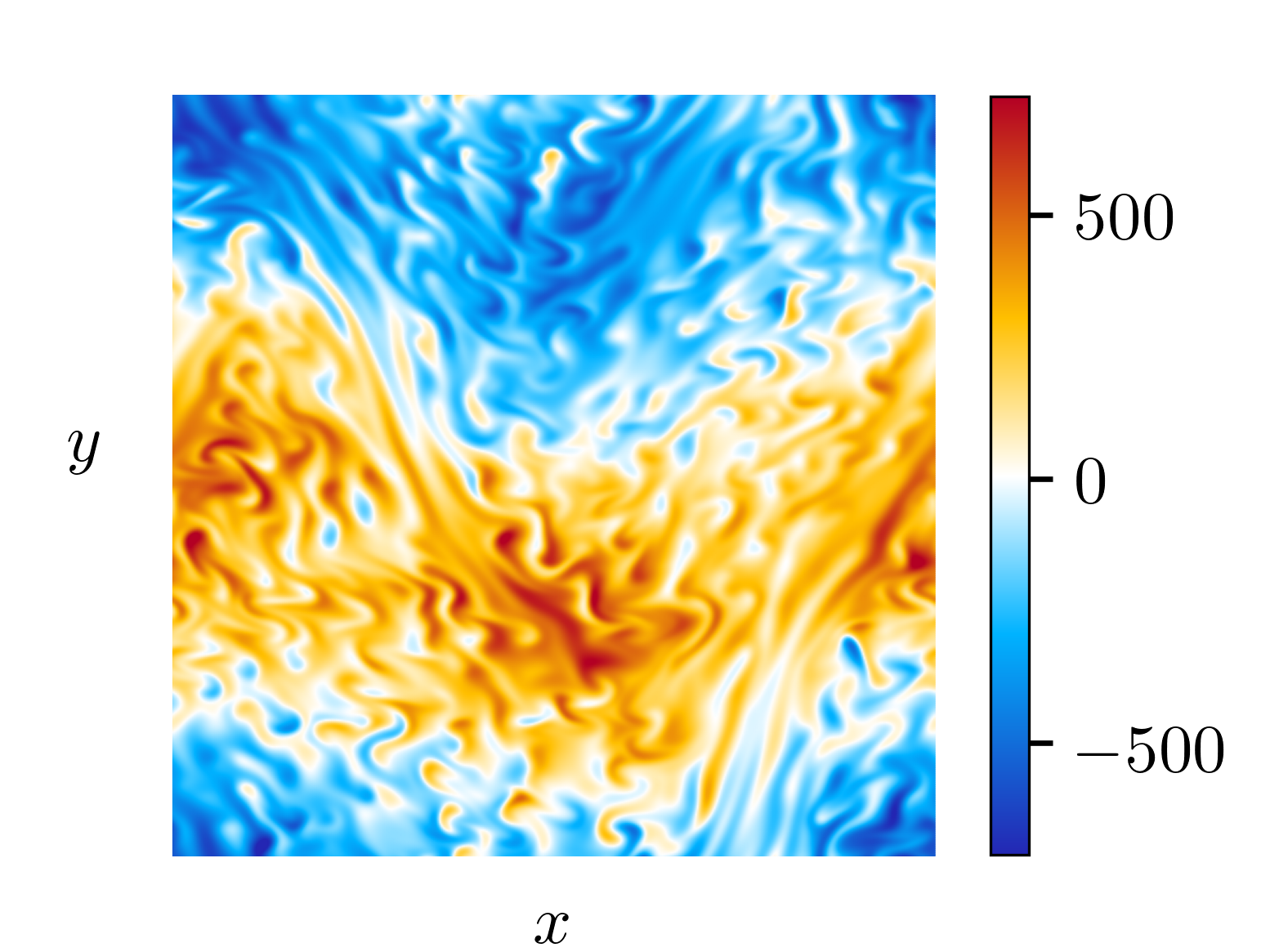}}
      \subfloat[][]{\includegraphics[width=0.4\textwidth]{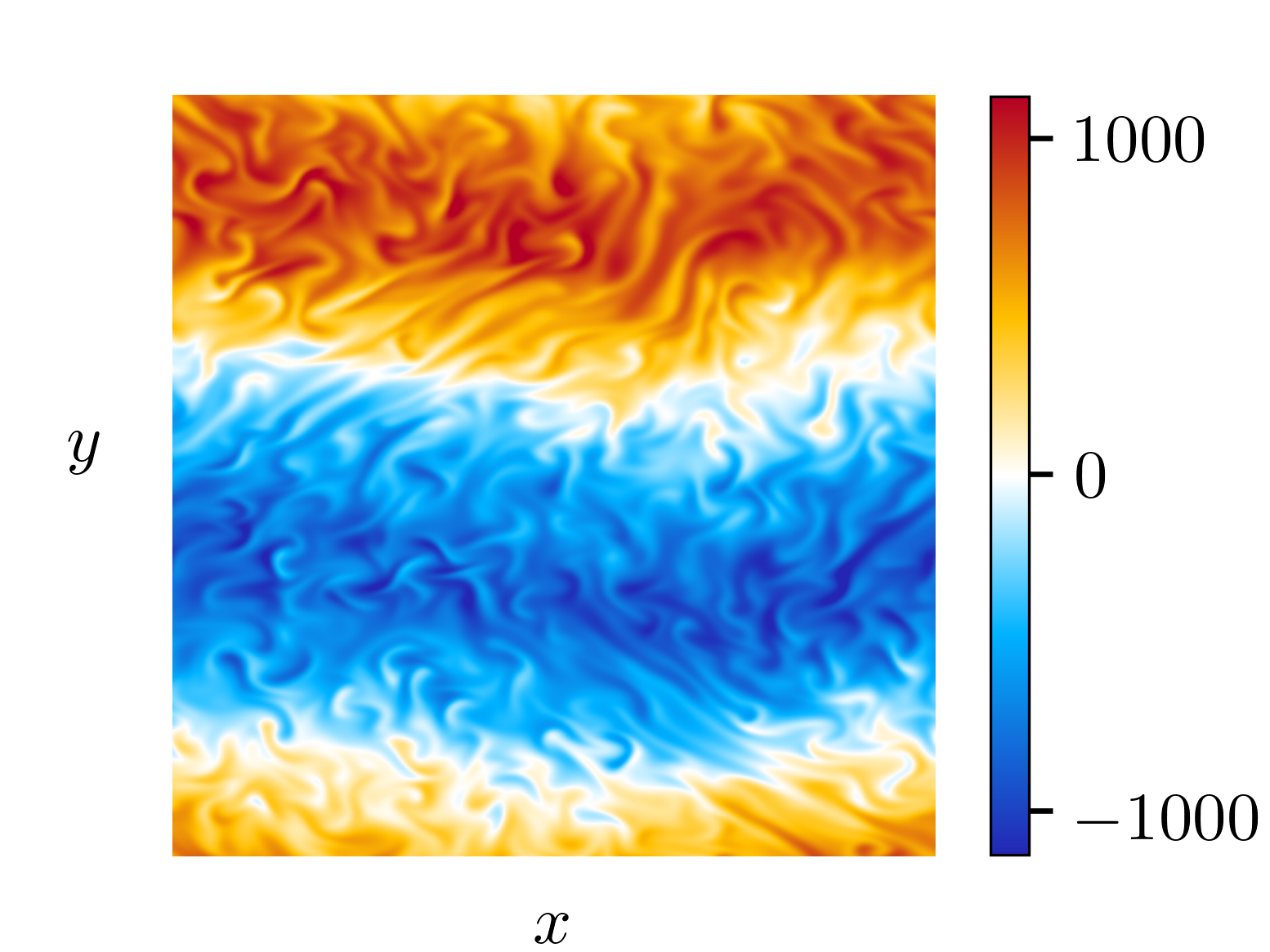}}
	 \caption{Instantaneous plots of the horizontal velocity components illustrating relaxation oscillations for $Q=2\times 10^5$ and $Ra=1.5\times 10^7$. The two plots on the left (a,c) are taken at time $t=5.200$, just prior to the peak in Reynolds number. The two plots on the right (b,d) are taken at time $t=5.228$, just after the Reynolds number reaches a peak. (a,b) $y$-velocity component; (c,d)  $x$-velocity component.}
     \label{F:RelaxationOscillationVelocity}
\end{center}
\end{figure}

Relaxation oscillations were observed in simulations with $Q=2\times 10^5$ and $10Ra_c \lesssim Ra \leq 30Ra_c$, where $30Ra_c$ was the highest supercritical Rayleigh number achieved for this value of $Q$. Relaxation oscillations were also found for a case with $Q=2\times10^6$ but no statistics were collected due to the long time integration required. The relaxation oscillations are characterized by a zonal flow magnitude that exhibits large oscillations in time. Figs.~\ref{F:RelaxationOscillation}(a) and (b) show the Reynolds number and Nusselt number, respectively, as a function of time for $Q=2\times10^5$ and $Ra=1.5\times10^7$. During times when the mean flow is strong (weak), the Nusselt number is generally smaller (larger) than the corresponding time-averaged value. Fig.~\ref{F:RelaxationOscillationVelocity} shows how the horizontal components of velocity change when the Reynolds number goes from increasing to decreasing. The $y$-component of the velocity decays rapidly and the $x$-component of velocity grows to be the dominant velocity component.

\section{Conclusion}

The direct numerical simulations reported here have shown that magnetoconvection with a tilted magnetic field (TMC) has both dynamic similarities, and key differences with magnetoconvection with a vertical field (VMC). For a meaningful comparison between these two cases we have chosen imposed field strengths that yield critical parameters that are identical to those used in an analogous VMC study \citep{mY19}; i.e.~the strength of the vertical component of the imposed magnetic fields are identical for our chosen values of $Q$. The simulations span dynamical regimes that are characterized as magnetically constrained, in which $Ha/Re_z \gg 1$, and those that might be considered magnetically influenced when $Ha/Re_z = O(1)$. Like VMC, we find that when $Ha/Re_z \gg 1$ and for sufficiently strong buoyancy forcing, the convective structures consist of anisotropic `columns' that align with the direction of the magnetic field. This alignment is no longer obvious when $Ha/Re_z \lesssim O(1)$.

The heat transport for TMC is qualitatively similar to VMC in which there is a rapid increase in the Nusselt number near the onset of convection that is associated with the formation of the thermal boundary layers, and the growth of $Nu$ with increasing $Ra/Ra_c$ slows beyond this regime. In contrast to VMC, we do not observe a clear power law scaling of the Nusselt number for TMC over our investigated range of parameter space. Ohmic dissipation dominates viscous dissipation in all simulations, including those in which $Ha/Re_z = O(1)$. However, viscous dissipation remains important and represents approximately 40$\%$ of the total dissipation as the Rayleigh number is increased.

Convective flow speeds, as characterized by the Reynolds number based on the vertical component of the velocity, $Re_z$, show behavior that is similar to VMC. However, it is unclear whether $Re_z$ exhibits power law behavior due to the limited range of parameter space accessible in the present simulations. We find that an asymptotic scaling of $Re_z \sim Q^{1/4}$ describes the $Q$-dependence of the convective flow speeds. This scaling is the same that was used in the asymptotic models of Refs.~\citep{kJ99c,kJ00}. Further investigation, particularly simulations with larger values of $Q$, is necessary to confirm the robustness of this scaling.

Mean flows, and associated mean magnetic fields, form for all values of $Q$ investigated. We find and study two distinct forms of mean flows, as characterized by the horizontal wavevector $\bf k$. Those mean flows in which $\mathbf{k}=(0,0)$ (i.e.~averaged over the horizontal plane) are dominated by the $x$-components, $\overline{u}$ and $\overline{b}_x$. The mean velocity field ($\overline{u}$) is generally aligned with the imposed external field in the sense that it flows in the positive direction above the mid-plane (i.e.~$z > 0$), and flows in the negative direction below the midplane ($z < 0$). The simulations show that the rms value of $\overline{u}$ is a decreasing function of $Q$; a balance analysis suggests that $\overline{u} = O\lb Q^{-1/2} \rb$, which is in agreement with the numerical findings. An identical scaling holds for $\overline{b}_x$, suggesting that this particular component of the mean fields becomes less significant dynamically as $Q$ is increased.

The energetically dominant mean flows are characterized by a meandering jet structure that tends to be aligned with the tilt of the imposed magnetic field. These mean flows are dominated by either $\mathbf{k}=(1,0)$ at lower values of $Ra/Ra_c$, or exhibit relaxation oscillations that shift between the $\mathbf{k}=(0,1)$  and $\mathbf{k}=(1,0)$ modes for sufficiently large $Ra/Ra_c$ and $Q > 2 \times 10^3$. We find that the `zonal' flows scale strongly with $Q$, i.e.~$\overline{v}^{y} \sim Q^{1/3}$; this scaling is the result of the $Q$-dependent Reynolds stresses and the fact that the zonal flow magnitude scales linearly with the horizontal dimensions of the simulation domain.  The zonal flows become energetically negligible for sufficiently large $Ra/Ra_c$, since then $Ha/Re_z \lesssim 1$ and the convection is no longer aligned with the imposed magnetic field. The relaxation oscillations evolve on the timescale of a large scale viscous diffusion unit and therefore require substantial computational resources to study; for this reason we were unable to study their characteristics in significant detail.

It is interesting to note that several phenomena observed here for convection in a tilted magnetic field are also observed for rotating convection \citep{lN19,lC20}. In convection with a tilted rotation vector, the fluid structures tend to align along the tilt axis of the rotation, much like how the fluid structures in this paper were aligned along the magnetic field. In addition, a strong shear flow arises in convection with a tilted rotation vector that closely resembles the mean flow found for convection in a tilted magnetic field. We anticipate that scalings for the strength of the mean flow and convective flow speeds can be found in tilted rotating convection, as was done in the present investigation for magnetoconvection.

The structure of the observed mean flows in the present work is a manifestation of the plane periodic geometry that was used. Obviously, confined geometries will yield different mean flows. However, the present work has shown that a robust inverse kinetic energy cascade is nevertheless present in TMC, and there is no a priori reason why such a cascade would not also be present in confined geometries such as cylinders or cubes. We might expect that the inverse cascade would manifest itself in the form of only large scale vortices in confined geometries, but future work is necessary to confirm this assertion. 

\section*{Acknowledgements}
The authors gratefully acknowledge funding from the National Science Foundation through grants EAR-1620649, EAR-1945270 and SPG-1743852. This work utilized the RMACC Summit supercomputer, which is supported by the National Science Foundation (awards ACI-1532235 and ACI-1532236), the University of Colorado Boulder, and Colorado State University. The Summit supercomputer is a joint effort of the University of Colorado Boulder and Colorado State University. This work used the Extreme Science and Engineering Discovery Environment (XSEDE) Stampede2 supercomputer at the Texas Advanced Computing Center (TACC) through allocation PHY180013. Volumetric renderings of the flow field were performed with VAPOR \citep{sL19}.

\clearpage
\newcommand{\jfm}{J. Fluid Mech.~}\newcommand{\mnras}{Mon. Not. Roy. Astron.
  Soc. }\newcommand{\jgr}{J. Geophys. Res.~}\newcommand{\araa}{Annu. Rev.
  Astron. Astrophys.~}\newcommand{\icarus}{Icarus }\newcommand{\aap}{Astron.
  Astrophys.~}\newcommand{\physscr}{Phys. Scripta }\newcommand{\ssr}{Space Sci.
  Rev.~}\newcommand{\pnas}{Proc. Nat. Acad. Sci.~}\newcommand{\ncom}{Nat.
  Comm.~}\newcommand{\njp}{New J. Phys.~}

\clearpage


\def\arraystretch{0.75}

\begin{table}
\begin{center}
\begin{tabular}{c c c c c c c}
$Q$ & $Ra$ & $Nu$ & $Re$ & $Re_z$ & $\Delta t$ & $N_x\times N_y \times N_z$ \\
\hline
2 $\times 10^3$ & 1.53 $\times 10^4$ & 1.012 $\pm$ 0.000 & 0.822 $\pm$ 0.000 & 0.719 $\pm$ 0.000 & $10^{-3}$ & $96\times 96\times 48$\\
2 $\times 10^3$ & 2 $\times 10^4$ & 1.30 $\pm$ 0.03 & 5.05 $\pm$ 0.19 & 4.16 $\pm$ 0.18 & $10^{-3}$ & $96\times 144\times 48$\\
2 $\times 10^3$ & 2.5 $\times 10^4$ & 1.54 $\pm$ 0.04 & 8.29 $\pm$ 0.24 & 6.61 $\pm$ 0.21 & $10^{-3}$ & $144\times 192\times 48$\\
2 $\times 10^3$ & 4 $\times 10^4$ & 2.19 $\pm$ 0.04 & 17.86 $\pm$ 0.24 & 11.34 $\pm$ 0.30 & $5 \times 10^{-4}$ & $144\times 192\times 48$\\
2 $\times 10^3$ & 6 $\times 10^4$ & 2.83 $\pm$ 0.05 & 30.98 $\pm$ 0.35 & 16.47 $\pm$ 0.36 & $2 \times 10^{-4}$ & $192\times 288\times 72$\\
2 $\times 10^3$ & 1 $\times 10^5$ & 3.75 $\pm$ 0.05 & 54.80 $\pm$ 0.37 & 25.12 $\pm$ 0.43 & $1 \times 10^{-4}$ & $288\times 288\times 144$\\
2 $\times 10^3$ & 2 $\times 10^5$ & 5.22 $\pm$ 0.07 & 103.43 $\pm$ 0.71 & 40.81 $\pm$ 0.66 & $4 \times 10^{-5}$ & $288\times 384\times 144$\\
2 $\times 10^3$ & 4 $\times 10^5$ & 7.06 $\pm$ 0.09 & 179.71 $\pm$ 0.79 & 63.67 $\pm$ 1.03 & $4 \times 10^{-5}$ & $384\times 576\times 144$\\
2 $\times 10^3$ & 6 $\times 10^5$ & 8.38 $\pm$ 0.09 & 229.58 $\pm$ 1.90 & 81.39 $\pm$ 1.14 & $2 \times 10^{-5}$ & $576\times 576\times 144$\\
2 $\times 10^3$ & 1 $\times 10^6$ & 10.35 $\pm$ 0.11 & 285.17 $\pm$ 2.17 & 110.31 $\pm$ 1.25 & $1 \times 10^{-5}$ & $768\times 768\times 192$\\
2 $\times 10^3$ & 2 $\times 10^6$ & 13.42 $\pm$ 0.12 & 214.17 $\pm$ 2.16 & 154.75 $\pm$ 1.39 & $5 \times 10^{-6}$ & $768\times 768\times 288$\\
\hline
2 $\times 10^5$ & 1.1 $\times 10^6$ & 1.024 $\pm$ 0.006 & 2.11 $\pm$ 0.29 & 2.01 $\pm$ 0.28 & $10^{-5}$ & $96\times 144\times 96$\\
2 $\times 10^5$ & 1.3 $\times 10^6$ & 1.17 $\pm$ 0.02 & 8.13 $\pm$ 0.52 & 7.16 $\pm$ 0.55 & $10^{-5}$ & $96\times 144\times 96$\\
2 $\times 10^5$ & 1.5 $\times 10^6$ & 1.31 $\pm$ 0.02 & 13.84 $\pm$ 0.54 & 10.73 $\pm$ 0.67 & $10^{-5}$ & $144\times 192\times 96$\\
2 $\times 10^5$ & 1.7 $\times 10^6$ & 1.46 $\pm$ 0.03 & 20.69 $\pm$ 0.62 & 14.33 $\pm$ 0.79 & $10^{-5}$ & $144\times 192\times 96$\\
2 $\times 10^5$ & 2 $\times 10^6$ & 1.72 $\pm$ 0.03 & 32.44 $\pm$ 0.63 & 19.53 $\pm$ 0.92 & $10^{-5}$ & $192\times 192\times 96$\\
2 $\times 10^5$ & 2.2 $\times 10^6$ & 1.87 $\pm$ 0.04 & 41.26 $\pm$ 0.64 & 22.65 $\pm$ 0.96 & $10^{-5}$ & $192\times 192\times 96$\\
2 $\times 10^5$ & 2.5 $\times 10^6$ & 2.10 $\pm$ 0.04 & 55.79 $\pm$ 0.70 & 27.05 $\pm$ 1.02 & $10^{-5}$ & $192\times 192\times 96$\\
2 $\times 10^5$ & 3 $\times 10^6$ & 2.45 $\pm$ 0.04 & 81.53 $\pm$ 0.76 & 34.23 $\pm$ 1.11 & $10^{-5}$ & $192\times 288\times 96$\\
2 $\times 10^5$ & 4 $\times 10^6$ & 3.11 $\pm$ 0.07 & 134.82 $\pm$ 1.32 & 47.60 $\pm$ 1.70 & $10^{-5}$ & $288\times 384\times 96$\\
2 $\times 10^5$ & 6 $\times 10^6$ & 4.24 $\pm$ 0.13 & 240.66 $\pm$ 1.88 & 70.43 $\pm$ 3.00 & $10^{-5}$ & $288\times 384\times 144$\\
2 $\times 10^5$ & 8 $\times 10^6$ & 5.18 $\pm$ 0.16 & 344.60 $\pm$ 2.37 & 90.74 $\pm$ 3.77 & $5\times 10^{-6}$ & $576\times 576\times 192$\\
2 $\times 10^5$ & 1 $\times 10^7$ & 6.55 $\pm$ 0.27 & 323.43 $\pm$ 72.48 & 112.83 $\pm$ 3.82 & $5\times 10^{-6}$ & $576\times 576\times 192$\\
2 $\times 10^5$ & 1.5 $\times 10^7$ & 8.30 $\pm$ 0.38 & 548.14 $\pm$ 114.34 & 154.27 $\pm$ 6.46 & $2\times 10^{-6}$ & $576\times 576\times 288$\\
2 $\times 10^5$ & 3 $\times 10^7$ & 12.67 $\pm$ 0.40 & 921.73 $\pm$ 195.51 & 251.40 $\pm$ 6.44 & $2\times 10^{-6}$ & $576\times 768\times 288$\\
\hline
2 $\times 10^6$ & 1.04 $\times 10^7$ & 1.015 $\pm$ 0.000 & 2.651 $\pm$ 0.000 & 2.575 $\pm$ 0.000 & $10^{-6}$ & $96\times 144\times 144$\\
2 $\times 10^6$ & 1.1 $\times 10^7$ & 1.07 $\pm$ 0.01 & 6.75 $\pm$ 0.37 & 6.54 $\pm$ 0.35 & $10^{-6}$ & $96\times 144\times 144$\\
2 $\times 10^6$ & 1.3 $\times 10^7$ & 1.21 $\pm$ 0.01 & 16.17 $\pm$ 0.75 & 13.33 $\pm$ 0.76 & $10^{-6}$ & $144\times 192\times 144$\\
2 $\times 10^6$ & 1.5 $\times 10^7$ & 1.37 $\pm$ 0.02 & 29.24 $\pm$ 0.91 & 20.16 $\pm$ 1.20 & $10^{-6}$ & $144\times 192\times 144$\\
2 $\times 10^6$ & 2 $\times 10^7$ & 1.81 $\pm$ 0.03 & 73.14 $\pm$ 1.03 & 36.13 $\pm$ 1.59 & $10^{-6}$ & $144\times 288\times 144$\\
2 $\times 10^6$ & 2.5 $\times 10^7$ & 2.25 $\pm$ 0.04 & 125.88 $\pm$ 1.29 & 51.13 $\pm$ 2.28 & $10^{-6}$ & $192\times 288\times 144$\\
2 $\times 10^6$ & 3 $\times 10^7$ & 2.64 $\pm$ .06 & 184.21 $\pm$ 1.91 & 63.98 $\pm$ 3.06 & $10^{-6}$ & $288\times 288\times 144$\\
2 $\times 10^6$ & 4 $\times 10^7$ & 3.41 $\pm$ .07 & 305.61 $\pm$ 2.88 & 89.59 $\pm$ 4.62 & $10^{-6}$ & $384\times 384\times 192$\\
2 $\times 10^6$ & 5 $\times 10^7$ & 4.14 $\pm$ .10 & 422.72 $\pm$ 5.20 & 113.64 $\pm$ 5.24 & $10^{-6}$ & $384\times 576\times 192$\\
2 $\times 10^6$ & 7 $\times 10^7$ & 5.46 $\pm$ .14 & 663.70 $\pm$ 3.09 & 156.39 $\pm$ 8.06 & $10^{-6}$ & $576\times 576\times 288$\\
\hline
\end{tabular}
\end{center}
\caption{Details of the simulations. $Q$ is the Chandraskehar number, $Ra$ is the Rayleigh number, $Nu$ is the Nusselt number, $Re$ is the Reynolds number, $Re_z$ is the Reynolds number based only on the vertical component of the velocity, $\Delta t$ is the timestep size and $N_x \times N_y \times N_z$ denotes the physical space resolution. The thermal Prandtl number is fixed at $Pr=1$ for all simulations.}
\label{tab:Data}
\end{table}

\end{document}